\documentclass[10pt,letterpaper]{article}
\usepackage[top=0.85in,left=2.75in,footskip=0.75in]{geometry}
\usepackage{changepage}
\usepackage[utf8x]{inputenc}
\usepackage{textcomp,marvosym}
\usepackage{amsmath,amssymb}
\usepackage{algorithm}
\usepackage[noend]{algpseudocode}
\usepackage{cite}
\usepackage{nameref,hyperref}
\usepackage{microtype}
\raggedright
\usepackage[aboveskip=1pt,labelfont=bf,labelsep=period,justification=raggedright,singlelinecheck=off]{caption}

\makeatletter
\renewcommand{\@biblabel}[1]{\quad#1.}
\makeatother
\date{}
\usepackage{lastpage,fancyhdr,graphicx}

\usepackage{longtable}
\usepackage{lscape}
\usepackage{array}

\begin{document}
\vspace*{0.2in}
\begin{flushleft}
{\Large
\textbf\newline{The Five Factor Model of personality \\ and evaluation of drug consumption risk} 
}
\newline
\\
Elaine Fehrman\textsuperscript{1},
Evgeny M. Mirkes\textsuperscript{2},
Awaz K. Muhammad\textsuperscript{2,4},
Vincent Egan\textsuperscript{3},
Alexander N. Gorban\textsuperscript{2*},\\
\bigskip
\textbf{1} Men's Personality Disorder and National Women's Directorate, Rampton Hospital, Retford, Nottinghamshire, United Kingdom\\
\textbf{2} Department of Mathematics, University of Leicester, Leicester, United Kingdom\\
\textbf{3} Department of Psychiatry and Applied Psychology, University of Nottingham, Nottingham, United Kingdom\\
\textbf{4} Department of Mathematics, University of Salahaddin, Erbil, Kurdistan Region, Iraq
\bigskip\\
* ag153@le.ac.uk

\end{flushleft}
\section*{Abstract}
The problem of evaluating an individual's risk of drug consumption and misuse is highly important. An online survey methodology was employed to collect data including Big Five personality traits (NEO-FFI-R), impulsivity (BIS-11), sensation seeking (ImpSS), and demographic information. The data set contained information on the consumption of 18 central nervous system psychoactive drugs.
Correlation analysis demonstrated the existence of groups of drugs with strongly correlated consumption patterns. Three correlation pleiades were identified, named by the central drug in the pleiade: ecstasy, heroin,  and benzodiazepines pleiades. An exhaustive search was performed to select the most effective subset of input features and data mining methods to classify users and non-users for each drug and pleiad. A number of classification methods were employed (decision tree, random forest, $k$-nearest neighbors, linear discriminant analysis, Gaussian mixture, probability density function estimation, logistic regression and na{\"i}ve Bayes) and the most effective classifier was selected for each drug. The quality of classification was surprisingly high with sensitivity and specificity (evaluated by leave-one-out cross-validation)  being greater than 70\% for almost all classification tasks.  The best results with sensitivity and specificity being greater than 75\% were achieved for cannabis, crack, ecstasy, legal highs, LSD, and volatile substance abuse (VSA).


\section*{Introduction}

After Popper, it is a commonplace opinion in philosophy of science that the `value' of definitions besides mathematics is generally low. Nevertheless, for many more practical needs, from jurisprudence to health planning, the definitions are necessary to impose theoretical boundaries of a subject despite of their incompleteness and temporariness. This highly applies to definitions of drugs and drug use. Following the standard definitions \cite{Kleiman11}, a {\em drug} is a `chemical that influences biological function (other than by providing nutrition or hydration)'. A {\em psychoactive drug} is a `drug whose influence is in a part on mental functions'. An {\em abusable   psychoactive drug} is a `drug whose mental effects are sufficiently pleasant or interesting or helpful that some people choose to take it for a reason other than to relieve a specific malady'. In our study we use the term `drug' for  abusable   psychoactive drug regardless of whether it's illicit or not.

Drug use is a risk behaviour that does not happen in isolation; it constitutes an important factor for increasing risk of poor health, along with earlier mortality and morbidity, and has significant consequences for society \cite{McGinnis93,Sutina13}. Drug consumption and addiction constitutes a serious problem globally. It includes numerous risk factors, which are defined as any attribute, characteristic, or event in the life of an individual that increases the probability of drug consumption. A number of factors are correlated with initial drug use including psychological, social, individual, environmental, and economic factors \cite{Cleveland08,Ventura14,WHO04}. These factors are likewise associated with a number of personality traits \cite{Dubey10,Bogg04}. While legal drugs such as sugar, alcohol and tobacco are probably responsible for far more premature death than illegal recreational drugs \cite{Beaglehole11}, the social and personal consequences of recreational drug use can be highly problematic \cite{Bickel14}.

Psychologists have largely agreed that the personality traits of the Five Factor Model (FFM) are the most comprehensive and adaptable system for understanding human individual differences \cite{Costa92}. The FFM comprises Neuroticism (N), Extraversion (E), Openness to Experience (O), Agreeableness (A), and Conscientiousness (C).

A number of studies have illustrated that personality traits are associated to drug consumption. Roncero et al \cite{Roncero14} highlighted the importance of the relationship between high N and the presence of psychotic symptoms following cocaine-induced drug consumption. Vollrath \& Torgersen \cite{Vollrath02} observed that the personality traits of N, E, and C are highly correlated with hazardous health behaviours. A low score of C, and high score of E or high score of N correlate strongly with multiple risky health behaviours. Flory et al \cite{Flory02} found alcohol use to be associated with lower A and C, and higher E. They found also that lower A and C, and higher O are associated with marijuana use. Sutina et al \cite{Sutina13} demonstrated that the relationship between low C and drug consumption is moderated by poverty; low C is a stronger risk factor for illicit drug usage among those with relatively higher socioeconomic status. They found that high N, and low A and C are associated with higher risk of drug use (including cocaine, crack, morphine, codeine, and heroin). It should be mentioned that high N is positively associated with many other addictions like Internet addiction, exercise addiction, compulsive buying, and study addiction \cite{Andreassen2013}.

An individual's personality profile plays a role in becoming a drug user. Terracciano et al \cite{Terracciano08} demonstrated that the personality profiles for the users and non-users of nicotine, cannabis, cocaine, and heroin are associated with a FFM of personality samples from different communities. They also highlight the links between the consumption of these drugs and low C. Turiano et al \cite{Turiano12} found a positive correlation between N and O, and drug use, while, increasing scores for C and A decreases risk of drug use. Previous studies demonstrated that participants who use drugs including alcohol and nicotine have a strong positive correlation between A and C and a strong negative correlation for each of these factors with N \cite{Stewart00,Haider02}. Three high-order personality traits are proposed as endophenotypes for substance use disorders: Positive Emotionality, Negative Emotionality, and Constraint \cite{Belcher2016}.

The statistical characteristics of groups of drug users and non-users have been studied by many authors (see, for example, Terracciano et al \cite{Terracciano08}). They found that the personality profile for the users and non-users of tobacco, marijuana, cocaine, and heroin are associated with a higher score on N and a very low score for C. Sensation seeking is also higher for users of recreational drugs \cite{Kopstein01}. The problem of risk evaluation for individuals is much more complex. This was explored very recently by Yasnitskiy et al \cite{Yasnitskiy15}, Valeroa et al \cite{Valeroa14} and Bulut \& Bucak \cite{Bulut14}. Both individual and environmental factors predict substance use and different patterns of interaction among these factors may have different implications \cite{Rioux2016}. Age is a very important attribute for diagnosis and prognosis of substance use disorders. In particular, early adolescent onset of substance use is a robust predictor of future substance use disorders \cite{Weissman2015}.

Valeroa et al \cite{Valeroa14} evaluated the individual risk of drug consumption for alcohol, cocaine, opiates, cannabis, ecstasy and amphetamines. Input data were collected using the Spanish version of the Zuckerman-Kuhlman Personality Questionnaire (ZKPQ). Two samples were used in this study. The first one consisted of 336 drug dependent psychiatric patients of one hospital. The second sample included 486 control individuals. The authors used a decision tree as a tool to identify the most informative attributes. The sensitivity of 40\% and the specificity of 94\% were achieved for the training set. The main purpose of this research was to test if predicting drug consumption was possible and to identify the most informative attributes using data mining methods. The  decision tree methods were applied to explore the differential role of personality profiles in drug consumer and control individuals. The two personality factors, Neuroticism and anxiety and the ZKPQ's Impulsivity, were found to be most relevant for drug consumption prediction. Low sensitivity (40\%) does not provide application of this decision tree to real life problems.

In our study we tested the associations with personality traits  for different types of drugs separately using the Revised NEO Five-Factor Inventory (NEO-FFI-R) \cite{McCrae04}, the Barratt Impulsiveness Scale version 11 (BIS-11) \cite{Stanford09}, and the Impulsivity Sensation-Seeking scale (ImpSS) \cite{Zuckerman94} to assess impulsivity and sensation-seeking respectively.

Bulut \& Bucak \cite{Bulut14} detected a risk rate for teenagers in terms of percentage who are at high risk without focusing on specific addictions. The attributes were collected by an original questionnaire, which included 25 questions. The form was filled in by 671 students. The first 20 questions asked about the teenagers' financial situation, temperament type, family and social relations, and cultural preferences. The last five questions were completed by their teachers and concerned the grade point average of the student for the previous semester according to a 5-point grading system, whether the student had been given any disciplinary punishment so far, if the student had alcohol problems, if the student smoked cigarettes or used tobacco products, and whether the student misused substances.
In Bulut et al's study there are five risk classes as outputs. The authors diagnosed teenagers risk to be a drug abuser using seven types of classification algorithms: $k$-nearest neighbor, ID3 and C4.5 decision tree based algorithms, na{\"i}ve Bayes classifier, na{\"i}ve Bayes/decision trees hybrid approach, one-attribute-rule, and projective adaptive resonance theory. The classification accuracy of the best classifier was reported as 98\%.

Yasnitskiy et al \cite{Yasnitskiy15}, attempted to evaluate the individual’s risk of illicit drug consumption and to recommend the most efficient changes in the individual’s social environment to reduce this risk. The input and output features were collected by an original questionnaire. The attributes consisted of: level of education, having friends who use drugs, temperament type, number of children in the family, financial situation, alcohol drinking and smoking, family relations (cases of physical, emotional and psychological abuse, level of trust and happiness in the family). There were 72 participants. A neural network model was used to evaluate the importance of attributes for diagnosis of the tendency to drug addiction. A series of virtual experiments was performed  for several test patients (drug users) to evaluate how it is possible to control the propensity for drug addiction.  The most effective change of social environment features was predicted for each patient. The recommended changes depended on the personal profile, and significantly varied for different patients. This approach produced individual bespoke advice for decreasing drug dependence.

In our study, the database was collected by an anonymous online survey methodology by Elaine Fehrman yielding 2051 respondents. The database is available online \cite{FehrmanData2016}.  Twelve  attributes are known  for each respondent: personality measurements which include N, E, O, A, and C scores from NEO-FFI-R, impulsivity (Imp.) from (BIS-11), sensation seeking (SS) from (ImpSS), level of education (Edu.), age, gender, country of residence, and ethnicity. The data set contains information on the consumption of 18 central nervous system psychoactive drugs including alcohol, amphetamines, amyl nitrite, benzodiazepines, cannabis, chocolate, cocaine, caffeine, crack, ecstasy, heroin, ketamine, legal highs, LSD, methadone, magic mushrooms (MMushrooms), nicotine, and Volatile Substance Abuse (VSA), and one fictitious drug (Semeron) which was introduced to identify over-claimers.  Participants selected for each drug either they never used this drug, used it over a decade ago, or in the last decade, year, month, week, or day.

Participants were asked about substances, which were classified as central nervous system depressants, stimulants, or hallucinogens. The depressant drugs comprised alcohol, amyl nitrite, benzodiazepines, tranquilizers, gamma-hydroxybutyrate solvents and inhalants, and opiates such as heroin and methadone/prescribed opiates. The stimulants consisted of amphetamines, nicotine, cocaine powder, crack cocaine, caffeine, and chocolate. Although chocolate contains caffeine, data for chocolate was measured separately, given that it may induce parallel psychopharmacological and behavioural effects in individuals congruent to other addictive substances \cite{Bruinsma99}. The hallucinogens included cannabis, ecstasy, ketamine, LSD, and magic mushrooms. Legal highs such as mephedrone, salvia, and various legal smoking mixtures were also measured.

We use four different definitions of `drug users' based on the recency of use. Firstly, two isolated categories (`Never used' and `Used over a decade ago') are placed into the class of non-users, and all other categories are merged to form the class of users. Secondly, we merge the categories `Used in last decade', `Used over a decade ago' and `Never used' into the  group of non-users and place four other categories (`Used in last year-month-week-day') into group of users. This classification  is called `year-based' one. Also `month-based' and `week-based'  user/non-user separations are considered.

The objective of the study was to assess the potential effect of big five personality traits, impulsivity, sensation-seeking, and demographic data on drug consumption for different drugs, groups of drugs and for different definitions of drug users. The study had two purposes: (i) to identify the association of personality profiles (i.e. NEO-FFI-R) with drug consumption and (ii) to predict the risk of drug consumption for each individual according to their personality profiles. Part of the results was presented in the preprint~\cite{Fehrman15}.

The sample was created by an anonymous online survey. It was found to be biased when compared with the general population, which was indicated from comparison to the data published by Egan, et al \cite{Egan00} and Costa Jr \& McCrae \cite{McCrae04}. Such a bias is usual for clinical cohorts \cite{Gurrera00,Terracciano08}.

Our study reveals that the personality profiles are strongly associated with belonging to groups of the users and non-users of the 18 drugs.  For analysis, we use the following  subdivision of the sample T-$score$: the interval 44-49 indicates a moderately low score, $(-)$, the interval 49-51 indicates  a neutral score $(0)$, and the interval 51-56 indicates  a moderately high $(+)$ score.
 We found that the N and O scores of drug users of all 18 drugs are moderately high $(+)$ or neutral $(0)$, except for crack usage for the week-based classification, for which the O score is moderately low $(-)$. The A and C scores are moderately low $(-)$  or neutral $(0)$ for all groups of drug users and all user/non-user separations.
For most groups of  illicit drug users the A and C scores  are moderately low $(-)$ except  two exclusions:   the A score is neutral $(0)$  in the year-based  classification for LSD users  and  in the week-based classification for LSD and magic mushrooms users.
The  A and C scores for groups of legal drugs users (i.e. alcohol, chocolate, caffeine, and nicotine) are neutral $(0)$,  apart from nicotine users, whose  C score  is moderately low $(-)$  for all bases of user/non-user separation.

The impact of the E score is drug specific. For example, for the week-based user/non-user separation the E scores are:
\begin{itemize}
\item The E score of users is moderately low $(-)$ for amphetamines, amyl nitrite,  benzodiazepines, heroin,  ketamine, legal highs,  methadone, and crack;
\item The E score of users is moderately high $(+)$ for  cocaine, ecstasy, LSD, magic mushrooms, and VSA;
\item The E score of users is neutral $(0)$ for  alcohol, caffeine, chocolate, cannabis, and nicotine.
\end{itemize}
For more details see Section `Comparison of personality traits means for drug users and non-users' in `Results'.

Usage of some drugs are significantly correlated. The structure of these correlations is analysed in Section `Correlation between usage of different drugs'. Two correlation measures are utilised: the Pearson Correlation Coefficient (PCC) and the Relative Information Gain (RIG).
 We found three groups of drugs with highly correlated use.  The central element is clearly identified for each group. These centres are: {\em heroin, ecstasy, and benzodiazepines}. It means that the drug consumption has a `modular structure'. The modular structure has clear reflection in the correlation graph. The idea to merge the correlated attributes into `modules' called as {\em correlation pleiades}  is popular in biology \cite{Terentjev31,Berg60,Mitteroecker07}.

The concept of correlation pleiades was introduced in biostatistics in 1931 \cite{Terentjev31}.  They  were used for identification of the modular structure in evolutionary physiology \cite{Terentjev31,Mitteroecker07,Berg60,Armbruster99}. According to Berg \cite{Berg60}, correlation pleiades are clusters of correlated traits. In our approach, we distinguish the core and the peripheral elements of correlation pleiades and allow different pleiads to have small intersections in their periphery. `Soft' clustering algorithms relax the restriction that each data object is assigned to only one cluster (like probabilistic \cite{Krishnapuram1993} or fuzzy \cite{Bezdek1981} clustering). See the book of R. Xu and D. Wunsch \cite{XuWunsch2009} for the modern review of hard and soft clustering. We refer to \cite{Omote2006} for a discussion of clustering in graphs with intersections .

The three groups of correlated drugs centered around heroin, ecstasy, and benzodiazepines are defined for the decade-, year-, month-, and week-based classifications:
\begin{itemize}
\item  The heroin pleiad  includes crack, cocaine, methadone, and heroin;
\item The ecstasy pleiad  consists of amphetamines, cannabis, cocaine, ketamine, LSD, magic mushrooms, legal highs, and ecstasy;
\item The benzodiazepines pleiad contains methadone, amphetamines,  cocaine, and benzodiazepines.
\end{itemize}

Analysis of the intersections between correlation pleiads of drugs can generate important question and hypotheses:
\begin{itemize}
\item Why is cocaine  a peripherical member of all pleiads?
\item Why does methadone belong to the  periphery of both the heroin and benzodiazepines pleiades?
\item Do these intersections reflect the structure of individual drug consumption or the structure of the groups of drug consumers?
\end{itemize}

Correlation analysis of the decade-based classification problems demonstrates that the consumption of legal drugs (i.e. alcohol, chocolate and caffeine) is not correlated with consumption of other drugs. The consumptions of seven illicit drugs (i.e. amphetamines, cannabis, cocaine, ecstasy, legal highs, LSD, and mushrooms) are symmetrically correlated (when the correlations are measured by relative information gain, which is not symmetric a priori). There are also many strongly asymmetric correlations. For example, knowledge of amphetamines, cocaine, ecstasy, legal highs, LSD, and magic mushroom consumption is useful for the evaluation of ketamine consumption, but  on the other hand, knowledge of ketamine consumption is significantly less useful for the evaluation of usage of the drugs listed above.

In this study, we evaluated the individual drug consumption risk separately, for each drug and pleiad of drugs. We also analyzed interrelations between the individual drug consumption risks for different drugs. We applied several data mining approaches: decision tree, random forest, $k$-nearest neighbors, linear discriminant analysis, Gaussian mixture, probability density function estimation, logistic regression and na{\"i}ve Bayes. The quality of classification was surprisingly high. We tested all the classifiers by  {\em Leave-One-Out Cross Validation}. The best results with sensitivity and specificity being greater than 75\% were achieved for cannabis, crack, ecstasy, legal highs, LSD, and VSA. Sensitivity and specificity greater than 70\%  were achieved for the following drugs: amphetamines, amyl nitrite, benzodiazepines, chocolate, caffeine, heroin, ketamine, methadone and nicotine. The poorest result was obtained for prediction of alcohol consumption. An exhaustive search was performed to select the most effective subset of input features, and data mining methods to classify users and non-users for each drug.

Users are defined for each correlation pleiad  of drugs as users of any of the drug from the pleiade.
We consider the classification problem for drug pleiades for the decade-, year-, month-, and week-based user/non-user separations. These problems are much better balanced  for short periods (the week-based user definition) than the classification problems for separate drugs. For example, there are  184 users for the heroin pleiad but only 29 heroin users in the database for the week-based definition of users. The quality of classification is high. For example,  for the month-based user/non-user separation of the heroin pleiad consumption, the best classifier is a decision tree with five features and sensitivity 74.18\% and specificity 74.11\%. A decision tree with seven attributes is the best classifier for the year-based classification problem of the ecstasy pleiad users/non-users and has sensitivity 80.65\% and specificity 80.72\%. In the week-based separation of the benzodiazepines pleiad users/non-users, the best classifier is a decision tree with five features, sensitivity 75.10\%, and specificity 75.76\%.

The creation of classifiers provided the capability to evaluate the risk of drug consumption in relation to individuals. The risk map is a useful tool for data visualisation and for  the generation of hypotheses for further study (see Section `Risk evaluation for the decade-based user/non-user separation').

The main results of the work are:
\begin{itemize}
\item Presentation and descriptive analysis of a database with information of 1885 respondents and usage of 18 drugs.
\item Demonstration that the personality traits (five factor model, impulsivity, and sensation seeking) together with simple demographic data give the possibility of predicting the risk of consumption of individual drugs with  sensitivity and specificity above 70\% for most  drugs.
\item The best  classifiers and most significant predictors  are found for each individual drug in question.
\item Revelation of three correlation pleiads of drugs, that are the clusters of drugs with correlated consumption centered around heroin, ecstasy, and benzodiazepines.
\item The best robust classifiers and most significant predictors are found for use of pleiads of drugs.
\item The risk map technology  is developed for the visualization of the probability of drug consumption.
\end{itemize}

\section*{Materials and Methods}
\subsection*{Database}
The database was collected by Elaine Fehrman between March 2011 and March 2012. In January 2011, the research proposal was approved by the University of Leicester’s Forensic Psychology Ethical Advisory Group, and subsequently received favourable opinion from the University of Leicester School of Psychology’s Research Ethics Committee (PREC).

 The data are available online \cite{FehrmanData2016}. An online survey tool from Survey Gizmo was employed to gather data which maximised anonymity, this being particularly relevant to canvassing respondents’ views, given the sensitive nature of drug use. All participants were required to declare themselves at least 18 years of age prior to informed consent being given.

The study recruited 2051 participants over a 12-month recruitment period. Of these persons, 166 did not respond correctly to a validity check built into the middle of the scale, so were presumed to being inattentive to the questions being asked. Nine of these persons were found to also have endorsed using a fictitious recreational drug, and which was included precisely to identify respondents who over-claim, as have other studies of this kind \cite{Hoare10}. This led a useable sample of 1885 participants (male/female = 943/942).

The snowball sampling methodology recruited a primarily (93.7\%) native English-speaking sample, with participants from the UK (1044; 55.4\%), the USA (557; 29.5\%), Canada (87; 4.6\%), Australia (54; 2.9\%), New Zealand (5; 0.3\%) and Ireland (20; 1.1\%). A total of 118 (6.3\%) came from a diversity of other countries, none of whom individually met 1\% of the sample or did not declare the country of location. Further optimizing anonymity, persons reported their age band, rather than their exact age; 18-24 years (643; 34.1\%), 25-34 years (481; 25.5\%), 35-44 years (356; 18.9\%), 45-54 years (294; 15.6\%), 55-64 (93; 4.9\%), and over 65 (18; 1\%). This indicates that although the largest age cohort band were in the 18 to 24 range, some 40\% of the cohort was 35 or above, which is an age range often missed in studies of this kind.

The sample recruited was highly educated, with just under two thirds (59.5\%) educated to, at least, degree or professional certificate level: 14.4\% (271) reported holding a professional certificate or diploma, 25.5\% (481) an undergraduate degree, 15\% (284) a master's degree, and 4.7\% (89) a doctorate. Approximately 26.8\% (506) of the sample had received some college or university tuition although they did not hold any certificates; lastly, 13.6\% (257) had left school at the age of 18 or younger.

Participants were asked to indicate which racial category was broadly representative of their cultural background. An overwhelming majority (91.2\%; 1720) reported being white, (1.8\%; 33) stated they were Black, and (1.4\%; 26) Asian. The remainder of the sample (5.6\%; 106) described themselves as `Other' or `Mixed' categories. This small number of persons belonging to specific non-white ethnicities precludes any analyses involving racial categories.

\subsubsection*{Personality measurements}
In order to assess personality traits of the sample, the Revised NEO Five-Factor Inventory (NEO-FFI-R) questionnaire was employed \cite{Costa92}. The NEO-FFI-R is a highly reliable measure of basic personality domains; internal consistencies are 0.84 (N); 0.78 (E); 0.78 (O); 0.77 (A), and 0.75 (C) Egan \cite{Egan11}. The scale is a 60-item inventory comprised of five personality domains or factors. The NEO-FFI-R is a shortened version of the Revised NEO-Personality Inventory (NEO-PI-R) \cite{Costa92}. The five factors are: N, E, O, A, and C with 12 items per domain.
The five traits can be summarized as:
\begin{enumerate}
\item \emph{Neuroticism} ({\bf N}) is a long-term tendency to experience negative emotions such as nervousness, tension, anxiety and depression;
\item \emph{Extraversion} ({\bf E}) is manifested in outgoing, warm, active, assertive, talkative, cheerful, and in search of stimulation characteristics;
\item \emph{Openness to experience} ({\bf O}) is a general appreciation for art, unusual ideas, and imaginative, creative, unconventional, and wide interests,
\item \emph{Agreeableness} ({\bf A}) is a dimension of interpersonal relations, characterized by altruism, trust, modesty, kindness, compassion and cooperativeness;
\item \emph{Conscientiousness} ({\bf C}) is a tendency to be organized and dependable, strong-willed, persistent, reliable, and efficient.
\end{enumerate}

All of these domains are hierarchically defined by specific facets \cite{McCrae91}. Egan et al \cite{Egan00} observe that the score O and E domains of the NEO-FFI instrument are less reliable than N, A, and C.

Participants were asked to read the 60 NEO-FFI-R statements and indicate on a five-point Likert scale how much a given item applied to them (i.e. 0 = `Strongly Disagree', 1 = `Disagree', 2 = `Neutral', 3 = `Agree', to 4 = `Strongly Agree').

We expected that drug usage is associated with high N, and low A and C. The darker dimension of personality can be described in terms of high N and low A and C, whereas much of the anti-social behaviour in non-clinical persons appears underpinned by high N and low C \cite{Jakobwitz06}. The so-called `\emph{negative urgency}' is the tendency to act rashly when distressed, and characterized by high N, low C, and low A \cite{Settles12}. The negative urgency is partially proved below for users of most of the illegal drugs. In addition, our findings suggest that O is higher for drug users.

The second measure used was the Barratt Impulsiveness Scale (BIS-11) \cite{Stanford09}. The BIS-11 is a 30-item self-report questionnaire, which measures the behavioural construct of impulsiveness, and comprises three subscales: motor impulsiveness, attentional impulsiveness, and non-planning. The `motor' aspect reflects acting without thinking, the `attentional' component poor concentration and thought intrusions, and the `non-planning' a lack of consideration for consequences \cite{Snowden11}. The scale's items are scored on a four-point Likert scale. This study modified the response range to make it compatible with previous related studies \cite{GMontes09}.
A score of five usually connotes the most impulsive response although some items are reverse-scored to prevent response bias. Items are aggregated, and the higher BIS-11 scores, the higher the impulsivity level \cite{Fossati01}. The BIS-11 is regarded a reliable psychometric instrument with good test-retest reliability (Spearman's rho is equal to 0.83) and internal consistency (Cronbach's alpha is equal to 0.83; \cite{Stanford09,Snowden11}.

The third measurement tool employed was the Impulsiveness Sensation-Seeking (ImpSS). Although the ImpSS combines the traits of impulsivity and sensation-seeking, it is regarded as a measure of a general sensation-seeking trait \cite{Zuckerman94}. The scale consists of 19 statements in true-false format, comprising eight items measuring impulsivity ({\bf Imp}), and 11 items gauging sensation-seeking ({\bf SS}). The ImpSS is considered a valid and reliable measure of high risk behavioural correlates such as, substance misuse \cite{McDaniel08}.

\subsubsection*{Drug use}
\label{DrugUse}
Participants were questioned concerning their use of 18 legal and illegal drugs (alcohol, amphetamines, amyl nitrite, benzodiazepines, cannabis, chocolate, cocaine, caffeine, crack, ecstasy, heroin, ketamine, legal highs, LSD, methadone, magic mushrooms, nicotine and volatile substance abuse (VSA)) and one fictitious drug (Semeron) which was introduced to identify over-claimers.

\begin{figure}
    \centering
    \includegraphics[width=0.95\textwidth]{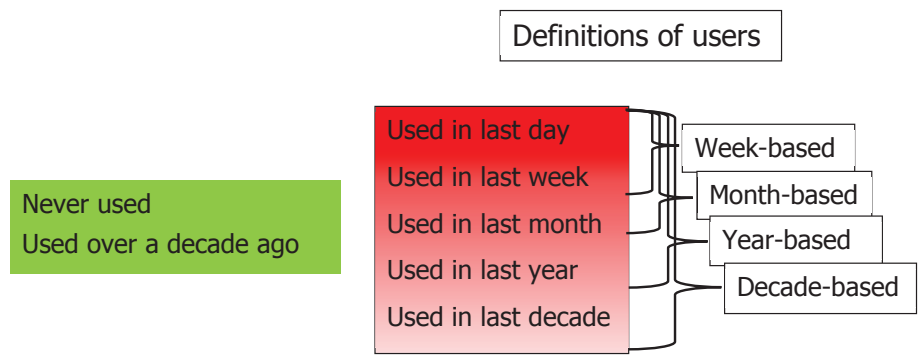}
    \caption {{\bf Categories of drug users.}  Categories with green background  always correspond to drug non-users. Four different definitions of drug users are presented.}
    \label{Categoriesfig:1}
\end{figure}

It was recognised at the outset of this study that drug use research regularly (and spuriously) dichotomises individuals as users or non-users, without due regard to their frequency or duration/desistance of drug use \cite{Ragan10}. In this study, finer distinctions concerning the measurement of drug use have been deployed, due to the potential for the existence of qualitative differences amongst individuals with varying usage levels. In relation to each drug, respondents were asked to indicate if they never used this drug, used it over a decade ago, or in the last decade, year, month, week, or day. This format captured the breadth of a drug-using career, and the specific recency of use. Different categories of drug users are depicted in Fig~\ref{Categoriesfig:1}.

Analysis of the classes of drug users shows that part of the classes are nested: participants which belong to the category `Used in last day' also belong to the categories `Used in last week', `Used in last month', `Used in last year' and `Used in last decade'. There are two special categories: `Never used' and `Used over a decade ago' (see Fig~\ref{Categoriesfig:1}). Data does not contain a definition of users and non-users groups. Formally only a participant in the class `Never used' can be called a non-user, but it is not a seminal definition because a participant who used a drug more than decade ago cannot be considered a drug user for most applications. There are several possible way to discriminate participants into groups of users and non-users for binary classification:

\begin{enumerate}
  \item Two isolated categories (`Never used' and `Used over a decade ago') are placed into the class of non-users with a green background in Fig~\ref{Categoriesfig:1} and all other categories into the class `users' as the simplest version of binary classification. This classification problem is called `\emph{decade-based}' user/non-user separation.

  \item The categories `Used in last decade', `Used over a decade ago' and `Never used' are merged to form a group of non-users and all other categories are placed into the  group of users. This classification problem is called `\emph{year-based}'.

  \item The categories `Used in last year', `Used in last decade', `Used over a decade ago' and `Never used' are combined to form a group of non-users and all three other categories are placed into the group of users. This classification problem is called `\emph{month-based}'.

  \item The categories `Used in last week' and `Used in last month' are merged to form a  group of users and all other categories are placed into the group of non-users. This classification problem is called `\emph{week-based}'.
\end{enumerate}

We begin our analysis from the decade-based user/non-user separation because it is a relatively well balanced classification problem,  that is, there are sufficiently many users in the united group `Used in last decade-year-month-week' for all drugs in the database. If the problem is not directly specified then it is the decade-based classification problem. We also perform analysis for the year-, month-, and week-based user/non-user separation. It is useful to group drugs with highly correlated usage for this purpose (see Section `Pleiades of drugs').

The proportion of drug users among all participants is different for different drugs and for different classification problems. The data set comprises 1885 individuals without any missing data. Table~\ref{tab:1a} shows the percentage of drug users for each drug and for each problem in the database. It is necessary to mention that the sample is biased to a higher proportion of drug users. This means that for the general population the fraction of an illegal drug users is expected to be significantly lower \cite{HomeOffice14}.

\begin{table}[!ht]
\begin{adjustwidth}{-2.25in}{0in}
\centering
\caption{The number and fraction of drug users}
\label{tab:1a}
\begin{tabular}{|l|c|c|c|c|}\hline
{\bf Drug} 	& \multicolumn{4}{c|}{ { \bf User definition based on}}                 \\\cline{2-5}
         	& {\bf Decade } &	{\bf Year }&	{ \bf Month }&	{\bf Week } \\\hline
Alcohol     &	1817; 96.39\% &	1749; 92.79\%&	1551; 82.28\%&  1264; 67.06\%\\\hline
Amphetamines&	679;  36.02\% &	436; 23.13\%&	238; 12.63\%&	163; 8.65\%\\\hline
Amyl nitrite&	370; 19.63\%  &	133; 7.06\%&	41; 2.18\%&	    17; 0.90\%\\\hline
Benzodiazepines&769; 40.80\%  &	535; 28.38\%&	299; 15.86\%&	179; 9.50\%\\\hline
Cannabis   &   1265; 67.11\%  &	999; 53.00\%&	788; 41.80\%&	648; 34.38\%\\\hline
Chocolate  & 1850; 98.14\%    &	1840; 97.61\%&	1786; 94.75\%&	1490; 79.05\%\\\hline
Cocaine    &687; 36.45\%      &	417; 22.12\%&	159; 8.44\%&	60; 3.18\%\\\hline
Caffeine   &	1848; 98.04\% &	1824; 96.76\%&	1764; 93.58\%&	1658; 87.96\%\\\hline
Crack      & 191; 10.13\%     &	79; 4.19\%  &	20; 1.06\%&	    11; 0.58\%\\\hline
Ecstasy    &751; 39.84\%      &	517; 27.43\%&	240; 12.73\%&	84; 4.46\%\\\hline
Heroin     &	212; 11.25\%  &	118; 6.26\%&	53; 2.81\%&	    29; 1.54\%\\\hline
Ketamine   &	350; 18.57\%  &	208; 11.03\%&	79; 4.19\%&	    37; 1.96\%\\\hline
Legal highs&	762; 40.42\%  &	564; 29.92\%&	241; 12.79\%&	131; 6.95\%\\\hline
LSD        &	557; 29.55\%  &	380; 20.16\%&	166; 8.81\%&	69; 3.66\%\\\hline
Methadone  &417; 22.12\%      &	320; 16.98\%&	171; 9.07\%&	121; 6.42\%\\\hline
MMushrooms&694; 36.82\%       &	434; 23.02\%&	159; 8.44\%&	44; 2.33\%\\\hline
Nicotine  &1264; 67.06\%      &	1060; 56.23\%&	875; 46.42\%&	767; 40.69\%\\\hline
VSA       &230; 12.20\%       &	95; 5.04\%  &	34; 1.80\%&	    21; 1.11\%\\\hline
\end{tabular}
\end{adjustwidth}
\end{table}

\subsection*{Data analysis}
The raw score for each factor of the NEO-FFI-R was converted into a T-Score based on normative data \cite{McCrae04}:
\begin{equation}
\label{eq:1}
T\textrm{-}score = 10 \left[ \frac{Raw\ score\ - Normative\ mean\ score}{Normative\ standard\ deviation}\right] +50
\end{equation}

Table~\ref{tab:1} presents  statistics: sample means, sample standard deviation (SD), and evaluation of the significance of the difference between the sample mean and the population mean. All differences between the population and sample means are statistically significant ($p<0.001$).

\begin{table}[!ht]
\begin{adjustwidth}{-2.25in}{0in}
\centering
\caption{ Descriptive statistics (Mean, 95\% CI, SD). \emph{p}-value is calculated by $t$-test.}
\label{tab:1}       
\begin{tabular}{|c|c|c|c|c|c|}\hline
{\bf Factors}& {\bf Sample mean} &{\bf 95\% CI for sample mean}&{ \bf SD}&{\bf Population mean}&{\bf \emph{p}-value}\\\hline
N&	59.64&	59.08, 60.20&	12.41&	50& \textless 0.001 \\\hline
E&	47.35&	46.88, 47.83&	10.48&	50&	\textless 0.001 \\\hline
O&	54.04&	53.55, 54.52&	10.75&	50&	\textless 0.001 \\\hline
A&	47.15&	46.62, 47.69&	11.88&	50&	\textless 0.001 \\\hline
C&	43.93&	43.25, 44.61&	11.06&	50&	\textless 0.001 \\\hline
\end{tabular}
\end{adjustwidth}
\end{table}

\begin{figure}
\centering
\includegraphics[scale=1]{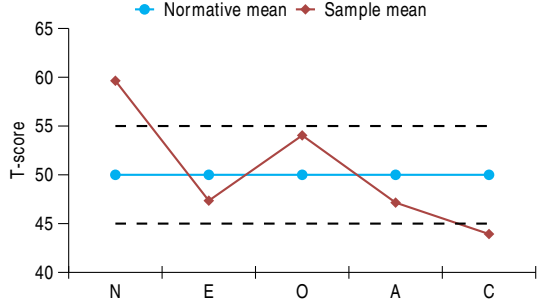}
\caption{{ \bf Mean T-score NEO-FFI-R}}
\label{MeanTscorefig:2}
\end{figure}

The means of the NEO-FFI-R T-scores based on normative data are depicted in Fig~\ref{MeanTscorefig:2}. Table~\ref{tab:1} and Fig~\ref{MeanTscorefig:2} illustrate that the sample is biased with respect to the population. Such a bias is usual for clinical cohorts, for example, the `problematic' or `pathological' groups \cite{Fridberg11} and the drug users \cite{Terracciano08,Flory02}. It is highlighted below that the sample in this study deviates from the population norm in the same direction as drug users deviate from sample mean (see Table~\ref{tab:5}). However, the deviances of mean of users groups are different for different drugs. Therefore, it is convenient to study deviations of users and non-users from the sample mean. We introduce sample based T-score for this purpose .

T-$score_{sample}$ is introduced for improving the visibility and simplicity of a comparison and is calculated using equation (\ref{eq:2}). The resulting T-$score_{sample}$ contains a mean of 50 and a standard deviation of 10.

\begin{equation}
\label{eq:2}
T\textrm{-}score_{sample} = 10 \left[ \frac{Raw\ score\ - Sample\ mean\ score}{Sample\ standard\ deviation}\right] +50
\end{equation}

Usually, the T-score is categorized into five categories to summarise an individual's personality score concerning each factor. The interval 20-35 indicates very low scores. The interval 35-45 indicates low scores. The interval 45-55 indicates average scores. The interval 55-65 indicates high scores. The interval 65-80 indicates very high scores. This study considers the mean T-$score_{sample}$ for two groups (users and non-users) instead of each individual's score.  A subdivision of the T-$score_{sample}$ interval is introduced as follows: the interval 44-49 indicates moderately low $(-)$, the interval 49-51 indicates neutral $(0)$, and the interval 51-56 indicates moderately high $(+)$.

The unification of the mean and variance of the T-$score_{sample}$ for all factors simplifies comparisons of groups (both users and non-users) for each drug. Any differences between the mean T-$score_{sample}$ for groups of users and non-users is usually used as a measure of the groups' dissimilarity in scores. The  NEO-FFI-R scores for groups of users and non-users for each drug were represented by the mean T-$score_{sample}$ of these groups for each factor. A $t$-test is employed to estimate the significance of the differences between the mean T-$score_{sample}$ for groups of users and non-users for each NEO-FFI-R factor and each drug. In this $t$-test, a  ${p}$-value is a probability of observing by chance the same or a greater difference of mean for two samples with the same mean. The 90\% level is chosen to select significant differences. The analysis was performed using SAS 9.4.

\subsection*{Input feature transformation}
\label{Input feature transformation}
There are many data mining methods to work with continuous data. It is necessary to quantify all categorical features to use these methods especially for features with many levels. To apply logistic regression to these data with categorical features and corresponding coefficients, we have to use dummy coding directly or indirectly. In this case we have $n-1$ coefficients for a feature with $n$ levels, meaning that we fit logistic regression in a 250 dimensional space (age contains six levels, gender contains two levels, education contains nine levels, country contains seven levels, ethnicity contains seven levels, N score contains 49 levels, E score contains 42 levels, O  score contains 35 levels, A  score contains 41 levels, C  score contains 41 levels, impulsivity contains 10 levels, and SS contains 11 levels: 5 + 1 + 8 + 6 + 6 + 48 + 41 + 34 + 41 + 41 + 9 + 10 = 250). After quantification we can fit a logistic regression model in a 12 dimensional space. This means that feature quantification can be used as an effective dimensionality reduction method.

\subsubsection*{Ordinal feature quantification}
\label{Ordinal feature quantification}
One of the widely used techniques to analyse categorical data is the calculation of polychoric correlation \cite{Lee95,Martinson71}. The matrix of polychoric coefficients is used  further  to calculate principal components (PCs), etc. The technique of polychoric correlation is based on the assumption that values of ordinal features result from the discretization of continuous random values with fixed thresholds. Furthermore, these latent continuous random values follow a normal distribution. Unfortunately, the polychoric correlation technique has two drawbacks: it defines the thresholds of discretization but not the values for each category, and the defined thresholds  differ for different pairs of attributes.

Consider the ordinal feature $ o $ with categories $o _{1} $,$ o_{2} $,...,$ o_{k} $ and with number of cases $ n_{i} $ of category $o_{i}  $. The empirical estimation of the probability of category $ o_{i} $  is $p_{i}=n _{i}/{N}$, where $N=\sum  n_{i} $. The sample estimation of thresholds are evaluating as:

\begin{equation}
t_{i}=\phi^{-1}\Bigg(\sum_{j=1}^{i}p_{j}\Bigg)\label{qqq1}
\end{equation}

The simplest method of the ordinal feature quantification is to use the thresholds~(\ref{qqq1}) and select the `average' value in each interval. There are several variants of the `average' value. In this study we use the value with average probability: if thresholds $ t_{i-1} $
and $ t_{i} $ define the interval of category $ o_{i} $, then average probability for this interval is
\begin{equation}
q_{i}=\phi^{-1}\Bigg(\sum_{j=1}^{i-1}p_{j}+\frac{p_{i}}{2}\Bigg)\label{qqq2}
\end{equation}

The polychoric coefficients, calculated on base of quantification~(\ref{qqq2}), have less likelihood than the polychoric coefficients calculated by using the maximum likelihood approach. The merit of this approach is the usage of the same thresholds for all pairs of attributes and explicit formulae for calculating the categories' values.

\subsubsection*{Nominal feature quantification}
\label{Nominal feature quantification}
We cannot use the techniques described above to quantify nominal features such as gender, country of location and ethnicity because the categories of these features are unordered. To quantify nominal features we implemented the technique of nonlinear CatPCA (Categorical Principal Component Analysis) \cite{Linting12}. This procedure includes four steps:
\begin{enumerate}
\item Exclude nominal features from the set of input features and calculate the informative PCs \cite{Gorban10,Gorban08,Gorban09,Pearson01} in the space of retained input features. To select informative components we use Kaiser's rule \cite{Guttman54,Kaiser60}.
\item Calculate the centroid of each category in projection on selected PCs.
\item Calculate the first PC of centroids.
\item The numerical value for each component is the projection of its centroid on this component.
\end{enumerate}

\begin{algorithm}                        
\caption{Nominal feature quantification} 
\label{alg1}                             
\begin{algorithmic}[1]                      
    \State Exclude nominal features from the set of input features and calculate the informative PCs \cite{Gorban10,Gorban08,Gorban09,Pearson01} in the space of retained input features. To select informative components we use Kaiser's rule \cite{Guttman54,Kaiser60}.
    \State Calculate the centroid of each category in projection on selected PCs.
    \State Calculate the first PC of centroids.
    \State The numerical value for each category is the projection of its centroid on this component.
\end{algorithmic}
\end{algorithm}

\begin{figure}
\centering
\includegraphics[scale=0.95]{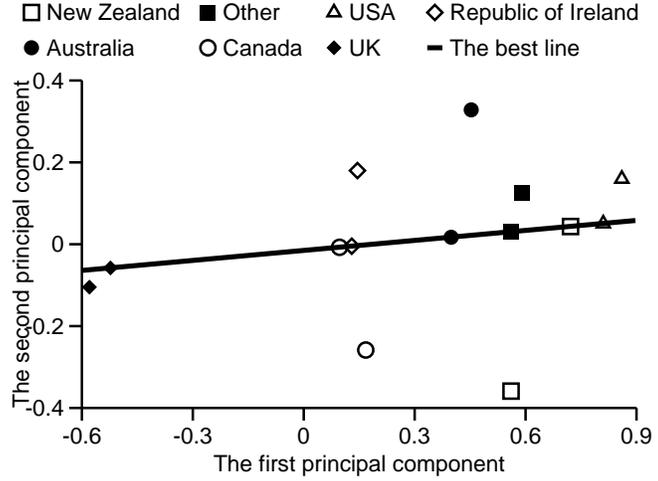}
\caption {{ \bf Nominal feature quantification of `Country' }}
\label{Quantificationfig:3}
\end{figure}

The process of nominal feature quantification for the feature `country' is depicted in Fig~\ref{Quantificationfig:3}.  Fig~\ref{Quantificationfig:3} shows that points corresponding to the UK category are located very far from any other points.

As an alternative variant of nominal feature quantification we use dummy coding \cite{Gujarati03} of nominal variables: `country' is transformed into seven binary features with values 1 (if `true') or 0 (if `false'): UK, Canada, USA, Other (country), Australia, Republic of Ireland and New Zealand; Ethnicity is transformed into seven binary features: Mixed-White/Asian, White, Other (ethnicity), Mixed-White/Black, Asian, Black and Mixed-Black/Asian.

\subsubsection*{Input feature ranking}
\label{Input feature ranking}
In this study, we used three different techniques for input feature ranking. The first technique was \textit{principal variables} \cite{McCabe84}. Principal variables are a set of input features which explain the maximal fraction of the data variance. The main idea of this approach is to select first the input feature which explains the maximal fraction of the data variance, then select the next feature which together with the previously selected features explains the maximal fraction of data variance, and so on.

The second technique was \textit{double Kaiser's selection}. Calculate PCs and select informative PCs by Kaiser's rule \cite{Guttman54,Kaiser60}. Kaiser's rule states the all PCs which  correspond to eigenvalues greater than the average are informative and all other PCs are uninformative.  We apply the covariance based PCs. that is, we calculate PCs as the normalized eigenvectors of the covariance matrix. For them, the Kaiser rule threshold is equal to the trace  of the covariance matrix divided by the number of attributes. The importance of an attribute is defined as the maximum of the absolute value of the corresponding coordinates in the informative PCs. In attribute selection we define the threshold of importance as the average value of coordinate which is equal to $1/\sqrt{n}$ for a unit length vector, where $n$ is the number of attributes. If the attribute importance is greater than the threshold of importance then this attribute is informative. Otherwise, the attribute is trivial. If there are trivial attributes then the worst attribute is the attribute with minimal value of importance. We removed the worst attribute and repeated the procedure. This procedure stops if there are no trivial attributes. This algorithm ranks attributes from worst to best.

The third technique was \textit{sparse PCA} \cite{Naikal11}. In this study, we used the simplest thresholding sparse PCA. The searching for each sparse PC contains several steps:
\begin{enumerate}
\item Define the number of features $ n $, variance of data $\sigma^2$, the Kaiser threshold for the components $h=\sigma^2/n$ and the Kaiser threshold for the coefficients $c=1/\sqrt{n}$.
\item Search for the usual PC and calculate the variance $\sigma_c^2$ explained by this component. The iterative algorithm gives the PCs in descending order of $\sigma_c^2$.
\item If $\sigma_c^2<h$ then all the informative components are found. Remove the last component and go to step 8.
\item Search the attribute with non-zero coefficients with least absolute value $c_{min}$.
\item If $c_{min}>c$ then there are no trivial attributes found. Go to step 7.
\item Set the value of the found coefficient to zero. Block changes to this attribute coefficient and search the PC under this condition. Go to step 4.
\item Subtract the projection onto the found component from the data and go to step 2.
\item Search the attributes with zero coefficients in each found component. These attributes are trivial. If there are trivial attributes then remove them from the set of attributes and go to step 1, else stop.
\end{enumerate}

\begin{algorithm}                        
\caption{Search of sparse PC} 
\label{alg2}                             
\begin{algorithmic}[1]                      
    \State Define the number of features $ n $, variance of data $\sigma^2$, the Kaiser threshold for the components $h=\sigma^2/n$ and the Kaiser threshold for the coefficients $c=1/\sqrt{n}$.
    \State Search for the usual PC and calculate the variance $\sigma_c^2$ explained by this component. The iterative algorithm gives the PCs in descending order of $\sigma_c^2$.
    \If {$\sigma_c^2<h$} \State All the informative components are found. Remove the last component and go to step 10.
    \EndIf
    \State Search the attribute with non-zero coefficients with least absolute value $c_{min}$.
    \If {$c_{min}>c$} \State There are no trivial attributes found. Go to step 9. \EndIf
    \State Set the value of the found coefficient to zero. Block changes to this attribute coefficient and search the PC under this condition. Go to step 5.
    \State Subtract the projection onto the found component from the data and go to step 2.
    \State Search the attributes with zero coefficients in each found component. These attributes are trivial.
    \If {there are trivial attributes} \State Remove trivial attributes from the set of attributes and go to step 1. \Else \State Stop \EndIf
\end{algorithmic}
\end{algorithm}

\subsection*{Risk evaluation methods}
\label{Risk evaluation methods}
In this study, we applied several classification methods which provide risk evaluation.
We used the set of input features after quantification described in the `Ordinal feature quantification' Section and in the `Nominal feature quantification' Section. It includes age, gender, education, N, E, O, A, C, Imp., and SS.

\subsubsection*{$k$ Nearest Neighbours ($k$NN)}
The basic concept of $k$NN is the class of an object is the class of the majority of its $k$ nearest neighbours \cite{Clarkson05}. This algorithm is very sensitive to distance definition. There are several commonly used variants of distance for $k$NN: Euclidean distance; Minkovsky distance; and distances calculated after some transformation of the input space.

In this study, we used three distances: the Euclidean distance, the Fisher's transformed distance \cite{Fisher36} and the adaptive distance \cite{Hastie96}. Moreover, we used a weighted voting procedure with weighting of neighbours by one of the standard kernel functions \cite{Li07}.

The $k$NN algorithm is well known \cite{Clarkson05}. The adaptive distance transformation algorithm is described in \cite{Hastie96}. $k$NN with Fisher's transformed distance is less known. The following parameters are used: $k$ is the number of nearest neighbours, $K$ is the kernel function, and $k_f$ is the number of neighbours which are used for the distance transformation. To define the risk of drug consumption we have to do the following steps:	
\begin{enumerate}
\item Find the $k_f$ nearest neighbours of the test point.
\item Calculate the covariance matrix of $k_f$ neighbours and Fisher's discriminant direction.
\item	Find the $k$ nearest neighbours of the test point using the distance along Fisher's discriminant direction among the $k_f$ neighbours found earlier.
\item Define the maximal distance from the test point to $k$ neighbours.
\item Calculate the membership for each class  as a sum of the points' weights. The weight of a point is the ratio: the value of the kernel function $K$ of distance from this point to the test point divided by the maximal distance defined at step 4.
\item Drug consumption risk is defined as the ratio of the positive class membership to the sum of memberships of all classes.
\end{enumerate}

\begin{algorithm}                        
\caption{$k$NN with Fisher's transformed distance} 
\label{alg3}                             
\begin{algorithmic}[1]                      
    \State Find the $k_f$ nearest neighbours of the test point.
    \State Calculate the covariance matrix of $k_f$ neighbours and Fisher's discriminant direction.
    \State Find the $k$ nearest neighbours of the test point using the distance along Fisher's discriminant direction among the $k_f$ neighbours found earlier.
    \State Define the maximal distance from the test point to $k$ neighbours.
    \State Calculate the membership for each class  as a sum of the points' weights. The weight of a point is the ratio: the value of the kernel function $K$ of distance from this point to the test point divided by the maximal distance defined at step 4.
    \State Drug consumption risk is defined as the ratio of the positive class membership to the sum of memberships of all classes.
\end{algorithmic}
\end{algorithm}

The adaptive distance version implements the same algorithm but uses another transformation on step 2 and another distance on step 3 \cite{Hastie96}. The Euclidean distance version simply defines $k_f=k$ and omits steps 2 and 3 of algorithm.
We tested 1,683 million versions of the $k$NN models per drug, which differ by:
\begin{itemize}
\item The number of nearest neighbours, which varies between 1 and 30;
\item The set of input features;
\item One of the three distances: Euclidean distance, adaptive distance and Fisher's distance;
\item The kernel function for adaptive distance transformation;
\item The kernel functions for voting.
\item Weight of class `users' is varied between 0.01 and 5.0.
\end{itemize}

\subsubsection*{Decision Tree (DT)}
The decision tree approach is a classifier that constructs a tree like structure, which can be used to choose between several courses of action. Binary decision trees are used in this study. A decision tree is comprised of nodes and leaves. Each node can have a child node. If a node has no child node, it is called a leaf or a terminal node. Any decision tree contains one root node, which has no parent node. Each non terminal node calculates its own Boolean expression (with the value `true' or `false'). According to the result of this calculation, the decision for a given sample would be delegated to the left child node ('true') or to the right child node ('false'). Each leaf (terminal node) has a label which shows how many samples of the training set belong to each class. The probability of each class is estimated as a ratio of the number of samples in this class to the total number of samples in the leaf.

There are many methods for developing a decision tree \cite{Breiman84,Quinlan87,Rokach10,Sofeikov14,Gelfand91,Dietterich96,Kearns99}. We use the methods based on information gain, Gini gain, and DKM gain. Let us consider one node and one binary input attribute which can take values 0 or 1. Let us use notation: $N$ is the number of cases in the node, $c$ is the number of categories of the target feature, $ n_{ij} $ is the number of $i$ category cases with input attribute value $j$ in the node, the number of $i$ category cases in the node is $ n_{i.}=n_{i0}+n_{i1} $the number of cases with the input attribute value $j$ in the node is $ n_{.j}=\sum_{i=1}^{c}n_{ij} $, where $ n_{j}=\big(n_{1j},...,n_{cj}\big) $ is the vector of frequencies with input attribute value $j$, and $n=\big(n_{1.}, ...,n_{c.}\big)$ is the vector of frequencies with any input attribute value. To form a tree we select the base function for information criterion among

$${Entropy}\big(\textit{m, M}\big)=-\sum_{i=1}^{c}\frac{m_{i}}{M}\log_{2}\frac{m_{i}}{M},$$
$${Gini}\big(\textit{m, M}\big)=1-\sum_{i=1}^{c}\bigg(\frac{m_{i}}{M}\bigg)^{2},$$ and
$${DKM}\big(\textit{m,M}\big)=2\sqrt{m_{0}m_{1}M^{-2}},$$
where $m$ is the vector of frequencies and $M$ is the sum of the elements of the vector $m$. The DKM can be applied to a binary target feature only. The value of the criterion is the gain of base function:

$${BG}=\textit{Base}(n,N)-\frac{n_{0}}{N}\textit{Base}(n_{.0},n_{0})-\frac{n_{1}}{N}\textit{Base}(n_{.1},n_{1}).$$

There are several approaches to use real valued inputs in decision trees. A commonly used approach is the binning of real valued attributes before forming the tree. In this study we implemented `on the fly' binning: the best threshold is searched in each node for each real valued attribute and then this threshold is used to bin these feature in this node. The best threshold depends on the split criteria used (information gain, Gini gain, or DKM gain).

Another possibility we employ is the use of Fisher's discriminant to define the best linear combination of the real valued features \cite{Fisher36} in each node. Pruning techniques are applied to improve the tree.

The specified minimal number of instances in the tree's leaf is used as a criterion to stop node splitting. Each leaf of the tree cannot contain fewer than a specified number of  instances.

We tested 166 million decision tree models (per drug), which differ by:
\begin{itemize}
\item The three split criterion (information gain, Gini gain or DKM gain);
\item The use of the real-valued features in the splitting criteria separately or in linear combination by Fisher's discriminant;
\item The set of  input features;
\item The minimal number of instances in the leaf, which varied between 3 and 30.
\item Weight of class `users' is varied between 0.01 and 5.0.
\end{itemize}

\subsubsection*{Linear Discriminant Analysis (LDA)}
We used Fisher's linear discriminant for the binary version of the problem \cite{Fisher36}, to separate users from non-users of each drug. We calculate the mean of points of the $i$th class, $\mu_{i}$, and covariance matrix of the $i$th class $\Sigma_{i}$ for both classes ($i=1,2$). Then we calculate the discriminating direction as $$\omega=\big(\Sigma_{1}+\Sigma_{2}\big)^{-1}\big(\mu_{1}-\mu_{2}\big).$$

Each point is projected onto the discriminating direction by calculating the dot product $(\omega,x_{i})$. The threshold to separate two classes is calculated by finding the maximum of relative information gain, Gini gain, or DKM gain. This method cannot be used for problems of risk evaluation.

In the  study we tested 8,192 LDA  models per drug, which differ by one of the three criteria (information gain, Gini gain or DKM gain) which were used to define the threshold and the set of input features.

\subsubsection*{Gaussian Mixture (GM)}
Gaussian mixture is a method of estimating probability under the assumption that each category of a target feature has a multivariate normal distribution \cite{Dinov08}. In each category we should estimate the covariance matrix and invert it. The primary probability of belonging to the $i$th category is:

$$p_{i}(x)=p_i^0(2\pi)^{-\frac{k}{2}}|\Sigma_{i}|^{-\frac{1}{2}} \textit{e}^{-\frac{1}{2}}(x-\mu_{i})^{\prime}\Sigma_i^{-1}(x-\mu_{i})$$

where $p_i^0$ is a prior probability of the $i$th category, $k$ is the dimension of the input space, $\mu_i$ is the mean point of the $i$th category, $x$ is the tested point, $\Sigma_i$ is the covariance matrix of the $i$th category and $|\Sigma_i |$ is its determinant. The final probability of belonging to the $i$th  category is calculated as $$p^f_{i}(x)=p_{i}(x)/\sum_{j}p_{j}(x).$$

The prior probabilities are estimated as the proportion of cases in the $i$th category. We also used s varied multiplier to correct priors for the binary problem.

In the study, we tested  1,024 million  Gaussian mixture models per drug, which differ by the set of input features and corrections applied to the prior probabilities.

\subsubsection*{Probability Density Function Estimation (PDFE)}
We implemented the radial-basis function method \cite{Buhmann03} for probability density function estimation \cite{Scott92}. The number of probability densities to estimate is equal to the number of categories of the target feature. Each probability density function is estimated separately by using nonparametric techniques. The prior probabilities are estimated from the database: $p_i=n_i/N$ where $n_i$ is the number of cases with $i$ category of the target feature and $N$ is the total number of cases in the database.

We also use the database to define the $k$ nearest neighbours of  each data point. These $k$ points are used to estimate the radius of the neighbourhood of each point as a maximum of the distance from the data point to each of its $k$ nearest neighbours. The centre of one of the kernel functions is placed at the data point \cite{Li07}. The integral of any kernel function over the whole space is equal to one. The total probability of the $i$th category is proportional to the integral of the sum of the kernel functions, which is equal to $n_i$. The total probability of each category has to be equal to the prior probability $p_i$. Thus,  the sum of the kernel functions has to be divided by $n_i$ and multiplied by $p_i$. This gives the probability density estimation for each category.

We tested  426,000 versions of the PDFE models per drug, which differ by:
\begin{itemize}
\item The number of nearest neighbours (varied between 5 and 30);
\item The set of the input features;
\item The kernel function which was placed at each data points.
\end{itemize}

\subsubsection*{Logistic Regression (LR)}
We implemented the weighted version of logistic regression \cite{Hosmer04}. This method can be used for binary problem only. The log likelihood estimate of the regression coefficients is used. The weights of categories are defined as the fraction of the $i$th category cases among all cases. Logistic regression gives only one result because there is no option to customize the method except by the set of input features. In our  study we tested 2,048 LR models per drug.

\subsubsection*{Na{\"i}ve Bayes (NB)}
We implemented the standard version of na{\"i}ve Bayes \cite{Russell95}. All attributes which contained $\leq$20 different values were interpreted as categorical and the standard contingency tables were calculated for such attributes. Calculated contingency tables are used to estimate the conditional probabilities. Attributes which contain more than 20 different values were interpreted as continuous.  The mean and the variance were calculated  for continuous attributes instead of the contingency tables. We calculated the isolated mean and variance for each value of the output attribute. The conditional probability of a specified outcome $o$ and a specified value of the attribute $x$ were calculated as the value of the probability density function for a normal distribution at point $x$ with matched mean and variance, which were calculated for the outcome $o$. This method has no customization options and was tested on different sets of input features. In the study we tested  2,048 NB models per drug.

\subsubsection*{Random Forest (RF)}
Random forests were proposed by Breiman \cite{Breiman01} for building a predictor ensemble with a set of decision trees that grow in randomly selected subspaces of the data \cite{Biau12}.  The random forests classification procedure consists of a collection of tree structured classifiers ${h\big(x,\Theta_k ),k=1,...}$, where the ${\Theta_k}$ are independent identically distributed random vectors and each tree casts a unit vote for the most popular class at input $x$” \cite{Breiman01}.

In a random forest, each tree is constructed using a different bootstrap sample from the original data \cite{Hastie09}. In standard trees, each node is split using the best split among all variables. In a random forest, each node is split using the best among a subset of predictors randomly chosen at that node \cite{Liaw02}.

Random forests try to improve on bagging by `de-correlating' the trees. Each tree has the same expectation \cite{Hastie09}.
The forest error rate depends on two things \cite{Breiman01}. The first is the correlation between any two trees in the forest. Increasing the correlation increases the forest error rate. The second is the strength of each individual tree in the forest. A tree with a low error rate is a strong classifier. Increasing the strength of the individual trees decreases the forest error rate. The random forest algorithm builds hundreds of decision trees and combines them into a single model \cite{Williams11}. In the case study we tested 2,048 RF models per drug.

\subsubsection*{Criterion for selecting the best method}\label{CriterionOfTheBestMethod}
A number of different criteria exist for the selection of the best classifier. The criterion we used was to pick the method such that the minimum vetween sensitivity and specificity was maximised.
If minimum between sensitivity and specificity is the same for two classifiers, then we select the classifier with the maximal sum of the sensitivity and specificity. Classifiers with sensitivity or specificity less than 50\% were not considered. There are several approaches to test the quality of classifier: usage of isolated test set, $n$-fold cross validation and Leave-One-Out Cross Validation \big(LOOCV\big) \cite{Arlot10}. LOOCV is used for all tests in this study. There are some problems with classifier quality estimation for the technique like decision tree and random forest. These problems are considered in details by Hastie et al \cite{Hastie09}.

\section*{Results}
The data set contains seven categories of drug uses: `Never used', `Used over a decade ago', `Used in last decade', `Used in last year', `Used in last month', and `Used in last week'. We form four problems based on four dichotomies of these classes (see `Drug use' Section): the decade-, year-, month-, and week-based user/non-user separations.
We identified the relationship between personality profiles (NEO-FFI-R) and drug consumption for the decade-, year-, month-, and week-based classification problems. We evaluated the risk of drug consumption for each individual according to their personality profiles. This evaluation was performed separately for each drug for the decade-based user/non-user separation. We also analysed interrelations between the individual drug consumption risks for different drugs.  Part of the results was presented in the preprint~\cite{Fehrman15}.
In addition, in the `Pleiades of drugs' Section we focus on use of correlation pleiades of drugs. We define three pleiades: heroin pleiade, ecstasy pleiade, and benzodiazepines pleiade with respect to the decade-, year-, month-, and week-based user/non-user separations.

\subsection*{Descriptive statistics}
 The descriptive statistics for five factors are presented in Table~\ref{tab:2}: means, standard deviations, normative data (means and standard deviations for the population following McCrae \& Costa (2004) \cite{McCrae04}), 95\% confidence intervals, kurtosis (kurtosis is a measure of flatness/`peakedness' of the distribution shape compared to normal distribution) and `skewness' (skewness is a measure of asymmetry of the distribution) for NEO-FFI-R for the full sample).

\begin{table}[!ht]
\begin{adjustwidth}{-2.25in}{0in}
\centering
\caption{ Descriptive statistics (Means, sample Standard Deviations (SD), 95\% Confidence Intervals (CI), normative mean and SD, kurtosis, skewness) for NEO-FFI-R for raw data}
\label{tab:2}       
\begin{tabular}{|c|c|c|c|c|c|c|c|}
\hline
{\bf Factors}&{\bf Mean} &{\bf 95\% CI for mean}&{\bf SD}&{\bf Normative mean}&{\bf Normative SD}&{\bf Kurtosis}&{\bf Skewness} \\\hline
N     &	23.92&	23.51, 24.33&  9.14&	16.83&	7.36&	    -0.55&	 0.11 \\ \hline
E     & 27.58&	27.27, 27.88&  6.77&	29.29&	6.46&		 0.06&	-0.27 \\ \hline
O     &	33.76&  33.47, 34.05&  6.58&	31.29&	6.12&		-0.27&	-0.30 \\ \hline
A     & 30.87&	30.58, 31.16&  6.44&	32.41&	5.42&		 0.13&	-0.26 \\ \hline
C     &	29.44&	29.12, 29.75&  6.97&	33.26&	6.3&		-0.17&	-0.38 \\ \hline
\end{tabular}
\end{adjustwidth}
\end{table}

Pearson's correlation coefficient (PCC or $r$) is employed as a measure of the strength of a linear association between two factors. PCC for all pairs of factors are presented in Table~\ref{tab:3}. Two pairs of factors do not have significant correlation: (1) N and O ($r$=0.017, $p$=0.471); (2) A and O ($r$=0.033, $p$=0.155). However, all other pairs of personality factors are significantly correlated in the sample.

\begin{table}[!ht]
\begin{adjustwidth}{-2.25in}{0in}
\centering
\caption{PCC for NEO-FFI-R for raw data}
\label{tab:3}
\begin{tabular}{|c|c|c|c|c|c|}\hline
{\bf Factors}& {\bf N} &{\bf E}&{\bf O}&{\bf A} &{\bf C} \\ \hline
N&&	-0.432\textsuperscript{**}&	0.017&-0.215\textsuperscript{**}& -0.398\textsuperscript{**} \\ \hline
E&-0.432\textsuperscript{**}&&0.236\textsuperscript{**}&0.159\textsuperscript{**}&0.318\textsuperscript{**}\\\hline
O&0.017&0.236\textsuperscript{**}&&0.033&-0.060\textsuperscript{*} \\\hline
A&-0.215\textsuperscript{**}&0.159\textsuperscript{**}&	0.033&&  0.249\textsuperscript{**} \\ \hline
C&-0.398\textsuperscript{**}&0.318\textsuperscript{**}&	-0.060\textsuperscript{*}&0.249\textsuperscript{**}&\\\hline
\end{tabular}
\begin{flushleft} $p$-value is the probability to observe by chance the same or greater correlation coefficient if data are uncorrelated: $^*p<$0.01; $\textsuperscript{**}p<$0.001.
\end{flushleft}
\end{adjustwidth}
\end{table}

\subsection*{Distribution of number of used drugs}
\begin{figure}[!ht]
\centering
\includegraphics[scale=0.67]{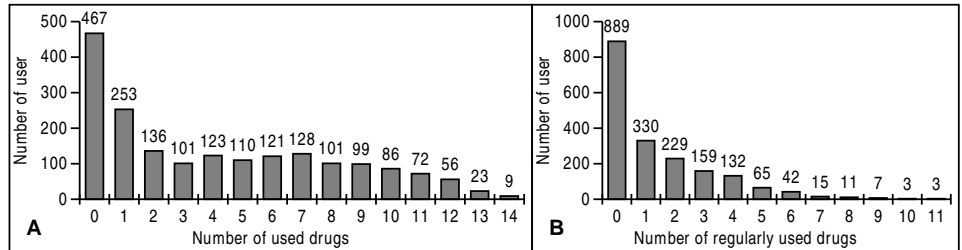}
\caption {{ \bf The histograms of the number of users:} A for the decade-based user/non-user separation),
 B for the month-based user/non-user separation}
\label{NumberOfUserfig:3a}
\end{figure}

The diagrams in Fig~\ref{NumberOfUserfig:3a} show the graph of the number of users versus the number of used illegal drugs for the decade-based (A) and month-based (B) user/non-user separations. In Fig~\ref{NumberOfUserfig:3a}A  we can see that the distribution of the number of users is bimodal with maxima in zero and 7 drugs. In Fig~\ref{NumberOfUserfig:3a}B distribution of number of regular users of illegal drugs looks like the exponential distribution.

\begin{figure}[!ht]
\centering
\includegraphics[scale=0.66]{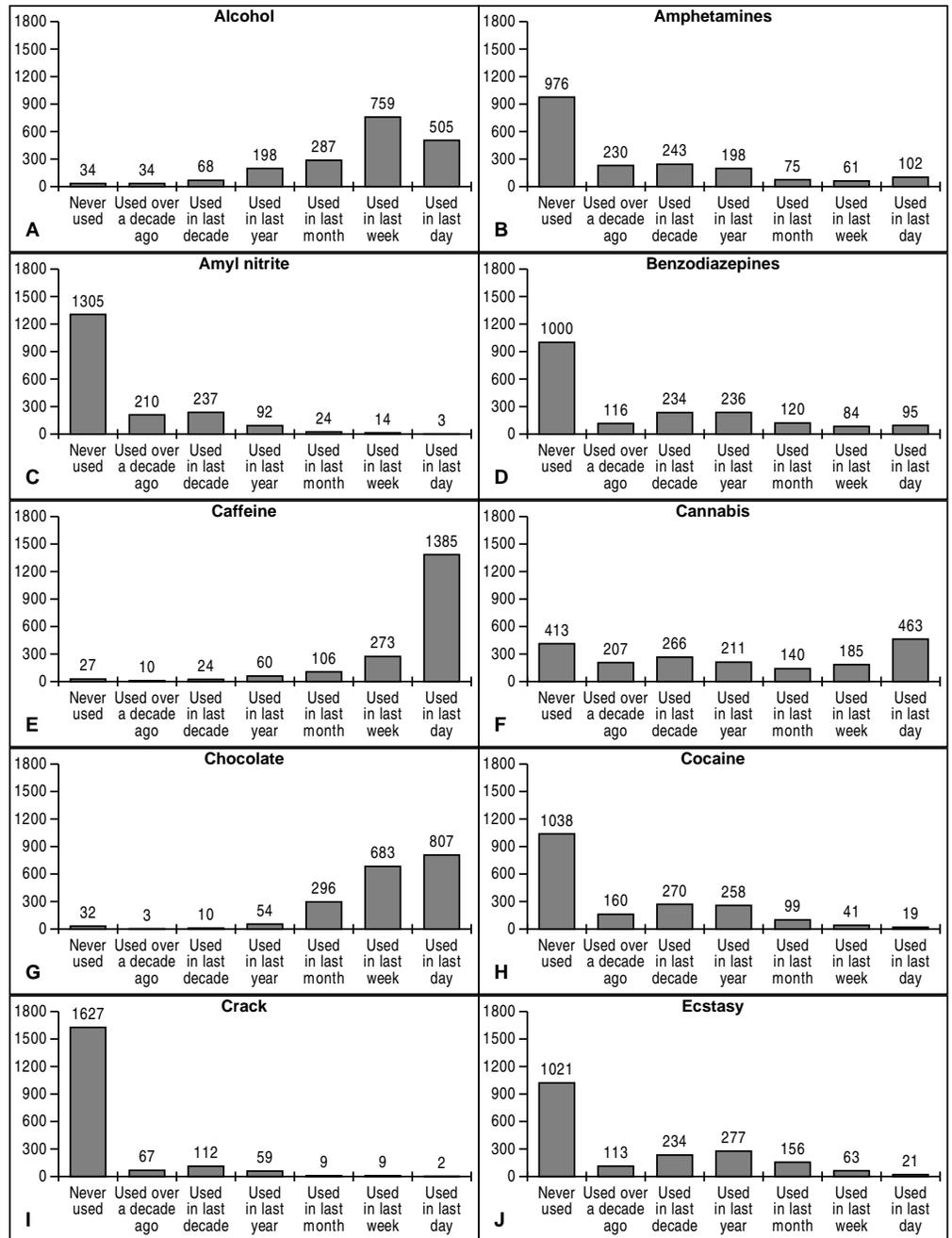}
\caption {{ \bf Distribution of drug usage:} A: Alcohol, B: Amphetamines, C: Amyl nitrite, D: Benzodiazepines, E: Cannabis, F: Chocolate, G: Cocaine, H: Caffeine, I: Crack, and J: Ecstasy}
\label{DrugUserDistrib1fig:3b}
\end{figure}

\begin{figure}[!ht]
\centering
\includegraphics[scale=0.66]{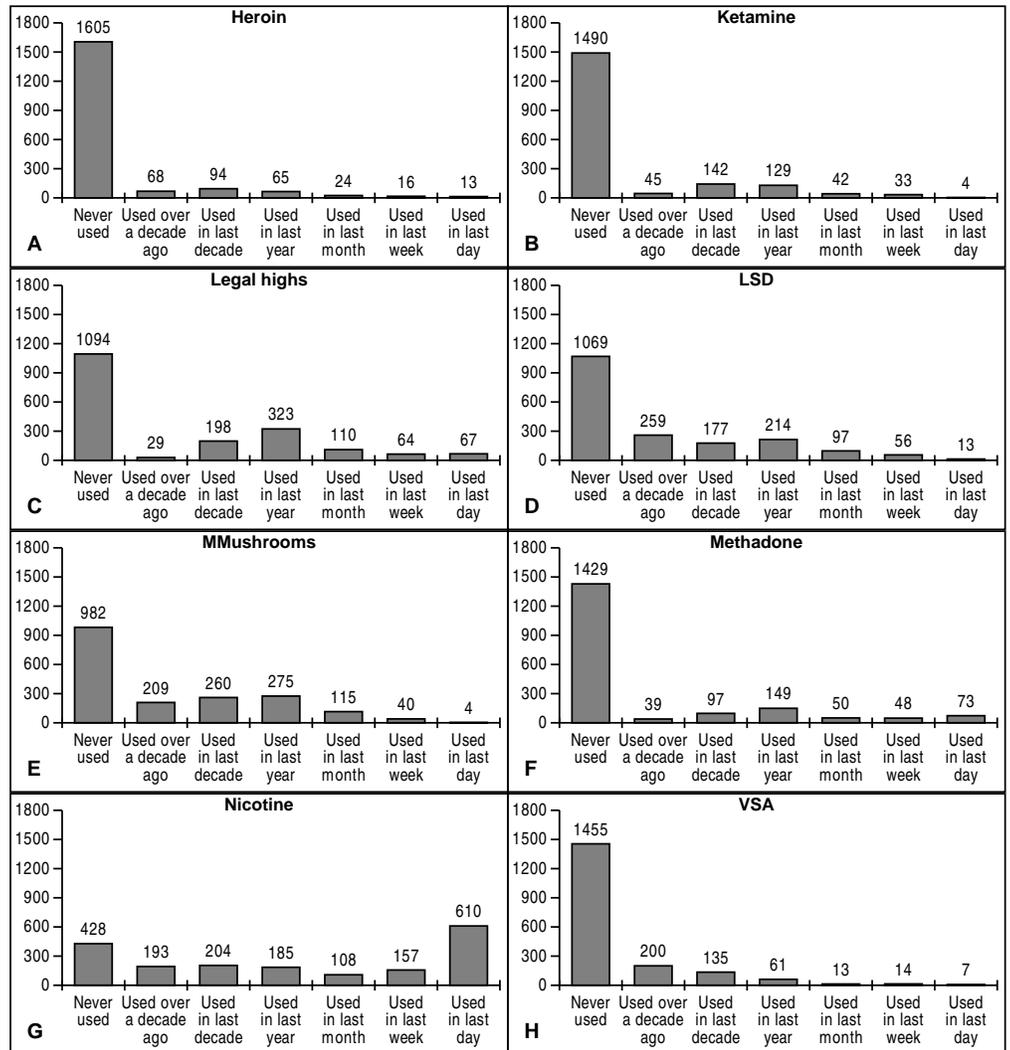}
\caption {{ \bf Distribution of drug usage:} A: Heroin, B: Ketamine, C: Legal highs, D: LSD, E: Methadone, F: Magic mushrooms, G: Nicotine, and H: VSA}
\label{DrugUserDistrib2fig:3c}
\end{figure}

The distributions of the number of users for each drug are presented in Fig~\ref{DrugUserDistrib1fig:3b} and Fig~\ref{DrugUserDistrib2fig:3c}.   Most of distributions have the exponential like shape but several have bimodal distributions. The distributions of the number of users for the three legal drugs have maximum at `Used in last day' or `Used in last week' (see Fig~\ref{DrugUserDistrib1fig:3b}A, E, and G). The distribution of the number of nicotine users (smokers) has three maxima: `Used in last day' for smokers, `Used in last decade' for smokers which broken smoking, and `Never used' (see Fig~\ref{DrugUserDistrib2fig:3c}G). All illegal drug users distribution have maximum in the category `Never used'. However, distribution of cannabis users has two maxima: the main in  the category `Used in last day' and the second in  the category `Never used' (see Fig~\ref{DrugUserDistrib1fig:3b}F).

\subsection*{Comparison of personality traits means for drug users and non-users}
Tables \ref{table4}, \ref{table4a}, \ref{table4b}, and \ref{table4c} demonstrate the mean T-$score_{sample}$ NEO-FFI-R factors for users and non-users for each drug with respect to the decade-, year-, month-, and week-based classification problems respectively (see \nameref{S1_Appendix}). Significant differences in personality factor scores are observed between these groups. The universal relationship between personality profile and risk of drug consumption can generally be described as follows: an increase in scores of N and O entails an increase in the risk of use, whereas an increase in the scores of A and C entails a decrease in risk of use. Thus for each drug, drug users scored higher on N and O, and lower on A and C when compared to drug non-users. The influence of the score of E is drug specific (non-universal).

The introduction of moderate subcategories of T-$score_{sample}$ enables to separates the drugs into five groups for the decade-based user/non-user separation as presented in Table~\ref{tab:5}. Each group can be coded by five moderate subcategories with the moderate profile (N, E, O, A, C). Firstly, the group with the profile $(0, 0, 0, 0, 0)$ includes the users of the legal drugs alcohol, chocolate and caffeine. Thus, the T-$score_{sample}$ for all factors for legal drug consumers does not significantly differ from the sample mean. Secondly, the group of drugs with the profile $(0, 0, +, -, -)$ includes the users of amyl nitrite, LSD, and magic mushrooms. Thirdly, nicotine users form their own group with the profile $(0, 0, +, 0, -)$. Fourthly, the largest group of drugs with the profile $(+, 0, +, -, -)$ includes the users of amphetamines, benzodiazepines, cannabis, cocaine, ecstasy, ketamine and legal highs. Finally, the group with the profile $(+,-, +, -, -)$ includes the users of crack, heroin, VSA and methadone.

\begin{table}[!ht]
\begin{adjustwidth}{-2.25in}{0in}
\centering
\caption{Moderate subcategories of T-$score_{sample}$ with respect to the sample mean for groups of users for the decade-based user/non-user separation: $(0)$ indicates a neutral score, $(+)$ indicates moderately high score, and $(-)$ indicates to moderately low score}
\label{tab:5}
\begin{tabular}{|l|c|c|c|c|c|}\hline
{\bf Drug}&{\bf N}&{\bf E}&{\bf O}&{\bf A}&{\bf C }\\\hline
Alcohol, Chocolate, Caffeine &   $0$   &  $0 $    &  $0$     &  $0$     &  $0$  \\\hline
Amyl nitrite, LSD, and Magic Mushrooms& $0$&	$0$&	$+$&	$-$&	$-$\\\hline
Nicotine                              &$0$&	$0$&	$+$&	$0$&	$-$\\\hline
Amphetamines, Benzodiazepines, Cannabis,&&&&&\\
Cocaine, Ecstasy, Ketamine, and Legal highs&$+$&	$0$&	$+$&$-$&$-$\\\hline
Crack, Heroin, VSA, and Methadone&$+$&	$-$&	$+$&	$-$&	$-$\\\hline
\end{tabular}
\end{adjustwidth}
\end{table}

All drugs for the year-based user/non-user classification are separated into eight groups as presented in Table~\ref{tab:5a}. Each group can be coded by five moderate subcategories with the moderate profile.
\begin{enumerate}
\item The group with the profile $(0, 0, 0, 0, 0)$ includes the users of the legal drugs alcohol, chocolate and caffeine. Thus, T-$score_{sample}$ for all factors for these legal drug consumers does not significantly differ from the sample mean.
\item The group of drugs with the profile $(0, 0, +, -, -)$ includes the users of magic mushrooms. \item The LSD users form their own group with the profile $(0, 0, +, 0, -)$.
\item The  group with the profile $(+, 0, +, -, -)$ includes the users of amphetamines, amyl nitrite, cannabis, cocaine, crack, legal highs and VSA.
\item The group of drugs with the profile $(+, -, +, -, -)$ includes the users of benzodiazepines, heroin, and methadone.
\item The ecstasy users form their own group with the profile $(0, +, +, -, -)$.
\item The ketamine users form their own group with the profile $(+, +, +, -, -)$.
\item The nicotine users form their own group with the profile $(+, 0, +, 0, -)$.
\end{enumerate}

Similarly, all drugs for the month-based user/non-user classification are separated into nine groups (Table~\ref{tab:5b}), and into 10 groups for the week-based user/non-user classification  (Table~\ref{tab:5c}).
\begin{table}[!ht]
\begin{adjustwidth}{-2.25in}{0in}
\centering
\caption{Moderate subcategories of T-$score_{sample}$ with respect to the sample mean for groups of users for the year-based user/non-user separation: $(0)$ indicates a neutral score, $(+)$ indicates moderately high score, and $(-)$ indicates moderately low score}
\label{tab:5a}
\begin{tabular}{|l|c|c|c|c|c|}\hline
{\bf Drug}&{\bf N}&{\bf E}&{\bf O}&{\bf A}&{\bf C }\\\hline
Alcohol, Chocolate, Caffeine &   $0$   &  $0 $    &  $0$     &  $0$     &  $0$  \\\hline
Magic Mushrooms &   $0$    &   $0$    &   $+$    &   $-$    &  $-$  \\\hline
LSD &    $0$   &   $0$    &$+$       &  $0$     &  $-$  \\\hline
Amphetamines, Amyl nitrite, Cannabis, &&&&&\\
Cocaine, Crack, Legal highs, and VSA &   $+$  & $0$      &  $+$     &  $-$     & $-$   \\\hline
Benzodiazepines, Heroin, and Methadone  &   $+$    &  $-$   &  $+$     &  $-$     & $-$   \\\hline
Ecstasy &   $0$   &    $+$   &   $+$  &  $-$     &  $-$  \\\hline
Ketamine&   $+$  & $+$  & $+$  &  $-$  &$-$ \\\hline
Nicotine&   $+$    &   $0$    &   $+$    &  $0$    &  $-$  \\\hline
\end{tabular}
\end{adjustwidth}
\end{table}

\begin{table}[!ht]
\begin{adjustwidth}{-2.25in}{0in}
\centering
\caption{Moderate subcategories of T-$score_{sample}$ with respect to the sample mean for groups of users for the month-based user/non-user separation: $(0)$ indicates a neutral score, $(+)$ indicates moderately high score, and $(-)$ indicates moderately low score}
\label{tab:5b}
\begin{tabular}{|l|c|c|c|c|c|}\hline
{\bf Drug}&{\bf N}&{\bf E}&{\bf O}&{\bf A}&{\bf C }\\\hline
Alcohol, Chocolate, Caffeine &   $0$   &  $0 $    &  $0$     &  $0$     &  $0$  \\\hline
Cannabis and Magic Mushrooms &   $0$    &   $0$    &   $+$    &   $-$    &  $-$  \\\hline
Nicotine &    $+$   &   $0$    &$+$       &  $0$     &  $-$  \\\hline
Amphetamines, Ketamine, and Legal highs &   $+$  & $0$      &  $+$     &  $-$     & $-$   \\\hline
Benzodiazepines, Heroin, and Methadone  &   $+$    &  $-$   &  $+$     &  $-$     & $-$   \\\hline
Ecstasy and  LSD&   $0$   &    $+$   &   $+$  &  $-$     &  $-$  \\\hline
Cocaine and VSA&   $+$  & $+$  & $+$  &  $-$  &$-$ \\\hline
Amyl nitrite &$0$ &$0$ &$0$ &$-$ &$-$ \\\hline
Crack&   $+$    &   $-$    &   $0$    &  $-$    &  $-$  \\\hline
\end{tabular}
\end{adjustwidth}
\end{table}

\begin{table}[!ht]
\begin{adjustwidth}{-2.25in}{0in}
\centering
\caption{Moderate subcategories of T-$score_{sample}$ with respect to the sample mean for groups of users for the week-based user/non-user separation: $(0)$ indicates a neutral score, $(+)$ indicates moderately high score and $(-)$ indicates moderately low score}
\label{tab:5c}
\begin{tabular}{|l|c|c|c|c|c|}\hline
{\bf Drug}&{\bf N}&{\bf E}&{\bf O}&{\bf A}&{\bf C }\\\hline
Alcohol, Chocolate, Caffeine &   $0$   &  $0 $    &  $0$     &  $0$     &  $0$  \\\hline
Cannabis &   $0$    &   $0$    &   $+$    &   $-$    &  $-$  \\\hline
LSD and Magic Mushrooms&    $0$   &   $+$    &$+$       &  $0$     &  $-$  \\\hline
Ketamine &   $0$  & $-$      &  $+$     &  $-$     & $-$   \\\hline
Amphetamines, Benzodiazepines, Heroin, Legal highs, and Methadone & $+$ & $-$ & $+$ & $-$ & $-$   \\\hline
Ecstasy and  VSA&   $0$   &    $+$   &   $+$  &  $-$     &  $-$  \\\hline
Cocaine &   $+$  & $+$  & $+$  &  $-$  &$-$ \\\hline
Nicotine &    $+$   &   $0$    &$+$       &  $0$     &  $-$  \\\hline
Amyl nitrite &$0$ &$-$ &$0$ &$-$ &$-$ \\\hline
Crack&   $+$    &   $-$    &   $-$    &  $-$    &  $-$  \\\hline
\end{tabular}
\end{adjustwidth}
\end{table}

The personality profiles are strongly associated with belonging to groups of the users and non-users of the 18 drugs. We found that the N and O score of drug users of all 18 drugs are moderately high $(+)$ or neutral $(0)$, and the A and C scores of drug users are moderately low $(-)$ or neutral $(0)$.  Detailed  results  can be seen in Tables~\ref{tab:5}, \ref{tab:5a}, \ref{tab:5b}, and \ref{tab:5c}.

The effect of the E score is drug specific. All drugs are divided in three groups with respect to the E score of users (in the year-, month-, and week-based classification problems)  (see Tables~\ref{tab:5}, \ref{tab:5a}, \ref{tab:5b} and \ref{tab:5c}). For example, for the week-based user/non-user separation the E score is:
\begin{itemize}
\item Moderately low  $(-)$ in groups of users of amphetamines, amyl nitrite,  benzodiazepines, heroin,  ketamine, legal highs,  methadone, and crack;
\item Moderately high $(+)$ in groups of users of cocaine, ecstasy, LSD, magic mushrooms, and VSA;
\item Neutral $(0)$  in groups of users of  alcohol, caffeine, chocolate, cannabis, and nicotine.
\end{itemize}

Tables~\ref{tab:6}, \ref{tab:6A}, \ref{tab:6B}, and  \ref{tab:6C} represent significant differences between the means of the personality traits for the group of users and the group of non-users for the decade-, year-, month-, and week-based classification problems.  A 99\% significance level is used ($p$-value is less than 0.01).

For example, three out of the six groups for the decade-based user/non-user separation correspond to legal drugs  (see Table~\ref{tab:6}). Chocolate does not have significant difference between users and non-users for all factors. Alcohol users and non-users have a  significant difference only in the C score, and caffeine users and non-users have a  significant difference only in the O score. Amyl nitrite, LSD, and magic mushrooms form a group with significant differences between users and non-users for three factors: O, A, and C. The next group contains amphetamines, cannabis, cocaine, crack, ecstasy, heroin, ketamine, legal highs, nicotine, and VSA and have significant differences between users and non-users for four factors: N, O, A, and C. The last group consists  of two drugs: benzodiazepines and methadone.  Groups of users differ significantly from groups of  non-users for these drugs in values of all factors.

\begin{table}[!ht]
\begin{adjustwidth}{-2.25in}{0in}
\centering
\caption{Significant differences of means for groups of users and non-users for the decade-based user/non-user separation.
Symbol ` $\Downarrow$ ' corresponds to significant difference where the mean in users group is less than mean in non-users group and symbol `$\Uparrow$ corresponds to significant difference where the mean in users group is greater than the mean in non-users group. Empty cells corresponds to insignificant differences. Difference is considered as significant if $p$-value is less than 0.01).} \label{tab:6} \begin{tabular}{|l|c|c|c|c|c|}\hline
{\bf Drug} & {\bf N} & {\bf E } & {\bf O} & {\bf A} & {\bf C}\\\hline
Chocolate  &  &  &  &  & \\\hline
Alcohol    &  &  &  &  & $\Downarrow$\\\hline
Caffeine   &  &  & $\Uparrow$ &  & \\\hline
Amyl nitrite, LSD, and Magic Mushrooms           &  &  & $\Uparrow$ & $\Downarrow$ & $\Downarrow$\\\hline
Amphetamines, Cannabis, Cocaine, Crack, Nicotine, &  &  &  &  & \\
 Ecstasy, Heroin, Ketamine, Legal highs, and VSA  & $\Uparrow$ &  & $\Uparrow$ & $\Downarrow$ & $\Downarrow$\\\hline
Benzodiazepines and Methadone  & $\Uparrow$ & $\Downarrow$ & $\Uparrow$ & $\Downarrow$ & $\Downarrow$\\\hline
\end{tabular}
\end{adjustwidth}
\end{table}

\begin{table}[!ht]
\begin{adjustwidth}{-2.25in}{0in}
\centering
\caption{Significant differences of means for groups of users and non-users for the year-based user/non-user separation.
Symbol ` $\Downarrow$ ' corresponds to significant difference where the mean of users group is less than the mean of non-users group and symbol `$\Uparrow$ corresponds to significant difference where the  mean of users group is greater than the mean of non-users group. Empty cells corresponds to insignificant differences. Difference is considered as significant if $p$-value is less than 0.01.} \label{tab:6A} \begin{tabular}{|l|c|c|c|c|c|}\hline
{\bf Drug} & {\bf N} & {\bf E } & {\bf O} & {\bf A} & {\bf C}\\\hline
Chocolate, Alcohol,  Caffeine &  &  &  &  & \\\hline
Amyl nitrite  &  &  &  & $\Downarrow$ & $\Downarrow$\\\hline
LSD           &  &  & $\Uparrow$ &  & $\Downarrow$\\\hline
Ketamine  and Magic Mushrooms      &  &  & $\Uparrow$ & $\Downarrow$ & $\Downarrow$\\\hline
VSA           & $\Uparrow$ &  & $\Uparrow$ &  & $\Downarrow$\\\hline
Amphetamines, Cannabis, Cocaine, &  &  &  &  & \\
Crack, Nicotine,Heroin, and Legal highs  & $\Uparrow$ &  & $\Uparrow$ & $\Downarrow$ & $\Downarrow$\\\hline
Ecstasy  &  & $\Uparrow$ & $\Uparrow$ & $\Downarrow$ & $\Downarrow$\\\hline
Benzodiazepines and Methadone  & $\Uparrow$ & $\Downarrow$ & $\Uparrow$ & $\Downarrow$ & $\Downarrow$\\\hline
\end{tabular}
\end{adjustwidth}
\end{table}

Three illicit drugs, amyl nitrite, crack, and cannabis, do not have significant differences between users and non-users for all factors for the month-based user/non-user classification problem. VSA users and non-users have significant difference only for O  for this problem.  Alcohol and cocaine users have  significant difference from non-users for E only. The group of  ketamine, LSD, and magic mushrooms users have significant differences from non-users for C and O. Heroin users have significant difference from non-users for C, A, and N factors. The next group contains amphetamines, caffeine, chocolate, legal highs, and nicotine: their user differ significantly from non-users for all factors except E.

The significance of the differences of the means for groups of users and non-users for the week-based user definition separates the drugs into 10 groups (see  Table~\ref{tab:6C}).

\begin{table}[!ht]
\begin{adjustwidth}{-2.25in}{0in}
\centering
\caption{Significant differences of means for groups of users and non-users for the month-based user/non-user separation.
Symbol ` $\Downarrow$ ' corresponds to significant difference where the mean of users group is less than the mean of non-users group and symbol `$\Uparrow$ corresponds to significant difference where the mean of users group is greater than the mean of non-users group. Empty cells corresponds to insignificant differences. Difference is considered as significant if $p$-value is less than 0.01.} \label{tab:6B} \begin{tabular}{|l|c|c|c|c|c|}\hline
{\bf Drug} & {\bf N} & {\bf E } & {\bf O} & {\bf A} & {\bf C}\\\hline
Amyl nitrite, Crack, and Cannabis &  &  &  &  & \\\hline
VSA           &  &  & $\Uparrow$ &  & \\\hline
Alcohol and Cocaine &  & $\Uparrow$ &  &  & \\\hline
Ketamine, LSD, and Magic Mushrooms      &  &  & $\Uparrow$ &  & $\Downarrow$\\\hline
Heroin        & $\Uparrow$ &  &  & $\Downarrow$ & $\Downarrow$\\\hline
Amphetamines, Caffeine, Chocolate, &  &  &  &  & \\
Legal highs, and Nicotine & $\Uparrow$ &  & $\Uparrow$ & $\Downarrow$ & $\Downarrow$\\\hline
Ecstasy  &  & $\Uparrow$ & $\Uparrow$ & $\Downarrow$ & $\Downarrow$\\\hline
Benzodiazepines and Methadone  & $\Uparrow$ & $\Downarrow$ & $\Uparrow$ & $\Downarrow$ & $\Downarrow$\\\hline
\end{tabular}
\end{adjustwidth}
\end{table}

\begin{figure}[!ht]
\centering
\includegraphics[width=1\textwidth]{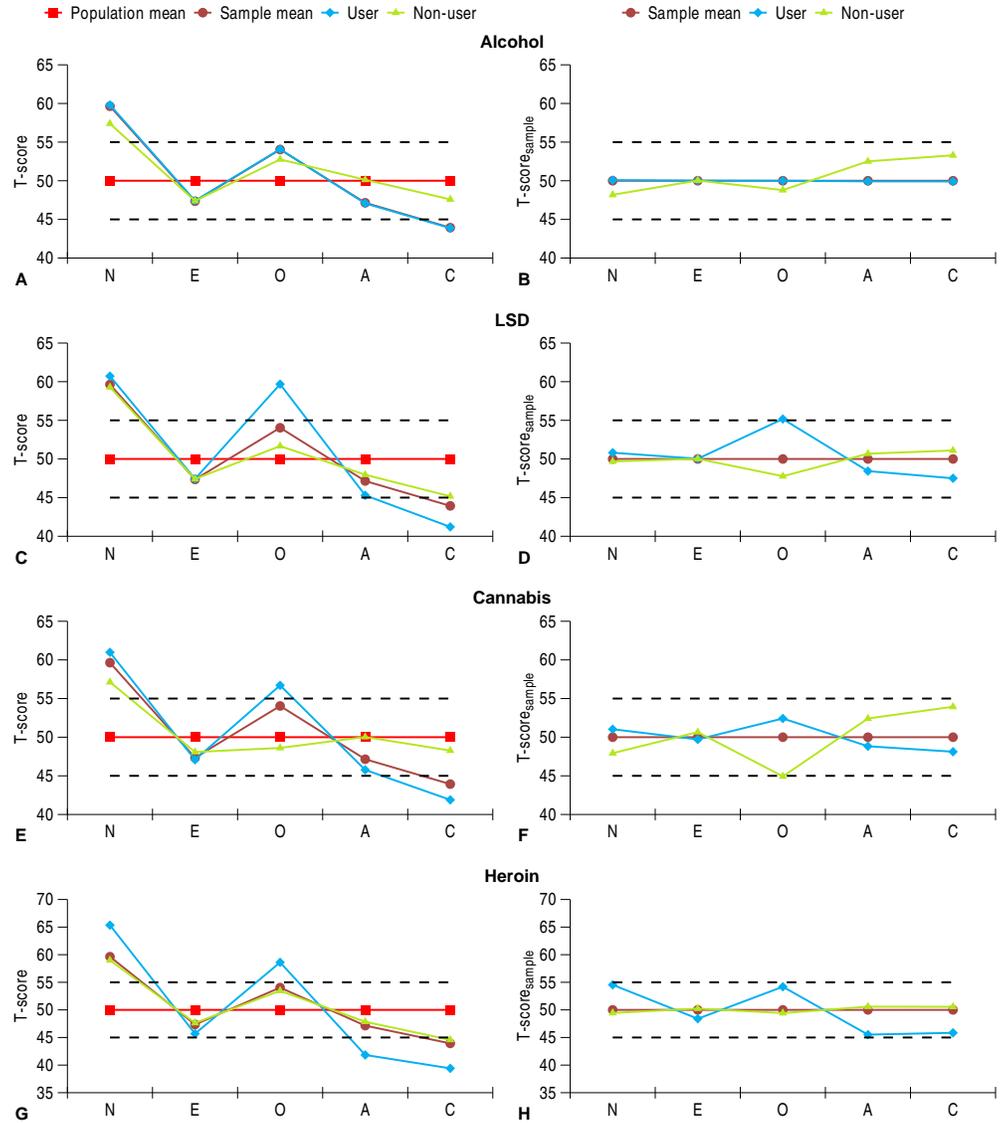}
\caption {{\bf Average personality profiles for the decade-based user/non-user separation.} T-scores with respect to the population norm mean (left column) and T-score$_{sample}$  with respect to the sample means (right column) for: A \& B: Alcohol, C \& D: LSD, E \& F: Cannabis, and G \& H: Heroin }
 \label{Averagepersonalityfig:4}
\end{figure}

\begin{figure}[!ht]
\centering
\includegraphics[width=1\textwidth]{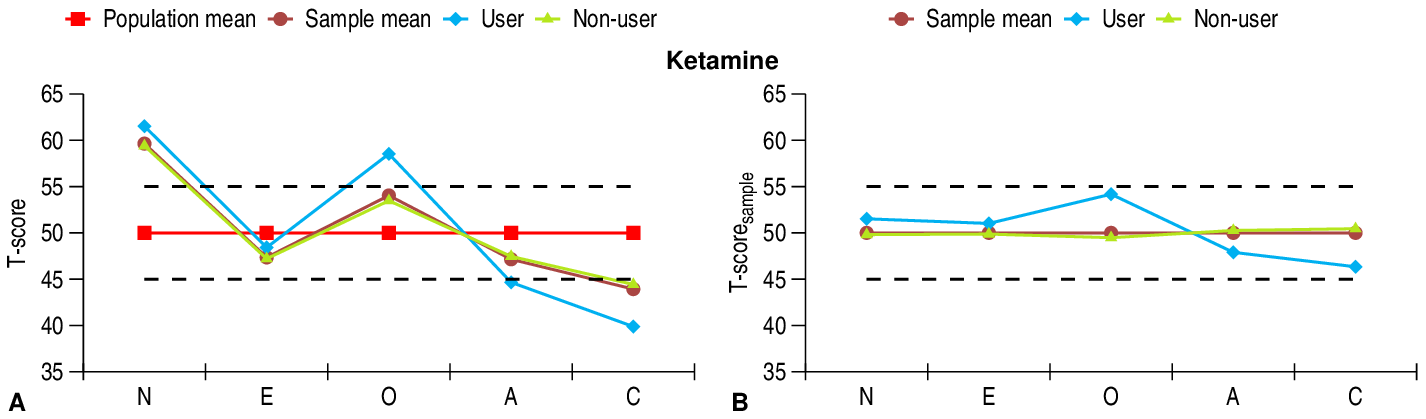}
\caption {{\bf Average personality profiles for Ketamine for the year-based user/non-user separation}. A: T-scores with respect to the population norm mean  and B: T-score$_{sample}$  with respect to the sample means}
 \label{Averageketamine}
\end{figure}

\begin{figure}[!ht]
\centering
\includegraphics[width=1\textwidth]{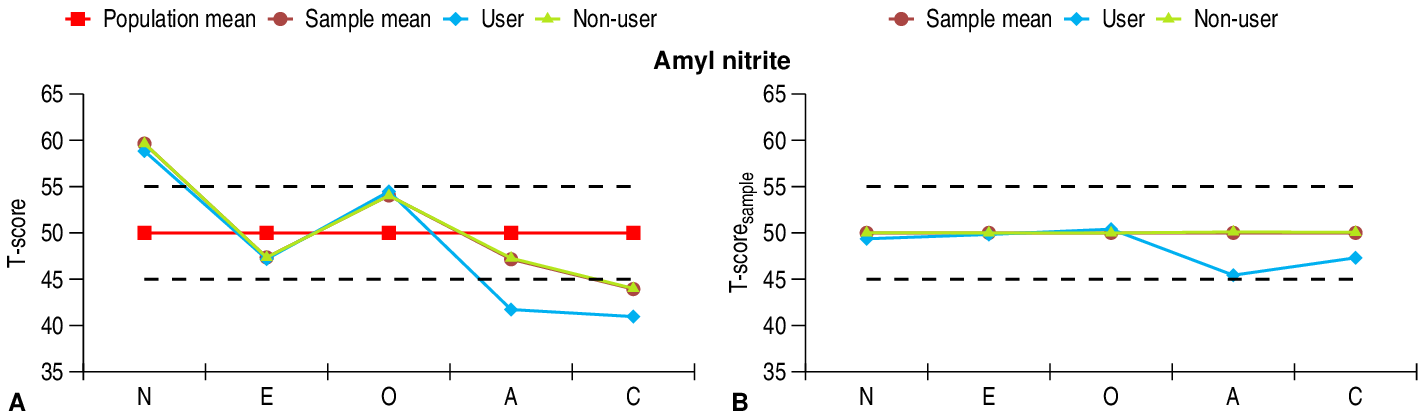}
\caption {{\bf Average personality profiles for Amyl nitrite for the month-based user/non-user separation.}  A:  T-scores with respect to the population norm mean  and B:  T-score$_{sample}$  with respect to the sample means}
 \label{AverageAmyl}
\end{figure}

\begin{figure}[!ht]
\centering
\includegraphics[width=1\textwidth]{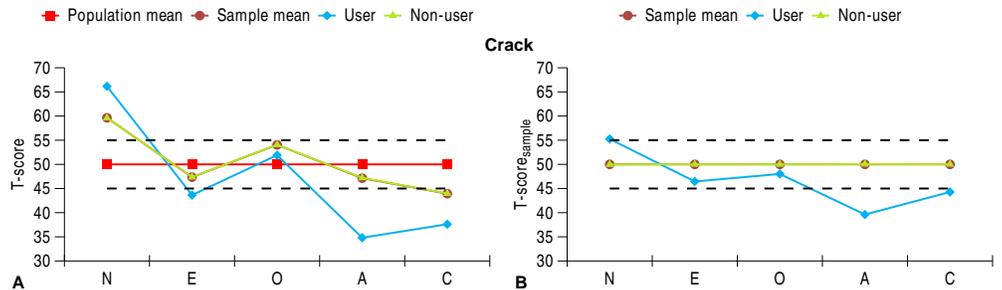}
\caption {{\bf Average personality profiles for Crack for the week-based user/non-user separation.}  A: T-scores with respect to the population norm mean and B: T-score$_{sample}$  with respect to the sample means}
 \label{AverageCrack}
\end{figure}

\begin{table}[!ht]
\begin{adjustwidth}{-2.25in}{0in}
\centering
\caption{Significant differences of means for groups of users and non-users for the week-based user/non-user separation.
Symbol ` $\Downarrow$ ' corresponds to significant difference where the mean of users group is less than the mean of non-users group and symbol `$\Uparrow$ corresponds to significant difference where the mean of users group is greater than the mean of non-users group. Empty cells corresponds to insignificant differences. Difference is considered as significant if $p$-value is less than 0.01.} \label{tab:6C} \begin{tabular}{|l|c|c|c|c|c|}\hline
{\bf Drug} & {\bf N} & {\bf E } & {\bf O} & {\bf A} & {\bf C}\\\hline
Amyl nitrite, Caffeine, Crack, Chocolate and Ketamine  &  &  &  &  & \\\hline
Alcohol  &  & $\Uparrow$ &  &  & \\\hline
Cocaine &  &  &  & $\Downarrow$ & $\Downarrow$\\\hline
Magic Mushrooms and VSA &  &  & $\Uparrow$ &  & \\\hline
Ecstasy and LSD &  &  & $\Uparrow$ &  & $\Downarrow$\\\hline
Cannabis        &  &  & $\Uparrow$ & $\Downarrow$ & $\Downarrow$\\\hline
Heroin        & $\Uparrow$ &  &  & $\Downarrow$ & $\Downarrow$\\\hline
Methadone      & $\Uparrow$ & $\Downarrow$ &  & $\Downarrow$ & $\Downarrow$\\\hline
Amphetamines and Nicotine  & $\Uparrow$ &  & $\Uparrow$ & $\Downarrow$ & $\Downarrow$\\\hline
Benzodiazepines and  Legal highs & $\Uparrow$ & $\Downarrow$ & $\Uparrow$ & $\Downarrow$ & $\Downarrow$\\\hline
\end{tabular}
\end{adjustwidth}
\end{table}
Mean values of factor scores for groups of drug users and non-users for the decade-based user/non-user separation are presented in Table~\ref{table4} and are depicted in Fig~\ref{Averagepersonalityfig:4}. A single drug was chosen to be plotted for each group, due to the fact that the shapes of the profile for all drugs in  one group are very similar. Nicotine is not plotted, since the factor scores of nicotine users are similar to those of the group consisting of amyl nitrite, LSD and magic mushrooms. Fig~\ref{Averagepersonalityfig:4} represents T-score graphs of the mean of personality factor scores for the groups of users and non-users with respect to the population norm mean (the left column) and with respect to the sample mean (the right column) for alcohol, LSD, cannabis, and heroin.
 Graphs of the same type are presented for the year-based classification problem for ketamine in Fig~\ref{Averageketamine}, for amyl nitrite for the month-based classification problem in  Fig~\ref{AverageAmyl} and for crack for the week-based classification problem in Fig~\ref{AverageCrack}. Mean values of factor scores for groups of drug users and non-users for the year-, month- and week-based user/non-user separations are presented in Tables~\ref{table4a}, \ref{table4b} and \ref{table4c}.

\subsection*{Correlation between usage of different drugs}
Usage of each drug is a binary variable (users or non-users) for all versions of user definition. Tables~\ref{tab:14} and \ref{tab:14a} contain PCCs, which are computed for each pair of the drug usages for the decade- and year-based user/non-user separations respectively (see \nameref{S2_Appendix}). The majority of the PCCs are significant, due to the fact that the sample size is 1885.

The correlation in 124 pairs of drug usages from a totality of 153 pairs  have for the decade-based  classification problem have \emph{p}-values less than 0.01 (\emph{p}-value is the probability to observe by chance the same or greater correlation coefficient for uncorrelated variables). It is necessary to employ a {\em multi-testing} approach when testing 153 pairs of drug usages in order to estimate the significance of the correlation \cite{Benjamini95}. We apply the most conservative technique, the Bonferroni correction, and  used the Benjamini-Hochberg (BH) step-up procedure \cite{Benjamini95} to control the False Discovery Rate (FDR) in order to estimate the genuine significance of these correlations. There are 115 significant  correlation coefficients with Bonferroni corrected \emph{p}-value 0.001. The BH step-up procedure with threshold of FDR equal to 0.01 defines 127 significant correlation coefficients.

\begin{figure}[!ht]
   \centering
   \includegraphics[scale=0.7]{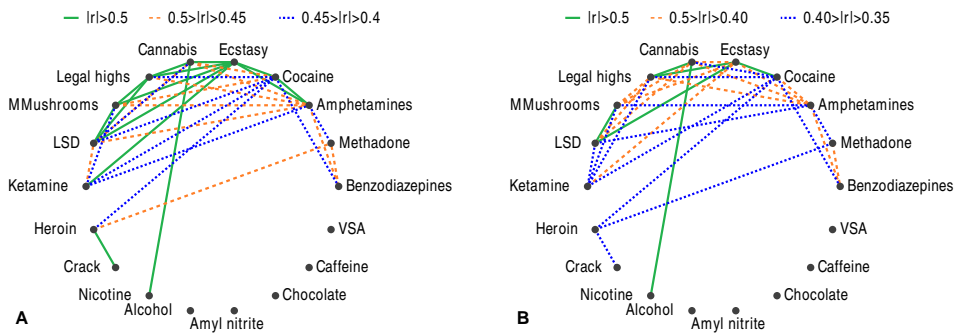}
   \caption {{ \bf Strong drug usage correlations:} A: for the decade-based classification  problem and B: for the year-based classification problem}
   \label{Strongdrugusfig:5}
\end{figure}

However, a significant correlation does not necessarily imply a strong association or causality. For example, the correlation coefficient for alcohol usage and amyl nitrate usage is significant (i.e. the $p$-value is equal to 0.0013) but the value of this coefficient is equal to 0.074, and thus cannot be considered as an important association. We consider correlations with absolute values of PCC  $|r|\ge 0.4$. Fig~\ref{Strongdrugusfig:5} sets out all significant identified correlations greater than 0.4. In this study for the decade-based classification problem we consider the correlation as weak if $|r|<0.4$, medium if $0.45>|r|\ge 0.4$, strong if $0.5>|r|\ge 0.45$ ,and very strong if $|r|\ge 0.5$.

The correlation coefficient is high for each pair  from the group: amphetamines, cannabis, cocaine, ecstasy, ketamine, legal highs, LSD, and magic mushrooms, excluding correlations between cannabis and ketamine usage (r=0.302) and between legal highs and ketamine usage (r=0.393) (Fig~\ref{Strongdrugusfig:5}A).
Crack, benzodiazepines, heroin, methadone, and nicotine usages are correlated with one, two, or three other drugs usage (see Fig~\ref{Strongdrugusfig:5}A). Alcohol, amyl nitrite, chocolate, and caffeine usage and VSA are uncorrelated or weakly correlated with usage of all other drugs.

The structure of correlations of the year-based user/non user separation are approximately the same as for the decade-based classification problem (see Fig~\ref{Strongdrugusfig:5}).  We consider correlations with absolute values of PCC  $|r|\ge 0.35$ for the year-based classification. Fig~\ref{Strongdrugusfig:5}B  sets out all identified significant correlations with $|r|>0.35$. The correlation can be interpreted as weak if $|r|<0.35$; medium if   $0.40>|r|\ge 0.35$; strong if $0.5>|r|\ge 0.40$; and very strong if $|r|\ge 0.5$. On base of this similarity of correlation structures we define pleiades for three central drugs: heroin, ecstasy, and benzodiazepines (as described in the `Pleiades of drugs' Section).

\begin{figure}[!ht]
 \centering
\includegraphics[width=1\textwidth]{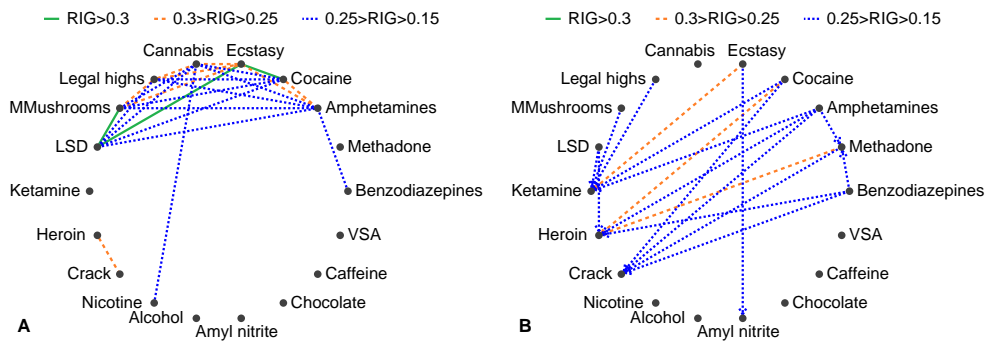}
\caption {{ \bf Pairs of drug usages with high RIG}: A: approximately symmetric RIG and B: significantly asymmetric RIG. In figure B arrow from cocaine usage to heroin usage, for example, means that knowledge of cocaine usage can decrease uncertainty in heroin usage.}
\label{RIGfig:6}
\end{figure}

{\em Relative Information Gain} (RIG) is widely used in data mining to measure dependence between categorical attributes \cite{Mitchell97}. The greater the value of RIG is, the stronger is the indicated correlation.  RIG is zero for independent attributes, is not symmetric and is a measure of mutual information.
For example, the value of RIG for drug 1 usage from usage of drug 2 is equal to a fraction of uncertainty (entropy) in drug 1 usage, which can be removed if the value of drug 2 usage is known. The significance of RIG for binary random variables is the same as for PCC. The majority of RIGs are significant, but have small values. Fig~\ref{RIGfig:6} presents all pairs with RIG \textgreater 0.15.

Fig~\ref{RIGfig:6}A shows `approximately symmetric' RIGs. Here, we call $\mathrm{RIG}(X|Y)$ approximately symmetric if
$$ \frac{| \mathrm{RIG}(X|Y)-\mathrm{RIG}(Y|X)|}{\min \left(\mathrm{RIG}(X|Y), \mathrm{RIG}(Y|X)\right)}<0.2.$$  RIG is approximately symmetric  for each pair from the following group: amphetamines, cannabis, cocaine, ecstasy, legal highs, LSD and magic mushrooms. This group is the same that in Fig~\ref{Strongdrugusfig:5} (except  ketamine).  Fig~\ref{RIGfig:6}B shows asymmetric RIGs. Asymmetric RIGs illustrate pattern significantly different from Fig~\ref{Strongdrugusfig:5}.

\subsection*{Input feature ranking}

It should be stressed that FFM, impulsivity, and sensation-seeking are all correlated.
To identify the most informative features we apply several methods which are described in the `Input feature ranking' Section. The results of the principal variables calculation are represented in Table~\ref{tab:8} for CatPCA quantification, and in Table~\ref{tab:9} for dummy coding of nominal features. Tables \ref{tab:8} and \ref{tab:9} contain the lists of attributes in order from best to worst. The results of Double Kaiser's ranking are shown in the same tables.

\begin{table}[!ht]
\begin{adjustwidth}{-2.25in}{0in}
\centering
\caption{Results of feature ranking. Data include country of residence and ethnicity quantified by CatPCA. FVE is the fraction of explained variance. CFVE is the cumulative FVE. The least informative features are lower located.}
\label{tab:8}
\begin{tabular}{|l|c|c||l|}\hline
\multicolumn{3}{|c||}{{\bf Principal variable ranking}}& { \bf Double Kaiser's ranking}\\\cline{1-3}
\multicolumn{1}{|c|} {\bf Attribute}&	{\bf FVE}&	{\bf CFVE} &\\\hline
Sensation-seeking&	0.192&	0.192&	Extraversion\\\hline
Neuroticism&	    0.153&	0.345&	Conscientiousness\\\hline
Agreeableness&  	0.106&	0.451&	Sensation-seeking\\\hline
Education&	        0.104&	0.555&	Neuroticism\\\hline
Openness &	        0.092&	0.647&	Impulsivity\\\hline
Conscientiousness&	0.088&	0.735&	Openness\\\hline
Extraversion&       0.076&	0.811&	Agreeableness\\\hline
Age&	            0.073&	0.884&	Age\\\hline
Impulsivity&	    0.055&	0.939&	Education\\\hline
Country&	        0.037	&0.976	&Country\\\hline
Gender	&           0.021&	0.997&	Gender\\\hline
Ethnicity&	        0.003&	1.000&	Ethnicity\\\hline
\end{tabular}
\end{adjustwidth}
\end{table}

\begin{table}[!ht]
\begin{adjustwidth}{-2.25in}{0in}
\centering
\caption{ Results of feature ranking. Data include dummy coded country of residence and ethnicity. FVE is the fraction of explained variance. CFVE is the cumulative FVE. The least informative features are lower located. }
\label{tab:9}
\begin{tabular}{|l|c|c||l|}\hline
\multicolumn{3}{|c||}{ { \bf Principal variable ranking}}&{ \bf Double Kaiser's ranking}\\\cline{1-3}
\multicolumn{1}{|c|}{\bf Attribute}&	{\bf FVE}&	{\bf CFVE}& \\\hline
Sensation-seeking&	0.186&	0.186&	Extraversion\\\hline
Neuroticism&	    0.149&	0.335&	Conscientiousness\\\hline
Agreeableness&	    0.103&	0.438&	Sensation-seeking\\\hline
Education&	        0.101&	0.539&	Neuroticism\\\hline
Openness&	        0.089&	0.627&	Impulsivity\\\hline
Conscientiousness&	0.086&	0.714&	Openness\\\hline
Extraversion&	    0.074&	0.787&	Agreeableness\\\hline
Age&	            0.071&	0.858&	Age\\\hline
Impulsivity&	    0.053&	0.911&	Education\\\hline
UK&	                0.027&	0.938&	UK\\\hline
Gender&         	0.020&	0.959&	USA\\\hline
USA	&               0.013&	0.972&	Gender\\\hline
White&           	0.010&	0.982&	Other (country)\\\hline
Other (country)&	0.005&	0.988&	White\\\hline
Canada&	            0.004&	0.991&	Other (ethnicity)\\\hline
Other (ethnicity)&	0.003&	0.994&	Canada\\\hline
Black&	            0.002&	0.995&	Asian\\\hline
Australia&	        0.002&	0.997&	Mixed-White/Black\\\hline
Asian&          	0.001&	0.998&	Australia\\\hline
Mixed-White\/Black&	0.001&	0.999&	Black\\\hline
Republic of Ireland&0.000&	1.000&	Mixed-White/Asian\\\hline
Mixed-White\/Asian&	0.000&	1.000&	Republic of Ireland\\\hline
New Zealand&	    0.000&	1.000&	New Zealand\\\hline
Mixed-Black\/Asian&	0.000&	1.000&	Mixed-Black/Asian\\\hline
\end{tabular}
\end{adjustwidth}
\end{table}

The results of application of sparse PCA are shown in Tables~\ref{tab:10} and \ref{tab:11}. As a result of feature selection we can exclude ethnicity from further consideration. There is a more intriguing effect regarding country of location. Only two countries are informative (in our sample): UK and USA. Furthermore, including country in personality measures does not add much to the  prediction of drug use. To understand the reasons for these two countries' importance in the prediction of drug consumption we compare the statistics for the subsamples: UK - non-UK and USA - non-USA. We calculated the $p$-value of coinciding distribution of personality measurements in each subsample.  We obtained the same results for both divisions into subsamples: all input features have significantly different distributions with a 99.9\% confidence level for UK and non-UK subsamples and likewise for USA – non-USA subsamples. This means that the UK and non-UK samples are biased. The same situation is found for the USA and non-USA samples.

To understand the importance of these features for evaluating of the risk of drug consumption  we perform a simple analysis of these two features for the classification of users and non-users for all drugs (see Table~\ref{tab:7}).
Table~\ref{tab:7} allows us to conclude that chocolate is UK specific drug, crack heroin, legal highs, LSD, magic mushrooms, methadone and VSA are USA specific drugs and all drugs excluding alcohol, amyl nitrite, caffeine and chocolate are mostly used out of UK. Unfortunately, this conclusion is mainly due to the composition of the dataset: participants from the UK (1044; 55.4\%), the USA (557; 29.5\%), Canada (87; 4.6\%), Australia (54; 2.9\%), New Zealand (5; 0.3\%) and Ireland (20; 1.1\%). A total of 118 (6.3\%) came from a diversity of other countries, none of whom individually formed 1\% of the sample, or did not declare the country of location.

Our goal is to predict the risk of drug consumption for an individual. This means that we have to consider individual specific factors. Occupation within a specific country can be considered as important risk factor but we do not have enough data for  countries other than UK and USA. We exclude the country feature from further study for these reasons. As a result, for further study we have 10 input features: age, Edu., N, E, O, A, C, Imp., SS, and gender.

\begin{table}[!ht]
\begin{adjustwidth}{-2.25in}{0in}
\centering
\caption{The result of sparse PCA feature ranking. Data include country of residence and ethnicity quantified by CatPCA. }
\label{tab:10}
\begin{tabular}{|c|c|l|}\hline
{ \bf Step}&	{\bf \# of components}&	{ \bf Removed attributes}\\\hline
1&	5&	Gender and Ethnicity\\\hline
2&	4&	No removed attributes. The retained set of attributes: age, Edu., \\
 &   &N, E, O, A, C, Imp. SS, and country\\\hline
\end{tabular}
\end{adjustwidth}
\end{table}

\begin{table}[!ht]
\begin{adjustwidth}{-2.25in}{0in}
\centering
\caption{ The result of sparse PCA feature ranking. Data include dummy coded country of residence and ethnicity.}
\label{tab:11}
\begin{tabular}{|c|c|l|}\hline
{ \bf Step}&	{ \bf \# of components}& { \bf Removed attributes}\\\hline
1&	8&	Canada	, Other (country), Australia, Republic of Ireland, New Zealand,\\
 &   & Mixed-White/Asian, White, Other (ethnicity), Mixed-White/Black, Asian, \\
 &   &Black and Mixed-Black/Asian\\\hline
2&	5&Gender, UK and USA\\\hline
3&	4&No removed attributes. The retained set of attributes: age, Edu., N, E, O,\\
 &   &A, C, Imp. and SS\\\hline
\end{tabular}
\end{adjustwidth}
\end{table}

\begin{table}[!ht]
\begin{adjustwidth}{-2.25in}{0in}
\centering
\caption{Analysis of UK and USA categories of feature country of residence as a classifier into users and non-users for the decade-based classification for all drugs.}
\label{tab:7}
\begin{tabular}{|l|c|c|c|c|c|c|c|c|}\hline
  { \bf Drug} &\multicolumn{2}{l|}{ { \bf UK means user}}&\multicolumn{2}{l}{{ \bf UK means non-user}}&\multicolumn{2}{|l|}{{ \bf USA means user}}&\multicolumn{2}{l|}{{ \bf USA means non-user}}\\ \cline{2-9}
       &{\bf Sens.} &{\bf Spec.}&{\bf Sens.}&{\bf Spec.}&{\bf Sens.}&{\bf Spec.}&{\bf Sens.}&{\bf Spec. }\\
       &{\bf (\%)} & {\bf (\%)}& {\bf (\%)}&  {\bf (\%)}&{\bf (\%)} &  {\bf (\%)} & {\bf (\%)} & {\bf (\%)} \\\hline
Alcohol     	&55&	38&	45&	62&	30&	81&	70&	19        \\\hline
Amphetamines    &32&    31& 68& 69&	48&	81&	52&	19    \\\hline
Amyl nitrite    &61&	46&	39&	54&	19&	68&	81&	32        \\\hline
Benzodiazepines &33&    29& 67& 71&49&84&	51&	16     \\\hline
Caffeine        &55&	16&	45&	84&	30&	86&	70&	14         \\\hline
Cannabis        &40&	12& 60& 88&42&95&	58&	5      \\\hline
Chocolate       &56&    63& 44& 37&	29&	54&	71&	46         \\\hline
Cocaine	        &38&	35&	62&  65 &	44&	79&	56&	21 \\\hline
Crack           &24&	41& 76&  59 &  59 &  74 &	41&	26 \\\hline
Ecstasy         &35&	31& 65&  69 &	45&	81&	55	&19    \\\hline
Heroin          &16&	40& 84&  60 &  68 &  75 &	32&	25 \\\hline
Ketamine        &42&	42&  58 &  58 &	37&	72&	63	&28    \\\hline
Legal highs     &30&	28&  70 &  72 &  51 &  85 &	49&	15 \\\hline
LSD             &20&    30&  80 &  70 &  56 &  82 &	44&	18 \\\hline
MMushrooms      &26&	27&  74 &  73 &  53 &  84 &	47&	16 \\\hline
Methadone       &17&	34&  83 &  66 &  67 &  81 &	33&	19 \\\hline
Nicotine        &46&	26&  54 &  74 &	37  &  85 &	63&	15 \\\hline
VSA             &20&	40&  80 &  60 &  63 &  75 &	37&	25 \\\hline
\end{tabular}
\end{adjustwidth}
\end{table}

\subsection*{Selection of the best classifiers for the decade based classification problem}

The first step for the risk evaluation is the construction of classifiers. We tested the eight methods described in the `Risk evaluation methods' Section and selected the best one. The results of the classifier selection are presented in Table~\ref{tab:12}. This table shows that for all drugs except alcohol, cocaine and magic mushrooms, the sensitivity and specificity are greater than 70\%, which  is an unexpectedly high accuracy.

Recall that we have 10 input features: age, Edu., N, E, O, A, C, Imp., SS, and gender; each of them is an important predictor for at least five drugs. However, there is no single most effective classifier which uses all input features. The maximal number of used attributes is six out of 10 and the minimal number is two. In Section `Criterion of the best method selection' the best method is defined as the method which maximises value of the minimum of sensitivity and specificity. If the minimum of sensitivity and specificity is the same for two classifiers then the classifier with the maximal sum of the sensitivity  and specificity is selected from them.
Table~\ref{tab:12} shows which different sets of attributes are used in the best user/non usser classifier for each different drugs.

\begin{table}[!ht]
\begin{adjustwidth}{-2.25in}{0in}
\centering
\footnotesize{
\caption {The best results of the drug users classifiers. Symbol `X' means the used input feature. Results are calculated by LOOCV.}
\label{tab:12}
\begin{tabular}{|l|c|c|c|c|c|c|c|c|c|c|c|c|c|c|} \hline
{ \bf Target feature} &{ \bf  Classifier} &{ \bf Age}&{ \bf Edu.}&{ \bf N} &{ \bf E} &{ \bf O} &{ \bf A} &{ \bf  C} &{ \bf Imp.} &{ \bf SS }&{ \bf Gender}&{ \bf  Sens.} &{ \bf Spec.}&{ \bf Sum}  \\
               &    &   & &  &  & & & & &  &&{ \bf (\%)}& { \bf (\%)}&{ \bf (\%)} \\   \hline
Alcohol        &LDA	& X&X &X & & & & &  &  X   & X &  75.34  & 63.24 &138.58 \\\hline
Amphetamines   &DT  & X&  &	X& &X& &X & X & X&   & 	81.30  &71.48  &152.77  \\\hline
Amyl nitrite   &DT	&	& &X &	& X& &X  & &X  &  &  73.51 &	87.86 &	161.37 \\\hline
Benzodiazepines& DT&X&	&X&	X&	&	&	&	X&	X&	X&	70.87&	71.51&	142.38\\ \hline
Cannabis       &DT&X&	X&	&&	X&	X&	X&	X&&	&79.29&	80.00&	159.29\\ \hline
Chocolate      &$k$NN&	X&	&	&	X&	&	&	X&&	&X&	72.43&	71.43&	143.86 \\ \hline
Cocaine	       &DT&X&	&&	&X&	X&	&	X&	X&	&	68.27&	83.06&	151.32\\ \hline
Caffeine       &$k$NN&X & X&	&&	X&	X&	&	X&	&	&	70.51&	72.97&	143.48\\ \hline
Crack	       &DT&	&&	&	X&	&	&	X&	&	&	&	80.63&	78.57&	159.20 \\ \hline
Ecstasy        &DT&	X&	&&	&&	&	&	&	X&	X&	76.17&	77.16&	153.33\\ \hline
Heroin	       &DT&X&	&	&&	&&	&	X&	&	X&	82.55&	72.98&	155.53 \\ \hline
Ketamine       &DT&	X&	&&	X&	&	X&	&	X&	X&	&	72.29&	80.98&	153.26 \\ \hline
Legal highs    &DT&	X&	&&	&	X&	X&	X&	&	X&	X&	79.53&	82.37&	161.90 \\ \hline
LSD	           &DT&X&	&X	&X	&X	&	&	&X	&	&X	&85.46&	77.56&	163.02 \\ \hline
Methadone      &DT&	X&	X&	&	X&	X&	&&	&&	X&	79.14&	72.48&	151.62\\ \hline
MMushrooms     &DT	&	&	&	&X	&&	&&	&	&X	&65.56&	94.79&	160.36 \\ \hline
Nicotine       &DT	&	&&X	&X	&	&	&X	&	&	&X	&71.28&	79.07&	150.35 \\ \hline
VSA            &DT&X&	X&	&	X&	&	X&	X&	&	X&	&	83.48&	77.64&	161.12 \\
\hline
\end{tabular}}
\end{adjustwidth}
\end{table}

\begin{figure}[!ht]
\centering
\includegraphics[scale=0.8]{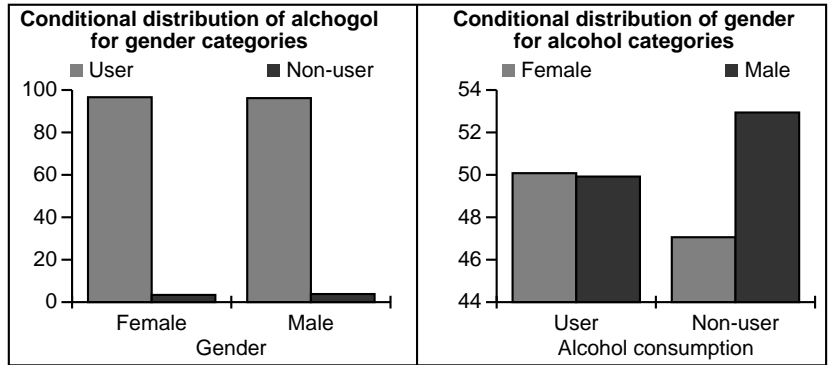}
\caption {{ \bf Conditional distribution for gender and alcohol.}}
\label{ConditionaldisAlcoholgenderfig:9}
\end{figure}

\begin{figure}[!ht]
\centering
\includegraphics[scale=0.8]{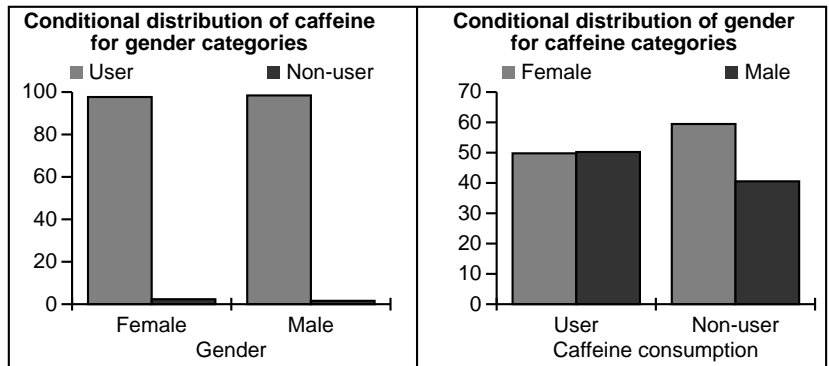}
\caption {{ \bf Conditional distribution for gender and caffeine.}}
\label{ConditionaldisCaffaingenderfig:10}
\end{figure}

\begin{figure}[!ht]
\centering
\includegraphics[scale=0.8]{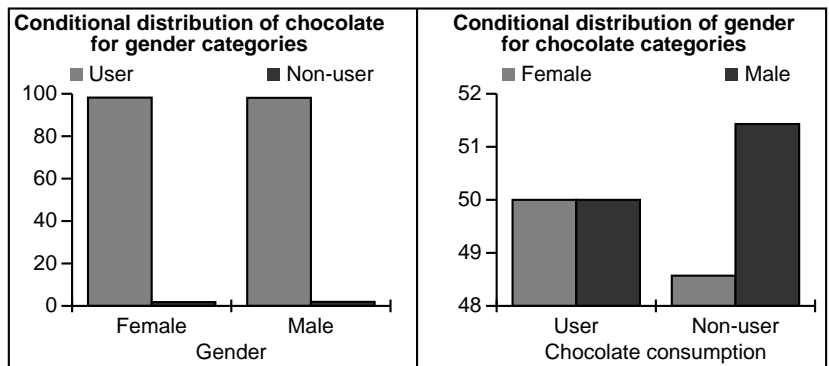}
\caption {{ \bf Conditional distribution for gender and chocolate.}}
\label{ConditionaldisChocolategenderfig:11}
\end{figure}

\begin{figure}[!ht]
\centering
\includegraphics[scale=0.8]{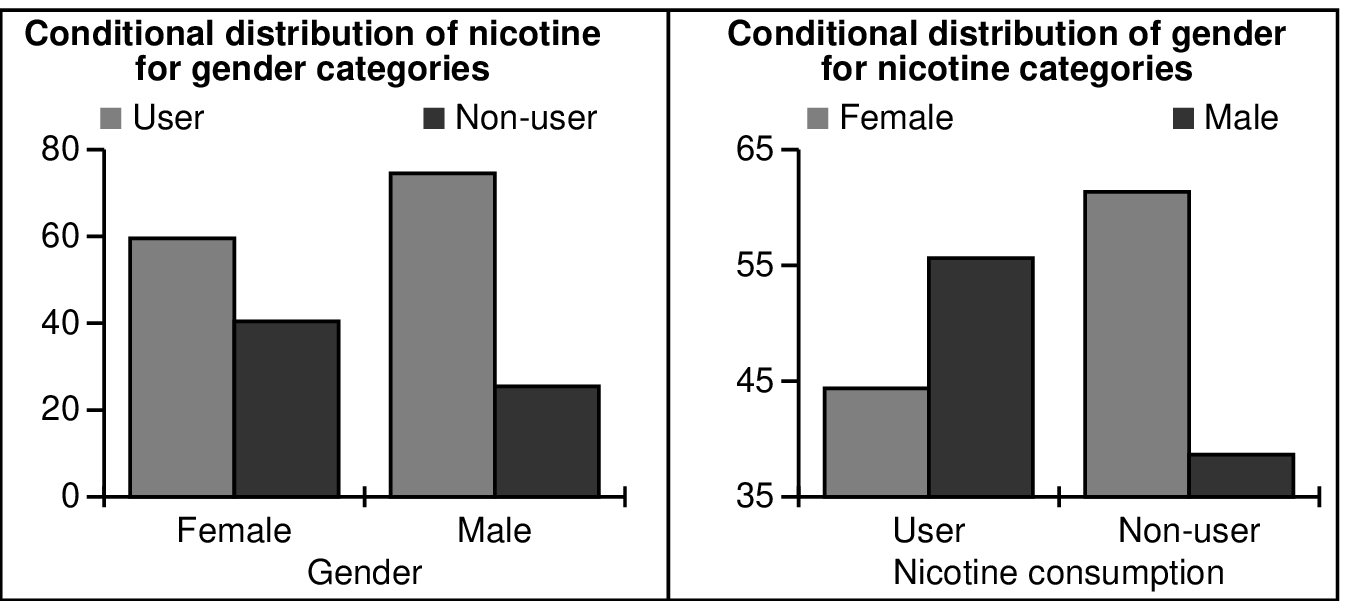}
\caption {{ \bf Conditional distribution for gender and nicotine.}}
\label{ConditionaldisNicotinegenderfig:19}
\end{figure}

The use of a feature in the best classifier can be interpreted as `ranking by fact'. We can note that this ranking by fact is very different from other rankings presented in Tables~\ref{tab:8} and \ref{tab:10}. For example, age is shown not to be the most informative measure in accordance with Tables~\ref{tab:8} and \ref{tab:10}, but it is used in the best classifiers for 14 drugs. The second most used input feature is gender, which is considered as non-informative by Sparse PCA (Table~\ref{tab:10}) and as one of the least informative by other methods (Table~\ref{tab:8}).
This means that consumption of these 10 drugs is gender dependent. We found some unexpected outcomes: for example, in the dataset the fraction of females who are alcohol users is greater than that fraction of males (Fig~\ref{ConditionaldisAlcoholgenderfig:9}) but the greater proportion of males drink coffee (Fig~\ref{ConditionaldisCaffaingenderfig:10}). The fraction of males who do not eat chocolate is greater than for females (Fig~\ref{ConditionaldisChocolategenderfig:11}). The conditional distributions for nicotine meets the common sense expectations (Fig~\ref{ConditionaldisNicotinegenderfig:19}).


The next most informative input features are E and SS which are used in the best classifiers for nine drugs. Features O, C, and Imp. are used in the best classifiers for eight drugs. Features N and A are used in the best classifiers for six drugs. Thus, personality factors are associated with drug use and each one impacts on specific drugs. Finally, Edu.  is used in the best classifiers for five drugs (see Table~\ref{tab:12}).

To predict the usage of the most drugs DT is the best classifier (see Table~\ref{tab:12}). LDA is the best classifier for alcohol use with five input features, and has sensitivity 75.34\% and specificity 63.24\%. $k$NN is the best classifier for chocolate and caffeine users. These $k$NN classifiers use four features for chocolate and five features for caffeine.

The drugs can be separated into disjoint groups by the number of attributes used for the best classifiers:
\begin{itemize}
  \item The group of classifiers with two input features contains classifiers for two drugs: crack and magic mushrooms. Both classifiers of this group use the E score.
  \item The group of classifiers with three input features includes classifiers for two drugs: ecstasy and heroin. Both classifiers in this group use age  and gender and do not use any NEO-FFI factors.
  \item The group of classifiers with four input features includes classifiers for three drugs: amyl nitrite, chocolate, and nicotine. All classifiers of this group use the C score.
  \item The group of classifiers with five input features includes classifiers for five drugs: alcohol, cocaine, caffeine, ketamine, and methadone. All classifiers of this group use age.
  \item The group of classifiers with six input features includes classifiers for six drug users: amphetamines, benzodiazepines, cannabis, legal highs, LSD, and VSA. All classifiers of this group use age.
\end{itemize}

It is important to stress that the {\em attributes which are not used in the best classifiers are not non-informative}. For example, for ecstasy consumption the best classifier is based on age, SS, and gender and has sensitivity 76.17\% and specificity 77.16\%. There exist a DT for usage of the same drug based on age, Edu., O, C, and SS with sensitivity 77.23\% and specificity 75.22\%, a DT based on age, Edu., E, O, and A with sensitivity 73.24\% and specificity 78.22\%, a LR classifier based on age, Edu., O, C, Imp., SS, and gender with sensitivity 74.83\% and specificity 74.52\%, and a $k$NN classifier based on age, Edu., N, E, O, C, Imp., SS, and gender with sensitivity 75.63\% and specificity 75.75\%. This means that for the  risk evaluation of ecstasy usage all input attributes are informative but the required information can be extracted from a smaller subset of the attributes.

\begin{figure}[!ht]
\centering
\includegraphics [scale=0.9]{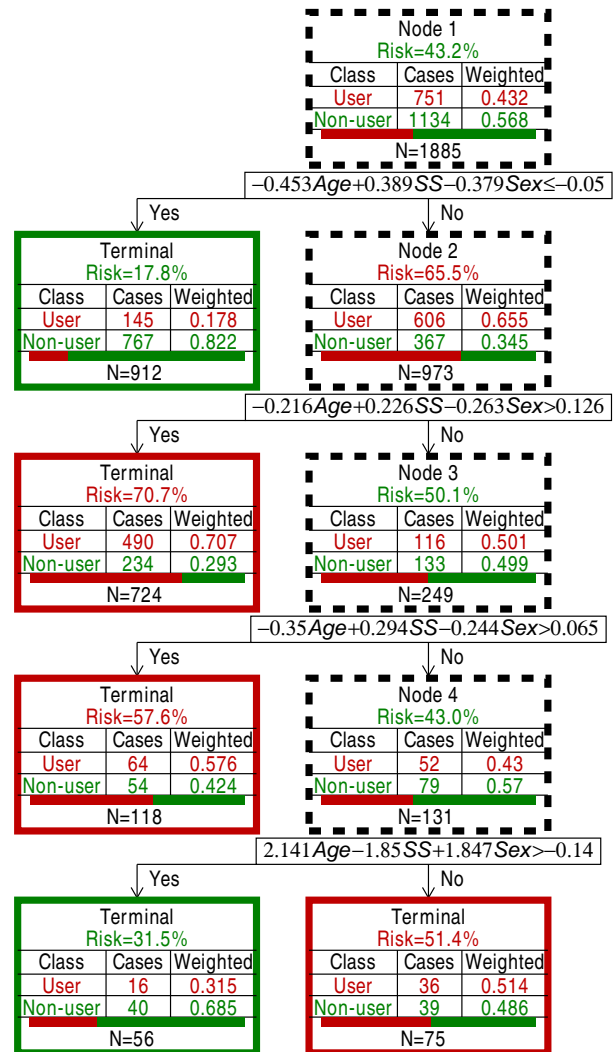}
\caption {{ \bf Decision tree for ecstasy}. Input features are: age, SS, and gender. Non-terminal nodes are depicted with dashed border. Values of age, SS, and gender are calculated by quantification procedures described in the `Input feature transformation' Section. Weight of each case of users class is 1.15 and of non-users class is 1. Column `Weighted' presents normalized weights: weight of each class is divided by sum of weights.}
\label{Treefig7}
\end{figure}

The results presented in Table~\ref{tab:12} were calculated by LOOCV. It should be stressed that the different methods of testing give different sensitivity and specificity. Common methods include calculation of test set errors (the holdout method), k-fold cross-validation, testing on the entire sample (if it is sufficiently large, so-called `na{\"i}ve' method), random sampling, and many others.
For example, a DT formed for the entire sample can have a sensitivity and specificity different from LOOCV \cite{Hastie09}. For illustration, consider the DT for ecstasy, depicted in the Fig~\ref{Treefig7}. It has sensitivity 78.56\% and specificity 71.16\%, calculated using the whole sample. The results of LOOCV for a tree with the same options are presented in the Table~\ref{tab:12}: sensitivity 76.17\% and specificity 77.16\%.

The role of SS is very important for most of the party drugs. In particular, the risk of ecstasy consumption can be evaluated with high accuracy on the basis of age, gender and SS (see Table~\ref{tab:12}, Fig~\ref{Treefig7}, and \ref{RiskMapsfig:8}), and does not need the personality traits from five factors model.

\subsection*{Pleiades of drugs}
\label{Pliad of drug users}

Consider correlations between drug usage for  the year- and decade-based definitions  (Fig~\ref{Strongdrugusfig:5}). It can be seen from Fig~\ref{Strongdrugusfig:5} that the structure of these correlations for the year- and decade-based definitions of drug users is approximately the same.
We found three groups of strongly correlated drugs, each containing several drugs which are pairwise strongly correlated. This means that drug consumption has a `modular structure'. Let us consider a modular structure for the three modules we found:
\begin{itemize}
  \item  Crack, cocaine, methadone, and heroin;
  \item Amphetamines, cannabis, cocaine, ketamine, LSD, magic mushrooms, legal highs, and ecstasy;
  \item Methadone, amphetamines, cocaine and benzodiazepines.
\end{itemize}
The modular structure has clear reflection in the correlation graph, Fig~\ref{Strongdrugusfig:5}.

The idea of merging correlated attributes into `modules' is popular in biology These modules are called  the `correlation pleiades' \cite{Terentjev31,Berg60,Mitteroecker07}.
The concept of correlation pleiades was introduced in biostatistics  in 1931 \cite{Terentjev31}.  Correlation pleiades were used  in evolutionary physiology for the  identification of the modular structure~\cite{Terentjev31,Mitteroecker07,Berg60,Armbruster99}.  Berg presented correlation data from three unspecialized and three specialized pollination species. According to Berg\cite{Berg60}, correlation pleiades are clusters of correlated traits.  This means that in the standard approach to clustering the pleiads do not intersect. The classical clustering methods are referred to as `hard' or `crisp' clustering, meaning that each data object is assigned to only one cluster.
This restriction is relaxed for fuzzy \cite{Bezdek1981} and probabilistic clustering \cite{Krishnapuram1993}. Such approaches are useful when the boundaries between clusters are not well separated.

In our study, correlation pleiades can be applied since the drugs can be grouped in clusters with highly correlated use   (see Fig~\ref{Strongdrugusfig:5}A and \ref{Strongdrugusfig:5}B):
\begin{itemize}
\item The \emph{Heroin pleiad (heroinPl)} includes crack, cocaine, methadone, and heroin;
\item The \emph{Ecstasy pleiad (ecstasyPl)} includes amphetamines, cannabis, cocaine, ketamine, LSD, magic mushrooms, legal highs, and ecstasy;
\item The \emph{Benzodiazepines pleiad (benzoPl)} includes methadone, amphetamines, and cocaine.
\end{itemize}

Fuzzy and probabilistic clustering may help to reveal more sophisticated relationships between objects and clusters. For example, analysis of the intersections between correlation pleiads of drugs can generate important question and hypotheses:
\begin{itemize}
\item Which patterns of behaviour are reflected by the existence of pleiades? (For example,  is the ecstasyPl just the group of party drugs united by traditions of use?)
\item Why is cocaine  a peripheral member of all pleiads?
\item Why does methadone belong  to the  periphery of  both the heroin and benzodiazepines pleiades?
\item Why does amphetamines belong  to the  periphery of both  the ecstasy and benzodiazepines pleiades?
\item Do these intersections reflect the structure of individual drug consumption or the structure of the groups of drug consumers?
\end{itemize}

We define groups of users and non-users for each pleiad.
 {\em A group of users for a pleiad includes the users of any individual drugs from the pleiad} (see Table~\ref{tab:13a}). A group of non-users contains all participants which are not included in the group of users. Table~\ref{tab:13a} presents the total number of users and their percentages in the database for three pleiades and for different definition of users (the decade-, year-, month-, or week-based user/non-user separation).

The class imbalance problem is well known \cite{Hastie09}. Users form a  small fraction of the dataset (significantly less than the half) for most of drugs (see Table~\ref{tab:1a}). The classes of users and non-users are more balanced  for pleiades of drugs  than for individual drugs (compare Table~\ref{tab:13a} and \ref{tab:1a}).
Table~\ref{tab:13a} shows that the number of drug users in the database for all three pleiades  are more balanced (close to 50\%) than the number of users of the corresponding individual drug  (Table~\ref{tab:1a}). For example, for the decade-based classification problem the number of benzoPl users is 1089 (57.77\%), while the number of benzodiazepines users is 769 (40.80\%) and the number of heroinPl users is 832 (44.14\%), while the number of heroin users is 212 (11.25\%).

\begin{table}[!ht]
\begin{adjustwidth}{-2.25in}{0in}
\centering
\caption{ Number of drug users for pleiades in the database}
\label{tab:13a}
\begin{tabular}{|l|c|c|c|c|}\hline
{\bf Pleiad} 	& \multicolumn{4}{c|}{ { \bf User definition based on}}\\\cline{2-5}
	           & {\bf Decade} &	{\bf Year}&	{ \bf Month}&	{\bf Week} \\\hline
HeroinPl&	832 (44.14\%) &	585 (31.03\%) &	309 (16.39\%)&	184 (9.76\%)\\\hline
EcstasyPl& 	1317 (69.87\%)&	1089 (57.77\%)&	921 (48.86\%)&	792 (42.02\%)\\\hline
BenzoPl&	1089 (57.77\%)&	830 (44.03\%) &	528 (28.01\%)&	363 (19.26\%)\\\hline
\end{tabular}
\end{adjustwidth}
\end{table}

\begin{table}[!ht]
\begin{adjustwidth}{-2.25in}{0in}
\centering
\caption{Significant differences of means for groups of users and non-users for each pleiad for decade- year-, month-, and week-based classification problem . Symbol ` $\Downarrow$ ' corresponds to significant difference where the mean in users group is less than mean in non-users group and symbol `$\Uparrow$ corresponds to significant difference where the mean in users group is greater than the mean in non-users group. Empty cells corresponds to insignificant differences. Difference is considered as significant if $p$-value is less than 0.01).} \label{tab:13ab}

\begin{tabular}{|l|c|c|c|c|c|}\hline
{\bf Pleiades od drugs} & {\bf N} & {\bf E } & {\bf O} & {\bf A} & {\bf C}\\\hline
\multicolumn{6}{|c|}{\bf{The decade-based user/non user separation}}\\\hline
HeroinPl, EcstasyPl, BenzoPl & $\Uparrow$ &  & $\Uparrow$ & $\Downarrow$ & $\Downarrow$\\\hline
\multicolumn{6}{|c|}{\bf{The year-based user/non user separation}}\\\hline
HeroinPl, EcstasyPl, BenzoPl & $\Uparrow$ &  & $\Uparrow$ & $\Downarrow$ & $\Downarrow$\\\hline
\multicolumn{6}{|c|}{\bf{The month-based user/non user separation}}\\\hline
HeroinPl, EcstasyPl & $\Uparrow$ &  & $\Uparrow$ & $\Downarrow$ & $\Downarrow$\\\hline
BenzoPl             & $\Uparrow$ & $\Downarrow$ & $\Uparrow$ & $\Downarrow$ & $\Downarrow$\\\hline
\multicolumn{6}{|c|}{\bf{The week-based user/non user separation}}\\\hline
HeroinPl, BenzoPl             & $\Uparrow$ & $\Downarrow$ & $\Uparrow$ & $\Downarrow$ & $\Downarrow$\\\hline
EcstasyPl                     & $\Uparrow$ &  & $\Uparrow$ & $\Downarrow$ & $\Downarrow$\\\hline
\end{tabular}
\end{adjustwidth}
\end{table}

The introduction of moderate subcategories of T-$score_{sample}$ for pleiades of drugs enables to separates the pleiades of drugs into two groups for the decade-, month-, and week-based user/non-user separation. While for year-based user/non-user separation it is only one group with the profile $(+, 0, +, -, -)$  includes the users of heroinPl, ecstasyPl and benzoPl.

For decade-based classification problem, the group with the profile $(+, 0, +, -, -)$  includes the users of heroinPl and benzoPl. The group with the profile $(0, 0, +, -, -)$  includes the users of EcstasyPl. 

For month- and week-based classification problem, the group with the profile $(+, -, +, -, -)$  includes the users of heroinPl and benzoPl. The group with the profile $(0, 0, +, -, -)$  includes the users of EcstasyPl.

\begin{figure}[!ht]
\centering
\includegraphics[width=1\textwidth]{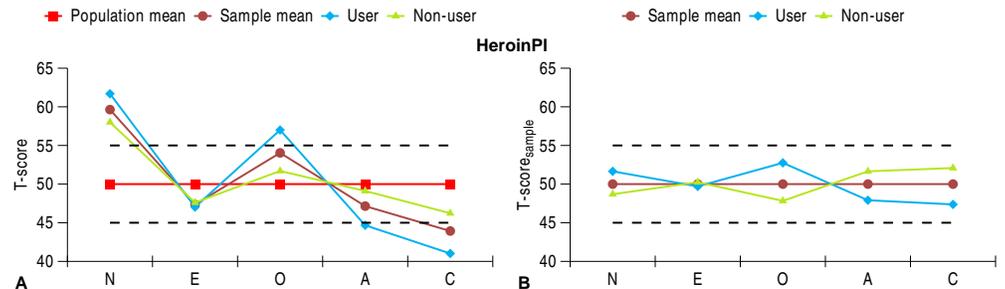}
\caption {{\bf Average personality profiles for HeroinPl for the Decade-based user/non-user separation}. A: T-scores with respect to the population norm mean  and B: T-score$_{sample}$  with respect to the sample means}
 \label{HeroinPlDecade}
\end{figure}

\begin{figure}[!ht]
\centering
\includegraphics[width=1\textwidth]{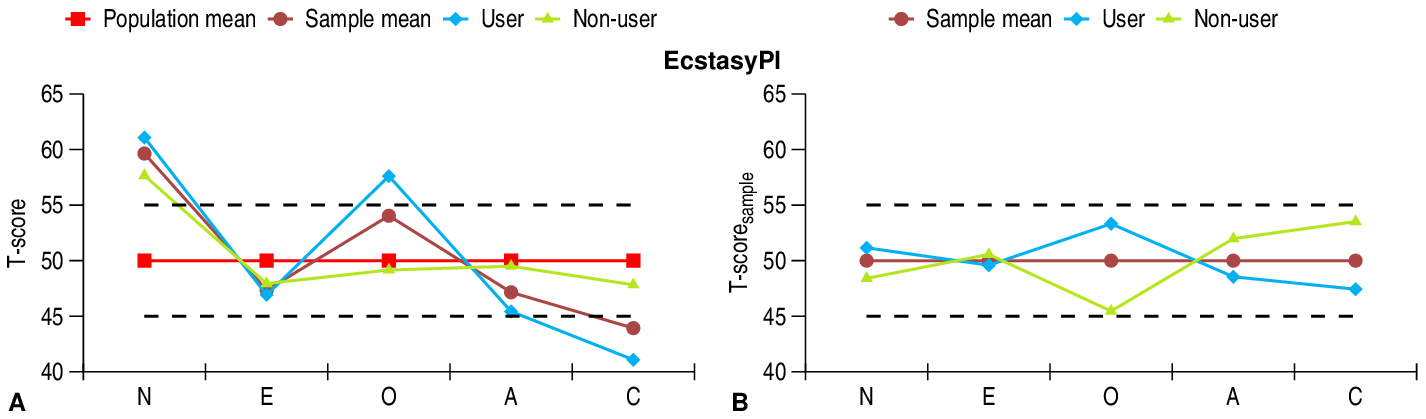}
\caption {{\bf Average personality profiles for EcstasyPl for the month-based user/non-user separation.}  A:  T-scores with respect to the population norm mean  and B:  T-score$_{sample}$  with respect to the sample means}
 \label{EcstasyPlYear}
\end{figure}

\begin{figure}[!ht]
\centering
\includegraphics[width=1\textwidth]{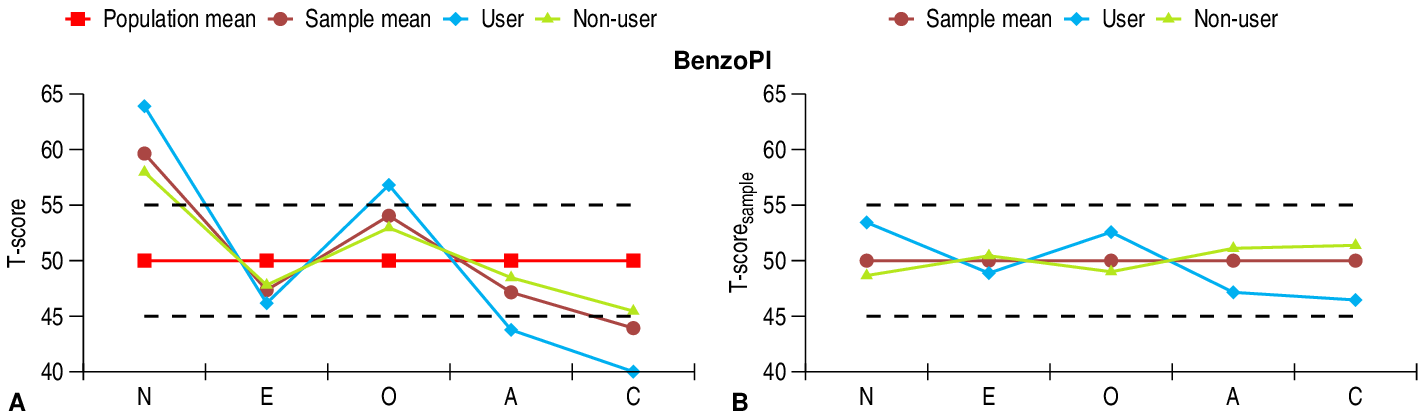}
\caption {{\bf Average personality profiles for BenzoPl for the month-based user/non-user separation.}  A: T-scores with respect to the population norm mean and B: T-score$_{sample}$  with respect to the sample means}
 \label{BenzoPlMonth}
\end{figure}

\begin{figure}[!ht]
\centering
\includegraphics[width=1\textwidth]{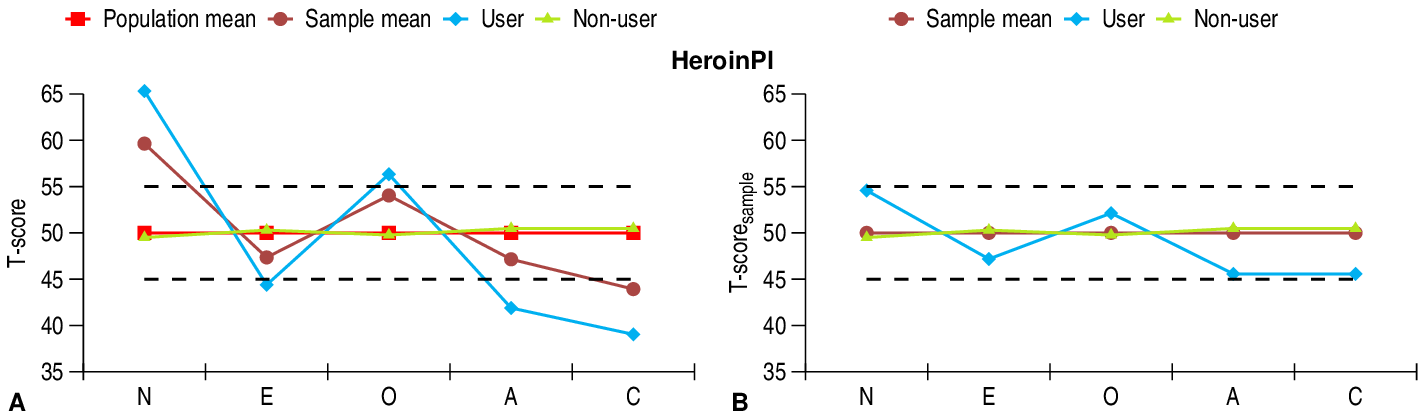}
\caption {{\bf Average personality profiles for  HeroinPl for the week-based user/non-user separation.}  A: T-scores with respect to the population norm mean and B: T-score$_{sample}$  with respect to the sample means}
 \label{HeroinPlWeek}
\end{figure}

The personality profile for pleiades of drugs also are strongly associated with belonging to groups of the users and non-users of each plead for decade-, year-, month-, and week-based classification problem (see Fig.~\ref{HeroinPlDecade}, Fig.~\ref{EcstasyPlYear}, Fig.~\ref{BenzoPlMonth} and ~\ref{HeroinPlWeek}).

We applied the eight methods described in Section `Risk evaluation methods' and selected the best one for each  pleiad for decade-, year-, month-, and week-based classification problem. The results of the classifier selection are depicted in Table~\ref{tab:12a} and the quality of classification is high.

The classification results are very satisfactory  for each pleiad for decade-, year-, month, and week based problem. We can compare the classifiers for one pleiad and for different problems (see Table~\ref{tab:12a}). For example,
\begin{itemize}
\item The best classifier for ecstasyPl  for the year-based user/non user separation is DT with seven attributes and has sensitivity 80.65\% and specificity 80.72\%.
\item The best classifier for heroinPl  for the month-based user/non user separation is DT with five attributes and has sensitivity 74.18\% and specificity 74.11\%.
\item The best classifier for benzoPl for the week-based user/non user separation is DT with five attributes and has sensitivity 75.10\% and specificity 75.76\%.
\end{itemize}

\begin{table}[!ht]
\begin{adjustwidth}{-2.25in}{0in}
\centering
\footnotesize{
\caption {The best results of the pleiad users classifiers. Symbol `X' means input feature used in the best classifier. Sensitivity and Specificity were calculated by LOOCV.}
\label{tab:12a}
\begin{tabular}{|l|c|c|c|c|c|c|c|c|c|c|c|c|c|c|c|} \hline
{ \bf Pleiades } &{ \bf  Classifier} &{ \bf Age}&{ \bf Edu.}&{ \bf N} &{ \bf E} &{ \bf O} &{ \bf A} &{ \bf  C} &{ \bf Imp.} &{ \bf SS }&{ \bf Gender}&\#&{ \bf  Sens.} &{ \bf Spec.}&{ \bf Sum}  \\
{\bf of drugs} &&&&&&&&&&&&&{ \bf (\%)}& { \bf (\%)}&{ \bf (\%)} \\   \hline

\multicolumn{16}{|c|}{\bf{The decade-based user/non user separation}}\\\hline
HeroinPl         &	DT&	 &	 	& &	X&	X&	 &	X&	 &	 &	X&	4&	71.23&	78.85&	150.07\\\hline
EcstasyPl        &  DT&	X&	 &	X&	X&	X&	 &	 &	 &	X&	X&	6&	80.63&	79.80&	160.44\\\hline
BenzoPl          &	DT&	 &	X&	 &	X&	X&	X&	 &	 &	X&	 &	5&	73.37&	72.45&	145.82\\\hline

\multicolumn{16}{|c|}{\bf{The year-based user/non user separation}}\\\hline
HeroinPl           &DT&	 &	X&	X&	 &	 &	X&	 &	 &	 &	X&	4&	73.69&	71.80&	145.49 \\\hline
EcstasyPl          &DT&	X&	X&	 &	 &	X&	 &	X&	X&	X&	X&	7&	80.65&	80.72&	161.37 \\\hline
BenzoPl	           &DT&	 &	X&	X&	 &	 &	 &	X&	 &	X&	 &	4&	73.93&	73.98&	147.91 \\\hline									
\multicolumn{16}{|c|}{\bf{The month-based user/non user separation}}\\\hline
HeroinPl&	DT  &	X&	 &	 &	X&	X&	 &	X&	 &	X&	 &	5&	74.18&	74.11&	148.29\\\hline
EcstasyPl&	PDFE&	X&	X&	 &	X&	X&	 &	X&	X&	X&	X&	8&	79.34&	79.50&	158.83\\\hline
BenzoPl	&	DT	&   X&	X&	 &	 &	 &	X&	X&	 &	 &	 &	4&	73.18&	73.11&	146.28\\\hline												
\multicolumn{16}{|c|}{\bf{The week-based user/non user separation}}\\\hline
HeroinPl&	DT&	X&	X&	X&	X&	X&	 &	 &	X&	X&	X&	8&	75.84&	73.91&	149.75\\\hline
EcstasyPl&	LR&	X&	X&	 &	X&	X&	X&	X&	 &	X&	X&	8&	77.68&	77.78&	155.45\\\hline
BenzoPl&	DT&	 &	X&	 &	 &	 &	X&	X&	 &	X&	X&	5&	75.10&	75.76&	150.86\\\hline
\end{tabular}}
\end{adjustwidth}
\end{table}

The comparison of Tables~\ref{tab:12} and~\ref{tab:12a} shows that the best classifiers for the ecstasy and benzodiazepines pleiades are more accurate than the best classifiers for consumption of the `central' drugs of the pleiades, ecstasy and benzodiazepines even for the decade-based user definition.  Classifiers for heroinPl may have slightly worse accuracy  but these classifiers are more robust because they solve classification problems which have more balanced classes. All other classifiers for pleiades of drugs are more robust too by the same reasons for all pleiades and definitions of users.

Tables~\ref{tab:12} and~\ref{tab:12a} for the decade-based user definition show that most of the classifiers for pleiades use more input features than the classifiers for individual drugs.
We can see from these tables that the accuracies of the classifiers for pleiades and for individual drugs do not differ drastically,  but the use of a greater number of input features indicates more robust classifiers.

It is important to stress that usually pleiades are assumed to be disjoint. We consider pleiades which are named by the central drug and have an intersection in the peripheral drugs. For example, heroin and ecstasy pleiades have cocaine as an intersection. This approach corresponds to the concept of `soft clustering'.

\subsection*{Risk evaluation for the decade-based user/non user separation}
\label{Risk evaluation}
\begin{figure}[!ht]
 \centering
\includegraphics[width=0.95\textwidth]{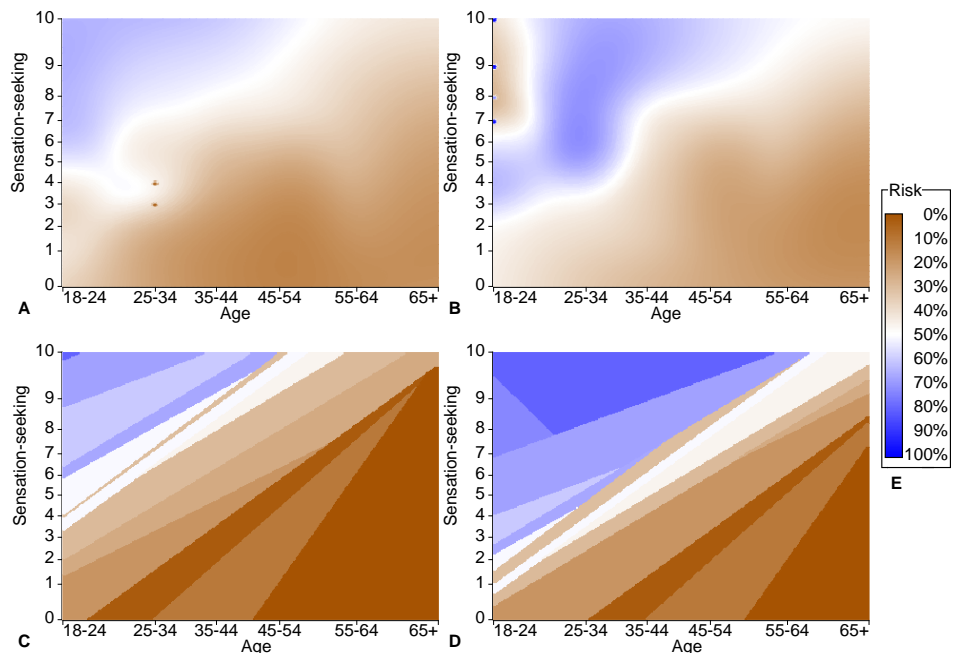}
\caption {{ \bf Risk map of ecstasy} consumption for: A \& C: female and B \& D: male; A \& B: PDFE-based map and C \& D: DT-based map; E: Legend of colours}
 \label{RiskMapsfig:8}
\end{figure}

The successful construction of a classifier provides an instrument for the evaluation of the risk of drug consumption for each individual, along with the creation of a map of risk \cite{Mirkes14a,Mirkes14b}. The risk map of ecstasy consumption on the basis of three input features is depicted in Fig~\ref{RiskMapsfig:8}. From the PDFE-based risk maps (Fig~\ref{RiskMapsfig:8}A and~\ref{RiskMapsfig:8}B) it can be observed that there is a considerable area of high risk (indicated in blue) for men aged between 25-34 years, but significantly less for females. However, young males with the highest SS scores have significantly less risk than females with the same profiles. DT-based risk maps (Fig~\ref{RiskMapsfig:8}C and~\ref{RiskMapsfig:8}D) illustrate qualitatively the same shapes. The risk maps provide a tool for the generation of hypotheses for further study. We can create risk maps for pleiades of drugs as well.

\section*{Summary}
Our study demonstrates strong correlations between personality profiles and the risk of drug use. This result supports  observations from some previous works  \cite{Sutina13,Haider02,Vollrath02,Flory02,Terracciano08,Turiano12,Stewart00,Roncero14}. Individuals involved in drug use are more likely to have higher scores for N, and low scores for A and C .  We analysed in detail the average differences in the groups of drug users and non-users for 18 drugs (Tables~\ref{tab:5}, \ref{tab:5a}, \ref{tab:5b}, and \ref{tab:5c}). In addition to this analysis, we achieved much more detailed knowledge about the relationship between the personality traits, biographic data and the use of individual drugs or drug clusters by an individual patient.

The analysed database contained 1885 participants and 12 features (input attributes). These features included five personality traits (NEO-FFI-R); impulsivity (BIS-11), sensation seeking (ImpSS), level of education, age, gender, country of residence, and ethnicity. The data set included information on the consumption of 18 central nervous system psychoactive drugs: alcohol, amphetamines, amyl nitrite, benzodiazepines, cannabis, chocolate, cocaine, caffeine, crack, ecstasy, heroin, ketamine, legal highs, LSD, methadone, mushrooms, nicotine, and VSA (output attributes).
There were limitations of this study since the collected sample was biased with respect to the general population, but it remained useful for risk evaluation.

We used three different techniques of feature ranking. After input feature ranking we excluded ethnicity and country of residence.  It was impossible to  completely exclude the possibility that ethnicity and country of residence may be important risk factors, but the dataset has not enough data for most of ethnicities and countries to prove the value of this information.  As a result, 10 input features remained: age, Edu., N, E, O, A, C, Imp., SS, and gender. Our aim was to predict the risk of drug consumption for an individual.

All input features are ordinal or nominal. To apply data mining methods which were developed for continuous input features we apply CatPCA technique to quantify data.

We used four different definitions of drug users which differ with regard to the recency of the last drug consumption: the decade-based,  year-based, month-based and  week-based user/non user separation (Fig.~\ref{Categoriesfig:1}). The day-based  classification problem is also possible but there is not enough data on drug use during the  last day for most drugs

Our findings allowed us to draw a number of important conclusions about the associations between personality traits and drug use. All five personality factors are relevant traits to be taken into account when assessing the  risk of drug consumption.
The mean scores for the groups of users of all 18 drugs are moderately high $(+)$ or neutral $(0)$ for N and O, and moderately low $(-)$ for A and C, except for crack usage for the week based classification problem which has a moderately low $(-)$ O score (see Table~\ref{tab:5c} and Fig~\ref{AverageCrack}). Users of legal drugs (alcohol, chocolate, caffeine, and nicotine) have neutral  A and C  scores $(0)$, except  nicotine users whose C  score  is moderately low $(-)$. For LSD users in the year based classification problem and for LSD and magic mushrooms users in the week based classification problem the A score  is neutral $(0)$.

The impact of the E score is drug specific. For example, for the decade-based user/non user definition the E score is negatively correlated with consumption of  crack, heroin, VSA, and methadone (E score is $(-)$ for their users). It is has no predictive value for other drugs for the decade-based classification (E score for users  is $(0)$),  whereas in the year-, month-, and week-based classification problems all three possible values of E score are observed (see Tables~\ref{tab:5}, \ref{tab:5a}, \ref{tab:5b} and \ref{tab:5c}).

We confirm the previous researcher finding that the higher scores for N and O  the lower scores for C and A lead to increased risk of drug use \cite{Belcher2016}. O score is marked by curiosity and open-mindedness (and correlated with intelligence), and it is therefore understandable why higher O may be sometimes associated with drug use \cite{Wilmoth12}. Flory et al \cite{Flory02} found that marijuana use to be associated with lower A and C, and higher O. These findings have been confirmed by our study. Our results improve the knowledge concerning the pathways leading to drug consumption.

It is known that significant predictors of  alcohol, tobacco and marijuana use may vary according to the drug in question \cite{Jones1998}.  Our study demonstrated that different attributes were important for different drugs.
We tested eight types of classifiers for each drug for the decade-based user definition. LOOCV was used to evaluate sensitivity and specificity.
 In this study we choose  the method which provide maximal value of minimum among sensitivity and specificity as the best one. The method with maximal sum of the sensitivity and specificity is selected as the best one  for two methods with the same minimum among sensitivity and specificity.
 There were classifiers with sensitivity and specificity greater than 70\% for the decade-based user/non user separation for all drugs except magic mushrooms, alcohol, and cocaine  (Table~\ref{tab:12}). This accuracy was unexpectedly high for this type of problem. The poorest result was obtained for the prediction of alcohol consumption.

The best set of input features was defined for each drug (Table~\ref{tab:12}). An exhaustive search was performed to select the most effective subset of input features, and the best data mining methods to classify users and non-users for each drug.   There were 10 input features. Each of them is an important factor for risk evaluation for the use of some drugs. However, there was no single most effective classifier using all input features. The maximal number of  attributes used in the best classifiers is six (out of 10) and the minimal number is two.

Table~\ref{tab:12} shows the best sets of attributes  for user/nonuser classification for different drugs and for the decade-based classification problem. This table together with its analogues for pleiads of drugs and all decade-year-month-week classification problems  (Table~\ref{tab:12a}) are important result of the analysis.

The DT for crack consumption used two features E and C only, and provided sensitivity of 80.63\% and specificity of 78.57\%. The DT for VSA consumption used age, Edu., E, A, C, and SS, and provided sensitivity 83.48\% and specificity 77.64\% (Table~\ref{tab:12}).

Age was a widely used feature which was employed in the best classifiers for 14 drugs for the decade-based classification problem.  Gender was used in the best methods for 10 drugs. We found some unexpected outcomes. For example, fraction of females which are alcohol users is greater than the fraction of males but the greater part of males consume caffeine (coffee).

Most of the features which are not used in the best classifiers are redundant but are not uninformative.
For example, the best classifier for ecstasy consumption used age, SS, and gender and had sensitivity 76.17\% and specificity 77.16\%. There exist a DT for prediction of usage of the same drug, which utilizes age, Edu., O, C, and SS with sensitivity 77.23\% and specificity 75.22\%, a DT with inputs age, Edu., E, O, and A with sensitivity 73.24\% and specificity 78.22\%, and an advanced $k$NN classifier with inputs age, Edu., N, E, O, C, Imp., SS, and gender with sensitivity 75.63\% and specificity 75.75\%. This means that for evaluating the risk of ecstasy usage all input attributes are informative but the required information can be extracted from a subset of attributes.

We demonstrated that there are three groups of drugs with strongly correlated consumption. That is, the drug usage has a `modular structure'. The idea to merge correlated attributes into `modules' is popular in biology. The modules are called the `correlation pleiades' \cite{Terentjev31,Berg60,Mitteroecker07} (see Section `Pleiades of drugs').
The modular structure contains three modules: the heroin pleiad, ecstasy pleiad, and benzodiazepines pleiad:
\begin{itemize}
\item The  \emph{Heroin pleiad (heroinPl)} includes crack, cocaine, methadone, and heroin.
\item The \emph{Ecstasy pleiad (ecstasyPl)} includes amphetamines, cannabis, cocaine, ketamine, LSD, magic mushrooms, legal highs, and ecstasy.
\item The \emph{Benzodiazepines pleiad (benzoPl)} includes contains methadone, amphetamines, and cocaine.
\end{itemize}
The modular structure has a clear reflection in the correlation graph, Fig~\ref{Strongdrugusfig:5}.
We define groups of users and non-users for each pleiad. In most of the databases the classes of users and non-users for most of the individual drugs are imbalanced (see Table~\ref{tab:1}) but the merging the users of all drugs in one class `drug users' does not seem to be the best solution because of physiological, psychological and cultural differences between usage of different drugs.
We propose instead to use correlation pleiades for the analysis of drug usage as a solution to the class imbalance problem because for all three pleiades the classes of users and non-users are better balanced (Table~\ref{tab:13a}) and the consumption of different drugs from the same pleiad is correlated.

We applied the eight methods described in the `Risk evaluation methods' Section and selected the best one for each problem for all pleiades. The results of the classifier selection are presented in Table~\ref{tab:12a} and the quality of the classification is high. The majority of the best classifiers for pleiades of drugs has a better accuracy than the classifiers for individual drug usage. (see Table \ref{tab:12} and~\ref{tab:12a}). The best classifiers for pleiades of drugs use more input features than the best  classifiers for the corresponding individual drugs. The classification problems for pleiades of drugs are more balanced. Therefore, we can expect that the classifiers for pleiades are more robust than the classifiers for individual drugs.

The user/non-user classifiers can be also used for forming of risk maps. Risk maps are useful tools for data visualisation and hypotheses generation.

\section*{Discussion }

These results are important as they examine the question of the relationship between drug use and personality comprehensively and engage the challenge of untangling correlated personality traits (the FFM, impulsivity, and sensation seeking \cite{Whiteside2001}), and clusters of substance misuse (the correlation pleiades).  The work acknowledged the breadth of a common behaviour which may be transient and leave no impact, or may significantly harm an individual.  We examined drug use behaviour comprehensively in terms of the many kinds of substances that may be used (from the legal and anodyne, to the deeply harmful), as well as the possibility of behavioural over-claiming.  We built into our study the wide temporality of the behaviour indicative of the chronicity of behaviour and trends and fashions (e.g. the greater use of LSD in the 1960s and 1970s, the rise of ecstasy in the 1980s, some persons being one-off experimenters with recreational drugs, and others using recreational substances on a daily basis).

We defined substance use in terms of behaviour rather than legality, as legislation in the field is variable.  Our data were gathered before  ‘ legal highs’ emerged as a health concern \cite{Gibbons2012} so we did not differentiate, for example, synthetic cannabinoids and cathinone-based stimulants; these substances have been since widely made illegal.  We were nevertheless able to accurately classify users of these substances (reciprocally, our data were gathered before cannabis decriminalisation in parts of North America, but again, we were able to accurately classify cannabis users).  We included control participants who had never used these substances, those who had used them in the distant past, up to and including persons who had used the drug in the past day, avoiding the procrustean data-gathering and classifying methods which may occlude an accurate picture of drug use behaviour and risk \cite{Nutt2007}.  Such rich data and the complex methods used for analysis necessitated a large and substantial sample.

The study was atheoretical regarding the morality of the behaviour, and did not medicalise or pathologise participants, optimising engagement by persons with heterogeneous drug-use histories.  Our study used a rigorous range of data-mining methods beyond those typically used in studies examining the association of drug use and personality in the psychological and psychiatric literature, revealing that decision tree methods were most commonly effective for classifying drug users.  We found that high N, low A, and low C are the most common personality correlates of drug use, these traits being sometimes seen in combination as an indication of higher-order stability and behavioural conformity, and, inverted, are associated with externalisation of distress \cite{Digman1997,DeYoung2002,DeYoung2008}.

 Low stability is also a marker of negative urgency \cite{Settles12} whereby persons act rashly when distressed.  Our research points to the importance of individuals acquiring emotional self-management skills anteceding distress as a means to reduce self-medicating drug-using behaviour, and the risk to health that injudicious or chronic drug use may cause.

\newpage

\section*{Supporting Information}
\paragraph*{S1 Appendix.}
\label{S1_Appendix}
{\bf Mean for groups of users and non-users.} In this section we present mean T-$score_{sample}$ for groups of users and non-users for decade, year, month, and week based user definitions respectively. Column $p$-value assesses the significance of differences of mean scores for groups of users and non-users: it is the probability of observing by chance the same or greater differences for mean scores if both groups have the same mean. Rows `\#' contain number of users and non-users for the drugs.
\begin{center}
\setlength\LTleft{-0.8in}
\setlength\LTcapwidth{11.7in}
\begin{adjustwidth}{-2.25in}{0in}
\begin{longtable}{|c|c|c|c|c|c|}
\caption{ Mean T-$score_{sample}$ and 95\% CI for it for groups of users and non-users with decade based definition of users}
\label{table4}
\\\hline
{\bf Factor} &\multicolumn{2}{c|}{\bf Users}&\multicolumn{2}{c|}{\bf Non-users}& {\bf $p$-value}\\\cline{2-5}
    &{  Mean T-score}& {  95\% CI for mean}& { Mean T-score}&{ 95\% CI for mean}&   \\ \hline
\endfirsthead
\multicolumn{6}{c} {\tablename\ \thetable.\textit{ Continued}} \\\hline
 {\bf Factor} &\multicolumn{2}{c|}{\bf Users}&\multicolumn{2}{c|}{\bf Non-users}& {\bf $p$-value}\\\cline{2-5}
   &Mean T-score& 95\% CI for mean& Mean T-score&95\% CI for mean &\\\hline
\endhead
\hline \multicolumn{6}{r}{\textit{\footnotesize{Continued on the next page}}} \\
\endfoot\hline
\endlastfoot
\multicolumn{6}{|c|}{\bf{Alcohol}}\\\hline
 \# &\multicolumn{2}{c}{1817}&\multicolumn{2}{|c|}{68}&    \\\hline
N    &50.13&	49.67, 50.59&	48.19&	45.77, 50.61&	0.116\\\hline
E    &50.06&	49.60, 50.52&	50.04&  47.61, 52.42&	0.988\\\hline
O    &50.04&	49.58, 50.51&	48.81&	46.45, 51.17&	0.318\\\hline
A    &49.93&	49.47, 50.39&	52.51&	50.26, 54.77&	0.036\\\hline
C    &49.94&	49.48, 50.40&	53.31	&51.05, 55.56&	0.006\\\hline

\multicolumn{6}{|c|}{\bf{Amphetamines}}\\\hline
 \# &\multicolumn{2}{c}{679}&\multicolumn{2}{|c|}{1206}& \\\hline
N	&51.71	&50.95, 52.46&	49.14&	48.58, 49.69&	0.001\\\hline
E	&49.71	&48.89,50.53&	50.26&	49.72, 50.80&	0.251\\\hline
O	&53.05	&52.34, 53.77&	48.28&	47.72, 48.84&	0.001\\\hline
A	&48.39	&47.60, 49.18&	50.94&	50.39, 51.48&	0.001\\\hline
C	&47.04	&46.29, 47.80&	51.76&	51.22, 52.30&	0.001\\\hline

\multicolumn{6}{|c|}{\bf{Amyl nitrite}}\\\hline
\# &\multicolumn{2}{c}{370}&\multicolumn{2}{|c|}{1515}& \\\hline

N&	50.78&	49.78, 51.79&	49.89&	49.37, 50.39&	0.122\\\hline
E&	50.97&	49.95, 51.99&	49.84&	49.33, 50.35&	0.052\\\hline
O&	51.45&	50.47, 52.43&	49.65&	49.14, 50.15&	0.002\\\hline
A&	48.69&	47.65, 49.72&	50.35&	49.84, 50.85&	0.004\\\hline
C&	48.08&	46.10, 49.07&	50.54&	50.04, 51.05&	0.001\\\hline

\multicolumn{6}{|c|}{\bf{Benzodiazepines}}\\\hline
 \# &\multicolumn{2}{c}{769}&\multicolumn{2}{|c|}{1116}& \\\hline

N&	52.83&	52.12, 53.54&	48.15&	47.59, 48.72&	0.001\\\hline
E&	49.07&	48.31, 49.83&	50.74&	50.19, 51.30&	0.001\\\hline
O&	52.66&	51.97, 53.34&	48.17&	47.59, 48.75&	0.001\\\hline
A&	48.28&	47.53, 49.02&	51.22&	50.67, 51.78&	0.001\\\hline
C&	47.70&	46.98, 48.41&	51.69&	51.12, 52.25&	0.001\\\hline

\multicolumn{6}{|c|}{\bf{Cannabis}}\\\hline
 \# &\multicolumn{2}{c}{1265}&\multicolumn{2}{|c|}{620}& \\\hline

N&	51.08&	50.52, 51.65&	47.98&	47.25, 48.71&	0.001\\\hline
E&	49.75&	49.17, 50.33&	50.70&	49.98, 51.41&	0.053\\\hline
O&	52.48&	51.96, 52.99&	44.95&	44.20,45.70&	0.001\\\hline
A&	48.84&	48.28,49.40&	52.42&	51.69,53.15&	0.001\\\hline
C&	48.15&	47.60,48.70&	53.92&	53.27,54.65&	0.001\\\hline

\multicolumn{6}{|c|}{\bf{Chocolate}}\\\hline
 \# &\multicolumn{2}{c}{1850}&\multicolumn{2}{|c|}{35}& \\\hline

N&	50.06&	49.60, 50.51&	50.29&	46.36, 54.21&	0.894\\\hline
E&	50.05&	49.59, 50.51&	50.80&	47.71, 53.89&	0.660\\\hline
O&	50.05&	49.59, 50.51&	47.37&	44.42, 50.32&	0.117\\\hline
A&	50.05&	49.59, 50.50&	48.66&	44.61, 52.70&	0.416\\\hline
C&	50.03&	49.58, 50.49&	51.57&	47.87, 55.27&	0.366\\\hline

\multicolumn{6}{|c|}{\bf{Cocaine}}\\\hline
 \# &\multicolumn{2}{c}{687}&\multicolumn{2}{|c|}{1198}& \\\hline

N&	51.85&	51.09, 52.60&	49.04&	48.48, 49.59&	0.001\\\hline
E&	50.33&	49.55, 51.12&	49.91&	49.35, 50.46&	0.374\\\hline
O&	52.60&	51.89, 53.30&	48.51&	47.94, 49.08&	0.001\\\hline
A&	47.71&	46.93, 48.50&	51.34&	50.81,51.88&	0.001\\\hline
C&	47.49&	46.76, 48.22&	51.53&	50.98, 52.09&	0.001\\\hline

\multicolumn{6}{|c|}{\bf{Caffeine}}\\\hline
\# &\multicolumn{2}{c}{1848}&\multicolumn{2}{|c|}{37}& \\\hline

N&	50.06&	49.61, 50.52&	50.08&	46.59,53.57&	0.991\\\hline
E&	50.13&	49.67, 50.59&	46.76&	43.82, 49.69&	0.043\\\hline
O&	50.11&	49.65, 50.57&	44.59&	41.67,47.52&	0.001\\\hline
A&	49.99&	49.54, 50.45&	51.11&	47.71, 54.51&	0.504\\\hline
C&	49.99&	49.53, 50.44&	53.59&	50.79, 56.40&	0.029\\\hline

\multicolumn{6}{|c|}{\bf{Crack}}\\\hline
 \# &\multicolumn{2}{c}{191}&\multicolumn{2}{|c|}{1694}& \\\hline

N&	53.08&	51.64, 54.52&	49.72&	49.25, 50.20&	0.001\\\hline
E&  48.80&	47.33, 50.27&	50.20&	49.73, 50.68&	0.066\\\hline
O&	52.91&	51.60, 54.21&	49.67&	49.19, 50.15&	0.001\\\hline
A&	47.02&	45.46, 48.58&	50.36&	49.89, 50.83&	0.001\\\hline
C&	46.19&	44.70, 47.68&	50.50&	50.03, 50.96&	0.001\\\hline

\multicolumn{6}{|c|}{\bf{Ecstasy}}\\\hline
\# &\multicolumn{2}{c}{751}&\multicolumn{2}{|c|}{1134}& \\\hline

N&	51.30&	50.58, 52.02&	49.24&	48.67, 49.82&	0.001\\\hline
E&	50.56&	49.80, 51.31&	49.73&	49.17, 50.30&	0.081\\\hline
O&	53.62&	52.96, 54.28&	47.60&	47.03, 48.17&	0.001\\\hline
A&	48.49&	47.75, 49.23&	51.03&	50.47, 51.60&	0.001\\\hline
C&	47.30&	46.59, 48.00&	51.89&	51.33, 52.45&	0.001\\\hline

\multicolumn{6}{|c|}{\bf{Heroin}}\\\hline
\# &\multicolumn{2}{c}{212}&\multicolumn{2}{|c|}{1673}& \\\hline

N&	54.60&	53.29, 55.92&	49.49&	49.01, 49.96&		0.001\\\hline
E&	48.42&	46.94, 49.90&	50.27&	49.80, 50.74&	   0.011\\\hline
O&	54.25&	53.04, 55.47&	49.46&	48.98, 49.94&		0.001\\\hline
A&	45.53&	44.00, 47.06&	50.59&	50.12, 51.05&		0.001\\\hline
C&	45.91&	44.55, 47.26&	50.59&	50.11, 51.06&		0.001\\\hline

\multicolumn{6}{|c|}{\bf{Ketamine}}\\\hline
\# &\multicolumn{2}{c}{350}&\multicolumn{2}{|c|}{1535}& \\\hline

N&	51.42&	50.40, 52.43&	49.75&	49.25, 50.26&	0.005\\\hline
E&	50.36&	49.23, 51.48&	49.99&	49.50, 50.49&	0.542\\\hline
O&	53.87&	52.90, 54.84&	49.12&	48.62, 49.62&	0.001\\\hline
A&	47.80&	46.67, 48.94&	50.53&	50.04, 51.01&	0.001\\\hline
C&	46.86&	45.81, 47.92&	50.79&	50.30, 51.28&	0.001\\\hline

\multicolumn{6}{|c|}{\bf{Legal highs}}\\\hline
\# &\multicolumn{2}{c}{762}&\multicolumn{2}{|c|}{1123}& \\\hline	

N&	51.49&	50.77, 52.22&	49.09&	48.52, 49.66&	0.001\\\hline
E&	49.71&	48.94, 50.47&	50.30&	49.75, 50.86&	0.206\\\hline
O&	54.30&	53.67, 54.92&	47.08&	46.51, 47.65&  0.001\\\hline
A&	48.61&	47.86, 49.36&	50.98&	50.42, 51.53&	0.001\\\hline
C&	46.98&	46.27, 47.72&	52.14&	51.60, 52.68&	0.001\\\hline

\multicolumn{6}{|c|}{\bf{LSD}}\\\hline
\# &\multicolumn{2}{c}{557}&\multicolumn{2}{|c|}{1328}& \\\hline

N&	50.87&	50.05, 51.69&	49.72&	49.18, 50.26&	0.023\\\hline
E&	50.07&	49.16, 50.98&	50.06&	49.54, 50.58&	0.986\\\hline
O&	55.25&	54.54, 55.96&	47.80&	47.27, 48.33&	0.001\\\hline
A&	48.44&	47.57, 49.31&	50.68&	50.16, 51.21&	0.001\\\hline
C&	47.54&	46.71, 48.36&	51.12&	50.59, 51.65&	0.001\\\hline

\multicolumn{6}{|c|}{\bf{Methadone}}\\\hline
\# &\multicolumn{2}{c}{417}&\multicolumn{2}{|c|}{1468}& \\\hline

N&	53.41&	52.47, 54.35&	49.11&	48.61, 49.62&	0.001\\\hline
E&	47.97&	46.88, 49.05&	50.66&	50.17, 51.15&	0.001\\\hline
O&	53.77&	52.86, 54.69&	48.93&	48.42, 49.44&	0.001\\\hline
A&	47.07&	46.03, 48.10&	50.86&	50.37, 51.35&	0.001\\\hline
C&	46.21&	45.22, 47.20&	51.15&	50.66, 51.64&	0.001\\\hline

 \multicolumn{6}{|c|}{\bf{Magic Mushrooms}}\\\hline
 \# &\multicolumn{2}{c}{694}&\multicolumn{2}{|c|}{1191}& \\\hline
	
N&	50.73&	49.98,51.47&	49.67&	49.10, 50.24&	0.027\\\hline
E&	50.15&	49.35,50.96&	50.01&	49.47, 50.55&	0.765\\\hline
O&	54.36&	53.71,55.01&	47.46&	46.90, 48.02&	0.001\\\hline
A&	48.55&	47.78,49.32&	50.88&	50.33, 51.43&	0.001\\\hline
C&	47.54&	46.81,48.27&	51.53&	50.97, 52.08&	0.001\\\hline

 \multicolumn{6}{|c|}{\bf{Nicotine}}\\\hline
 \# &\multicolumn{2}{c}{1264}&\multicolumn{2}{|c|}{621}& \\\hline

N&	50.97&	50.41, 51.52&	48.22&	47.45, 48.99&	0.001\\\hline
E&	49.98&	49.41, 50.54&	50.24&	49.48, 50.99&	0.599\\\hline
O&	51.47&	50.92, 52.01&	47.01&	46.26, 47.77&	0.001\\\hline
A&	49.20&	48.65, 49.75&	51.69&	50.92, 52.46&	0.001\\\hline
C&	48.64&	48.08, 49.20&	52.95&	52.24, 53.67&	0.001\\\hline

 \multicolumn{6}{|c|}{\bf{VSA}}\\\hline
 \# &\multicolumn{2}{c}{230}&\multicolumn{2}{|c|}{1655}& \\\hline

N&	52.88&	51.57, 54.20&	49.67&	49.19,50.15&	0.001\\\hline
E&	48.96&	47.45, 50.47&	50.22&	49.74,50.69&	0.075\\\hline
O&	54.20&	53.00, 55.41&	49.42&	48.93,49.90&	0.001\\\hline
A&	47.30&	45.92, 48.68&	50.40&	49.92,50.87&	0.001\\\hline
C&	45.22&	43.88, 46.56&	50.73&	50.26,51.20&	0.001 \\\hline
\end{longtable}
\end{adjustwidth}
\end{center}

\begin{center}
\setlength\LTleft{-0.8in}
\setlength\LTcapwidth{11.7in}
\begin{adjustwidth}{-2.25in}{0in}
\begin{longtable}{|c|c|c|c|c|c|}
\caption{ Mean T-$score_{sample}$ and 95\% CI for it for groups of users and non-users with year based definition}
\label{table4a}
\\\hline
{\bf Factor} &\multicolumn{2}{c|}{\bf Users}&\multicolumn{2}{c|}{\bf Non-users}& {\bf $p$-value}\\\cline{2-5}
    &{  Mean T-score}& {  95\% CI for mean}& { Mean T-score}&{ 95\% CI for mean}&   \\ \hline
\endfirsthead
\multicolumn{6}{c} {\tablename\ \thetable.\textit{ Continued}} \\\hline
 {\bf Factor} &\multicolumn{2}{c|}{\bf Users}&\multicolumn{2}{c|}{\bf Non-users}& {\bf $p$-value}\\\cline{2-5}
   &Mean T-score& 95\% CI for mean& Mean T-score&95\% CI for mean &\\\hline
\endhead
\hline \multicolumn{6}{r}{\textit{\footnotesize{Continued on the next page}}} \\
\endfoot\hline
\endlastfoot
\multicolumn{6}{|c|}{\bf{Alcohol}}\\\hline
 \# &\multicolumn{2}{c}{1749}&\multicolumn{2}{|c|}{136}&    \\\hline
N &	49.96 &	49.49, 50.43 &	50.55 &	48.87, 52.24 &	0.5012\\\hline
E&  50.12&  49.64, 50.59 &	48.52 &	46.89, 50.14 &	0.0642\\\hline
O &	50.07 &	49.60, 50.54 &	49.11 &	47.46, 50.76 &	0.2712\\\hline
A &	49.95 &	49.48, 50.42 &	50.70 &	49.03, 52.37 &	0.3921\\\hline
C &	49.94 &	49.47, 50.41 &	50.73 &	49.03, 52.44 &	0.3782 \\\hline

\multicolumn{6}{|c|}{\bf{Amphetamines}}\\\hline
 \# &\multicolumn{2}{c}{436}&\multicolumn{2}{|c|}{1449}& \\\hline
N &	52.40 &	51.43, 53.36 &	49.28 &	48.77, 49.78 &	0.0001\\\hline
E &	49.48 &	48.45, 50.50 &	50.16 &	49.66, 50.66 &	0.2431\\\hline
O &	53.83 &	52.92, 54.74 &	48.85 &	48.34, 49.35 &	0.0001\\\hline
A &	47.47 &	46.45, 48.48 &	50.76 &	50.27, 51.26 &	0.0001\\\hline
C &	45.91 &	44.94, 46.88 &	51.23 &	50.74, 51.72 &	0.0001\\\hline

\multicolumn{6}{|c|}{\bf{Amyl nitrite}}\\\hline
\# &\multicolumn{2}{c}{133}&\multicolumn{2}{|c|}{1752}& \\\hline

N&	51.57&	49.85, 53.29&	49.88&	49.41, 50.35&	0.0633\\\hline
E&	50.25&	48.44, 52.06&	49.98&	49.51, 50.45&	0.7777\\\hline
O&	51.97&	50.36, 53.59&	49.85&	49.38, 50.32&	0.0136\\\hline
A&	46.35&	44.64, 48.07&	50.28&	49.81, 50.74&	0.0001\\\hline
C&	46.92&	45.25, 48.60&	50.23&	49.77, 50.70&	0.0002\\\hline

\multicolumn{6}{|c|}{\bf{Benzodiazepines}}\\\hline
 \# &\multicolumn{2}{c}{535}&\multicolumn{2}{|c|}{1350}& \\\hline

N&	53.69&	52.84, 54.54&	48.54&	48.03, 49.05&	0.0001\\\hline
E&	48.91&	47.97, 49.85&	50.43&	49.92, 50.94&	0.0053\\\hline
O&	53.00&	52.18, 53.83&	48.81&	48.28, 49.34&	0.0001\\\hline
A&	47.55&	46.64, 48.46&	50.97&	50.46, 51.48&	0.0001\\\hline
C&	47.06&	46.19, 47.94&	51.16&	50.65, 51.68&	0.0001\\\hline

\multicolumn{6}{|c|}{\bf{Cannabis}}\\\hline
 \# &\multicolumn{2}{c}{999}&\multicolumn{2}{|c|}{886}& \\\hline

N&	51.10&	50.46, 51.74&	48.76&	48.14, 49.38&	0.0001\\\hline
E&	49.73&	49.08, 50.39&	50.30&	49.69, 50.91&	0.2177\\\hline
O&	53.68&	53.13, 54.23&	45.85&	45.22, 46.48&	0.0001\\\hline
A&	48.77&	48.13, 49.40&	51.39&	50.76, 52.02&	0.0001\\\hline
C&	47.32&	46.70, 47.95&	53.02&	52.42, 53.62&	0.0001\\\hline

\multicolumn{6}{|c|}{\bf{Chocolate}}\\\hline
 \# &\multicolumn{2}{c}{1840}&\multicolumn{2}{|c|}{45}& \\\hline

N&	50.01&	49.56, 50.47&	49.45&	46.04, 52.86&	0.7445\\\hline
E&	49.99&	49.53, 50.45&	50.49&	47.75, 53.24&	0.7156\\\hline
O&	50.04&	49.58, 50.50&	48.27&	45.59, 50.94&	0.1945\\\hline
A&	50.02&	49.56, 50.48&	49.17&	45.72, 52.62&	0.6259\\\hline
C&	49.97&	49.51, 50.42&	51.41&	47.81, 55.01&	0.4254\\\hline

\multicolumn{6}{|c|}{\bf{Cocaine}}\\\hline
 \# &\multicolumn{2}{c}{417}&\multicolumn{2}{|c|}{1468}& \\\hline

N&	52.16&	51.19, 53.13&	49.39&	48.88, 49.89&	0.0001\\\hline
E&	50.93&	49.88, 51.98&	49.74&	49.24, 50.23&	0.0439\\\hline
O&	52.79&	51.87, 53.71&	49.21&	48.70, 49.72&	0.0001\\\hline
A&	46.84&	45.81, 47.86&	50.90&	50.41, 51.39&	0.0001\\\hline
C&	46.81&	45.86, 47.75&	50.91&	50.40, 51.41&	0.0001\\\hline

\multicolumn{6}{|c|}{\bf{Caffeine}}\\\hline
\# &\multicolumn{2}{c}{1824}&\multicolumn{2}{|c|}{61}& \\\hline

N&	49.99&	49.53, 50.45&	50.36&	47.75, 52.96&	0.7827\\\hline
E&	50.10&	49.64, 50.56&	46.92&	44.34, 49.50&	0.0181\\\hline
O&	50.10&	49.64, 50.56&	47.07&	44.45, 49.69&	0.0265\\\hline
A&	49.96&	49.50, 50.42&	51.15&	48.61, 53.69&	0.3605\\\hline
C&	49.91&	49.45, 50.37&	52.79&	50.14, 55.43&	0.0362\\\hline

\multicolumn{6}{|c|}{\bf{Crack}}\\\hline
 \# &\multicolumn{2}{c}{79}&\multicolumn{2}{|c|}{1806}& \\\hline

N&	54.06&	51.78, 56.35&	49.82&	49.36, 50.28&	0.0005\\\hline
E&	49.24&	47.20, 51.28&	50.03&	49.57, 50.50&	0.4548\\\hline
O&	52.90&	50.72, 55.08&	49.87&	49.41, 50.33&	0.0084\\\hline
A&	46.65&	44.17, 49.12&	50.15&	49.69, 50.61&	0.0070\\\hline
C&	45.08&	42.92, 47.25&	50.22&	49.76, 50.67&	0.0001\\\hline

\multicolumn{6}{|c|}{\bf{Ecstasy}}\\\hline
\# &\multicolumn{2}{c}{517}&\multicolumn{2}{|c|}{1368}& \\\hline

N&	50.71&	49.82, 51.61&	49.73&	49.21, 50.25&	0.0624\\\hline
E&	51.51&	50.60, 52.42&	49.43&	48.91, 49.95&	0.0001\\\hline
O&	54.10&	53.32, 54.89&	48.45&	47.92, 48.97&	0.0001\\\hline
A&	48.55&	47.64, 49.45&	50.55&	50.03, 51.07&	0.0002\\\hline
C&	47.02&	46.17, 47.87&	51.13&	50.61, 51.65&	0.0001\\\hline

\multicolumn{6}{|c|}{\bf{Heroin}}\\\hline
\# &\multicolumn{2}{c}{118}&\multicolumn{2}{|c|}{1767}& \\\hline

N&	55.37&	53.63, 57.12&	49.64&	49.18, 50.10&	0.0001\\\hline
E&	47.68&	45.68, 49.69&	50.15&	49.69, 50.62&	0.0192\\\hline
O&	53.45&	51.80, 55.11&	49.77&	49.30, 50.24&	0.0001\\\hline
A&	44.37&	42.25, 46.50&	50.38&	49.92, 50.83&	0.0001\\\hline
C&	45.45&	43.52, 47.39&	50.30&	49.84, 50.77&	0.0001\\\hline

\multicolumn{6}{|c|}{\bf{Ketamine}}\\\hline
\# &\multicolumn{2}{c}{208}&\multicolumn{2}{|c|}{1677}& \\\hline

N&	51.52&	50.20, 52.84&	49.81&	49.33, 50.29&	0.0175\\\hline
E&	51.02&	49.48, 52.57&	49.87&	49.40, 50.34&	0.1619\\\hline
O&	54.18&	52.88, 55.49&	49.48&	49.00, 49.96&	0.0001\\\hline
A&	47.90&	46.42, 49.38&	50.26&	49.79, 50.73&	0.0030\\\hline
C&	46.34&	44.91, 47.76&	50.45&	49.98, 50.93&	0.0001\\\hline

\multicolumn{6}{|c|}{\bf{Legal highs}}\\\hline
\# &\multicolumn{2}{c}{564}&\multicolumn{2}{|c|}{1321}& \\\hline	

N&	51.27&	50.41, 52.12&	49.46&	48.93, 49.99&	0.0004\\\hline
E&	50.02&	49.11, 50.92&	49.99&	49.48, 50.51&	0.9659\\\hline
O&	54.49&	53.76, 55.22&	48.08&	47.55, 48.62&	0.0001\\\hline
A&	48.09&	47.23, 48.96&	50.81&	50.29, 51.34&	0.0001\\\hline
C&	46.56&	45.71, 47.41&	51.47&	50.95, 51.98&	0.0001\\\hline

\multicolumn{6}{|c|}{\bf{LSD}}\\\hline
\# &\multicolumn{2}{c}{380}&\multicolumn{2}{|c|}{1505}& \\\hline

N&	49.98&	48.97, 50.99&	50.00&	49.50, 50.51&	0.9691\\\hline
E&	50.72&	49.63, 51.81&	49.82&	49.32, 50.31&	0.1403\\\hline
O&	56.29&	55.48, 57.11&	48.41&	47.91, 48.91&	0.0001\\\hline
A&	49.05&	47.98, 50.12&	50.24&	49.74, 50.74&	0.0484\\\hline
C&	47.74&	46.71, 48.77&	50.57&	50.07, 51.07&	0.0001\\\hline

\multicolumn{6}{|c|}{\bf{Methadone}}\\\hline
\# &\multicolumn{2}{c}{320}&\multicolumn{2}{|c|}{1565}& \\\hline

N&	53.74&	52.65, 54.84&	49.23&	48.75, 49.72&	0.0001\\\hline
E&	47.75&	46.46, 49.03&	50.46&	49.99, 50.94&	0.0001\\\hline
O&	53.81&	52.76, 54.86&	49.22&	48.73, 49.71&	0.0001\\\hline
A&	46.53&	45.33, 47.73&	50.71&	50.23, 51.19&	0.0001\\\hline
C&	46.01&	44.87, 47.15&	50.82&	50.33, 51.30&	0.0001\\\hline

 \multicolumn{6}{|c|}{\bf{Magic Mushrooms}}\\\hline
 \# &\multicolumn{2}{c}{434}&\multicolumn{2}{|c|}{1451}& \\\hline
	
N&	50.33&	49.40, 51.26&	49.90&	49.38, 50.42&	0.4311\\\hline
E&	50.71&	49.66, 51.76&	49.79&	49.29, 50.28&	0.1179\\\hline
O&	55.72&	54.96, 56.49&	48.29&	47.78, 48.80&	0.0001\\\hline
A&	48.51&	47.48, 49.53&	50.45&	49.95, 50.95&	0.0009\\\hline
C&	47.30&	46.36, 48.24&	50.81&	50.30, 51.32&	0.0001\\\hline

 \multicolumn{6}{|c|}{\bf{Nicotine}}\\\hline
 \# &\multicolumn{2}{c}{1060}&\multicolumn{2}{|c|}{825}& \\\hline

N&	51.16&	50.55, 51.76&	48.51&	47.85, 49.18&	0.0001\\\hline
E&	49.80&	49.17, 50.42&	50.26&	49.61, 50.91&	0.3094\\\hline
O&	51.91&	51.31, 52.51&	47.55&	46.89, 48.20&	0.0001\\\hline
A&	49.04&	48.42, 49.65&	51.24&	50.58, 51.90&	0.0001\\\hline
C&	47.98&	47.38, 48.57&	52.60&	51.95, 53.25&	0.0001\\\hline

 \multicolumn{6}{|c|}{\bf{VSA}}\\\hline
 \# &\multicolumn{2}{c}{95}&\multicolumn{2}{|c|}{1790}& \\\hline

N&	53.81&	51.67, 55.95&	49.80&	49.34, 50.26&	0.0004\\\hline
E&	49.55&	47.32, 51.79&	50.02&	49.56, 50.49&	0.6832\\\hline
O&	53.59&	51.58, 55.60&	49.81&	49.35, 50.27&	0.0004\\\hline
A&	47.48&	45.35, 49.61&	50.13&	49.67, 50.60&	0.0174\\\hline
C&	45.31&	43.18, 47.44&	50.25&	49.79, 50.71&	0.0001\\\hline

\end{longtable}
\end{adjustwidth}
\end{center}

\begin{center}
\setlength\LTleft{-0.8in}
\setlength\LTcapwidth{11.7in}
\begin{adjustwidth}{-2.25in}{0in}
\begin{longtable}{|c|c|c|c|c|c|}
\caption{ Mean T-$score_{sample}$ and 95\% CI for it for groups of users and non-users with month based problem}
\label{table4b}
\\\hline
{\bf Factor} &\multicolumn{2}{c|}{\bf Users}&\multicolumn{2}{c|}{\bf Non-users}& {\bf $p$-value}\\\cline{2-5}
    &{  Mean T-score}& {  95\% CI for mean}& { Mean T-score}&{ 95\% CI for mean}&   \\ \hline
\endfirsthead
\multicolumn{6}{c} {\tablename\ \thetable.\textit{ Continued}} \\\hline
 {\bf Factor} &\multicolumn{2}{c|}{\bf Users}&\multicolumn{2}{c|}{\bf Non-users}& {\bf $p$-value}\\\cline{2-5}
   &Mean T-score& 95\% CI for mean& Mean T-score&95\% CI for mean &\\\hline
\endhead
\hline \multicolumn{6}{r}{\textit{\footnotesize{Continued on the next page}}} \\
\endfoot\hline
\endlastfoot
\multicolumn{6}{|c|}{\bf{Alcohol}}\\\hline
 \# &\multicolumn{2}{c}{1551}&\multicolumn{2}{|c|}{334}&    \\\hline
N&	49.85&	49.35, 50.35&	50.69&	49.61, 51.78&	0.1669\\\hline
E&	50.62&	50.12, 51.11&	47.14&	46.11, 48.17&	0.0001\\\hline
O&	50.22&	49.73, 50.72&	48.96&	47.87, 50.06&	0.0401\\\hline
A&	50.10&	49.60, 50.60&	49.55&	48.50, 50.60&	0.3575\\\hline
C&	50.15&	49.66, 50.65&	49.28&	48.22, 50.35&	0.1460\\\hline

 \multicolumn{6}{|c|}{\bf{Amphetamines}}\\\hline
 \# &\multicolumn{2}{c}{238}&\multicolumn{2}{|c|}{1647}&    \\\hline

N&	52.78&	51.49, 54.07&	49.60&	49.12, 50.08&	0.0001\\\hline
E&	49.07&	47.63, 50.51&	50.13&	49.66, 50.61&	0.1665\\\hline
O&	52.94&	51.69, 54.20&	49.57&	49.09, 50.06&	0.0001\\\hline
A&	46.57&	45.21, 47.92&	50.50&	50.02, 50.97&	0.0001\\\hline
C&	45.06&	43.75, 46.37&	50.71&	50.24, 51.19&	0.0001\\\hline

\multicolumn{6}{|c|}{\bf{Amyl nitrite}}\\\hline
\# &\multicolumn{2}{c}{41}&\multicolumn{2}{|c|}{1844}& \\\hline

N&	49.36&	46.12, 52.61&	50.01&	49.56, 50.47&	0.6911\\\hline
E&	49.83&	46.38, 53.28&	50.00&	49.55, 50.46&	0.9218\\\hline
O&	50.40&	47.06, 53.74&	49.99&	49.53, 50.45&	0.8083\\\hline
A&	45.43&	41.87, 49.00&	50.10&	49.65, 50.56&	0.0122\\\hline
C&	47.31&	44.60, 50.01&	50.06&	49.60, 50.52&	0.0490\\\hline

\multicolumn{6}{|c|}{\bf{Benzodiazepines}}\\\hline
 \# &\multicolumn{2}{c}{299}&\multicolumn{2}{|c|}{1586}& \\\hline

N&	55.26&	54.11, 56.41&	49.01&	48.53, 49.48&	0.0001\\\hline
E&	48.06&	46.79, 49.33&	50.37&	49.89, 50.84&	0.0010\\\hline
O&	52.68&	51.59, 53.77&	49.49&	49.00, 49.99&	0.0001\\\hline
A&	46.72&	45.47, 47.97&	50.62&	50.14, 51.10&	0.0001\\\hline
C&	46.54&	45.40, 47.69&	50.65&	50.17, 51.14&	0.0001\\\hline

\multicolumn{6}{|c|}{\bf{Cannabis}}\\\hline
 \# &\multicolumn{2}{c}{788}&\multicolumn{2}{|c|}{1097}& \\\hline

N&	50.72&	49.99, 51.45&	49.48&	48.91, 50.06&	0.7142\\\hline
E&	50.06&	49.31, 50.80&	49.96&	49.40, 50.52&	0.5738\\\hline
O&	54.34&	53.74, 54.95&	46.88&	46.30, 47.46&	0.3233\\\hline
A&	48.66&	47.93, 49.38&	50.97&	50.39, 51.54&	0.5477\\\hline
C&	47.26&	46.55, 47.97&	51.97&	51.41, 52.53&	0.3800\\\hline

\multicolumn{6}{|c|}{\bf{Chocolate}}\\\hline
 \# &\multicolumn{2}{c}{1786}&\multicolumn{2}{|c|}{99}& \\\hline

N& 49.98&	49.52, 50.44&	50.40&	48.19, 52.60&	0.0002\\\hline
E& 50.03&	49.57, 50.49&	49.43&	47.38, 51.49&	0.1295\\\hline
O& 50.05&	49.58, 50.52&	49.13&	47.36, 50.90&	0.0008\\\hline
A &50.04&	49.57, 50.50&	49.36&	47.19, 51.53&	0.0001\\\hline
C& 49.95&	49.49, 50.41&	50.90&	48.82, 52.97&	0.0002\\\hline

\multicolumn{6}{|c|}{\bf{Cocaine}}\\\hline
\# &\multicolumn{2}{c}{159}&\multicolumn{2}{|c|}{1726}& \\\hline
N&	52.87&	51.29, 54.46&	49.74&	49.27, 50.21&	0.3787\\\hline
E&	51.34&	49.50, 53.18&	49.88&	49.41, 50.34&	0.0078\\\hline
O&	52.53&	51.00, 54.06&	49.77&	49.29, 50.24&	0.0560\\\hline
A&	45.75&	44.05, 47.46&	50.39&	49.93, 50.86&	0.7537\\\hline
C&	47.23&	45.71, 48.75&	50.25&	49.78, 50.73&	0.4728\\\hline

\multicolumn{6}{|c|}{\bf{Caffeine}}\\\hline
 \# &\multicolumn{2}{c}{1764}&\multicolumn{2}{|c|}{121}& \\\hline

N&	50.05&	49.59, 50.52&	49.21&	47.37, 51.05&	0.0092\\\hline
E&	50.16&	49.69, 50.63&	47.68&	45.93, 49.43&	0.8390\\\hline
O&	50.12&	49.65, 50.58&	48.29&	46.47, 50.11&	0.0001\\\hline
A&	49.98&	49.51, 50.45&	50.27&	48.50, 52.04&	0.0001\\\hline
C&	49.95&	49.49, 50.42&	50.67&	48.77, 52.56&	0.0001\\\hline

\multicolumn{6}{|c|}{\bf{Crack}}\\\hline
 \# &\multicolumn{2}{c}{20}&\multicolumn{2}{|c|}{1865}& \\\hline

N&	57.86&	52.26, 63.46&	49.92&	49.46, 50.37&	0.1931\\\hline
E&	45.97&	41.30, 50.64&	50.04&	49.59, 50.50&	0.0852\\\hline
O&	50.89&	47.21, 54.57&	49.99&	49.54, 50.45&	0.6165\\\hline
A&	42.98&	36.34, 49.62&	50.08&	49.62, 50.53&	0.0380\\\hline
C&	45.14&	40.37, 49.91&	50.05&	49.60, 50.51&	0.0448\\\hline

\multicolumn{6}{|c|}{\bf{Ecstasy}}\\\hline
\# &\multicolumn{2}{c}{240}&\multicolumn{2}{|c|}{1645}& \\\hline

N&	49.53&	48.18, 50.89&	50.07&	49.59, 50.55&	0.4649\\\hline
E&	52.24&	50.83, 53.65&	49.67&	49.20, 50.15&	0.0008\\\hline
O&	54.41&	53.23, 55.59&	49.36&	48.88, 49.84&	0.0001\\\hline
A&	48.10&	46.75, 49.45&	50.28&	49.80, 50.76&	0.0029\\\hline
C&	47.27&	45.96, 48.59&	50.40&	49.92, 50.88&	0.0001\\\hline

\multicolumn{6}{|c|}{\bf{Heroin}}\\\hline
\# &\multicolumn{2}{c}{53}&\multicolumn{2}{|c|}{1832}& \\\hline

N&	56.70&	54.05, 59.34&	49.81&	49.35, 50.26&	0.0001\\\hline
E&	45.58&	42.06, 49.10&	50.13&	49.67, 50.58&	0.0130\\\hline
O&	52.48&	49.63, 55.34&	49.93&	49.47, 50.39&	0.0823\\\hline
A&	42.18&	39.00, 45.35&	50.23&	49.77, 50.68&	0.0001\\\hline
C&	43.36&	40.35, 46.37&	50.19&	49.74, 50.65&	0.0001\\\hline

\multicolumn{6}{|c|}{\bf{Ketamine}}\\\hline
\# &\multicolumn{2}{c}{79}&\multicolumn{2}{|c|}{1806}& \\\hline
N&	51.29&	49.14, 53.45&	49.94&	49.48, 50.41&	0.2270\\\hline
E&	49.62&	46.74, 52.49&	50.02&	49.56, 50.47&	0.7851\\\hline
O&	54.79&	52.59, 56.98&	49.79&	49.33, 50.25&	0.0001\\\hline
A&	46.90&	44.15, 49.66&	50.14&	49.68, 50.59&	0.0237\\\hline
C&	45.03&	42.50, 47.56&	50.22&	49.76, 50.67&	0.0001\\\hline

\multicolumn{6}{|c|}{\bf{Legal highs}}\\\hline
\# &\multicolumn{2}{c}{241}&\multicolumn{2}{|c|}{1644}& \\\hline	

N&	52.02&	50.68, 53.36&	49.70&	49.23, 50.18&	0.0015\\\hline
E&	49.10&	47.59, 50.61&	50.13&	49.66, 50.60&	0.2002\\\hline
O&	54.37&	53.22, 55.53&	49.36&	48.88, 49.84&	0.0001\\\hline
A&	46.83&	45.50, 48.16&	50.46&	49.99, 50.94&	0.0001\\\hline
C&	45.30&	44.01, 46.60&	50.69&	50.21, 51.16&	0.0001\\\hline

\multicolumn{6}{|c|}{\bf{LSD}}\\\hline
\# &\multicolumn{2}{c}{166}&\multicolumn{2}{|c|}{1719}& \\\hline

N&	50.55&	48.97, 52.12&	49.95&	49.48, 50.42&	0.4717\\\hline
E&	51.28&	49.53, 53.04&	49.88&	49.41, 50.34&	0.1279\\\hline
O&	57.28&	56.16, 58.41&	49.30&	48.83, 49.77&	0.0001\\\hline
A&	48.92&	47.35, 50.48&	50.10&	49.63, 50.58&	0.1533\\\hline
C&	47.10&	45.59, 48.60&	50.28&	49.81, 50.75&	0.0001\\\hline

\multicolumn{6}{|c|}{\bf{Methadone}}\\\hline
\# &\multicolumn{2}{c}{171}&\multicolumn{2}{|c|}{1714}& \\\hline

N&	54.53&	53.00, 56.06&	49.55&	49.08, 50.02&	0.0001\\\hline
E&	46.86&	45.07, 48.65&	50.31&	49.85, 50.78&	0.0003\\\hline
O&	52.89&	51.37, 54.40&	49.71&	49.24, 50.18&	0.0001\\\hline
A&	46.18&	44.50, 47.87&	50.38&	49.92, 50.85&	0.0001\\\hline
C&	45.44&	43.83, 47.05&	50.45&	49.99, 50.92&	0.0001\\\hline

 \multicolumn{6}{|c|}{\bf{Magic Mushrooms}}\\\hline
 \# &\multicolumn{2}{c}{159}&\multicolumn{2}{|c|}{1726}& \\\hline
N&	49.91&	48.41, 51.42&	50.01&	49.53, 50.48&	0.9064\\\hline
E&	50.31&	48.50, 52.12&	49.97&	49.51, 50.44&	0.7205\\\hline
O&	56.92&	55.77, 58.07&	49.36&	48.89, 49.83&	0.0001\\\hline
A&	48.57&	46.94, 50.19&	50.13&	49.66, 50.60&	0.0699\\\hline
C&	46.85&	45.25, 48.45&	50.29&	49.82, 50.76&	0.0001\\\hline

 \multicolumn{6}{|c|}{\bf{Nicotine}}\\\hline
 \# &\multicolumn{2}{c}{875}&\multicolumn{2}{|c|}{1010}& \\\hline
N&	51.11&	50.43, 51.79&	49.04&	48.44, 49.64&	0.0001\\\hline
E&	49.98&	49.29, 50.66&	50.02&	49.42, 50.62&	0.9237\\\hline
O&	51.86&	51.19, 52.52&	48.39&	47.79, 48.99&	0.0001\\\hline
A&	49.02&	48.34, 49.71&	50.85&	50.25, 51.44&	0.0001\\\hline
C&	47.69&	47.03, 48.34&	52.01&	51.41, 52.60&	0.0001\\\hline

 \multicolumn{6}{|c|}{\bf{VSA}}\\\hline
 \# &\multicolumn{2}{c}{34}&\multicolumn{2}{|c|}{1851}& \\\hline
N&	51.34&	47.48, 55.20&	49.98&	49.52, 50.43&	0.4793\\\hline
E&	51.80&	48.00, 55.60&	49.97&	49.51, 50.42&	0.3378\\\hline
O&	54.65&	51.65, 57.66&	49.91&	49.46, 50.37&	0.0032\\\hline
A&	45.91&	42.12, 49.71&	50.08&	49.62, 50.53&	0.0336\\\hline
C&	47.22&	43.63, 50.81&	50.05&	49.60, 50.51&	0.1209\\\hline
\end{longtable}
\end{adjustwidth}
\end{center}

\begin{center}
\setlength\LTleft{-0.8in}
\setlength\LTcapwidth{11.7in}
\begin{adjustwidth}{-2.25in}{0in}
\begin{longtable}{|c|c|c|c|c|c|}
\caption{ Mean T-$score_{sample}$ and 95\% CI for it for groups of users and non-users with week based definition}
\label{table4c}
\\\hline
{\bf Factor} &\multicolumn{2}{c|}{\bf Users}&\multicolumn{2}{c|}{\bf Non-users}& {\bf $p$-value}\\\cline{2-5}
    &{  Mean T-score}& {  95\% CI for mean}& { Mean T-score}&{ 95\% CI for mean}&   \\ \hline
\endfirsthead
\multicolumn{6}{c} {\tablename\ \thetable.\textit{ Continued}} \\\hline
 {\bf Factor} &\multicolumn{2}{c|}{\bf Users}&\multicolumn{2}{c|}{\bf Non-users}& {\bf $p$-value}\\\cline{2-5}
   &Mean T-score& 95\% CI for mean& Mean T-score&95\% CI for mean &\\\hline
\endhead
\hline \multicolumn{6}{r}{\textit{\footnotesize{Continued on the next page}}} \\
\endfoot\hline
\endlastfoot
\multicolumn{6}{|c|}{\bf{Alcohol}}\\\hline
 \# &\multicolumn{2}{c}{1264}&\multicolumn{2}{|c|}{621}&    \\\hline

N&	49.82&	49.28, 50.36&	50.37&	49.56, 51.19&	0.2673\\\hline
E&	50.90&	50.35, 51.44&	48.17&	47.38, 48.97&	0.0001\\\hline
O&	50.08&	49.54, 50.63&	49.83&	49.03, 50.62&	0.6023\\\hline
A&	50.05&	49.50, 50.60&	49.89&	49.09, 50.69&	0.7454\\\hline
C&	50.19&	49.64, 50.74&	49.61&	48.81, 50.42&	0.2441\\\hline

 \multicolumn{6}{|c|}{\bf{Amphetamines}}\\\hline
 \# &\multicolumn{2}{c}{163}&\multicolumn{2}{|c|}{1722}&    \\\hline

N&	52.86	&51.27, 54.45&	49.73&	49.26, 50.20&	0.0003\\\hline
E&	48.50&	46.78, 50.21&	50.14&	49.68, 50.61&	0.0695\\\hline
O&	52.87&	51.27, 54.47&	49.73&	49.26, 50.20&	0.0003\\\hline
A&	46.77&	45.09, 48.44&	50.31&	49.84, 50.77&	0.0001\\\hline
C&	45.29&	43.72, 46.85&	50.45&	49.98, 50.91&	0.0001\\\hline

\multicolumn{6}{|c|}{\bf{Amyl nitrite}}\\\hline
\# &\multicolumn{2}{c}{17}&\multicolumn{2}{|c|}{1868}& \\\hline
N&	49.64&	43.98, 55.29&	50.00&	49.55, 50.46&	0.8922\\\hline
E&	45.85&	39.05, 52.64&	50.04&	49.59, 50.49&	0.2106\\\hline
O&	49.02&	43.02, 55.02&	50.01&	49.56, 50.46&	0.7320\\\hline
A&	44.36&	38.95, 49.77&	50.05&	49.60, 50.50&	0.0409\\\hline
C&	44.81&	39.74, 49.89&	50.05&	49.59, 50.50&	0.0447\\\hline

\multicolumn{6}{|c|}{\bf{Benzodiazepines}}\\\hline
 \# &\multicolumn{2}{c}{179}&\multicolumn{2}{|c|}{1706}& \\\hline

N&	56.56&	55.08, 58.05&	49.31&	48.85, 49.77&	0.0001\\\hline
E&	46.15&	44.61, 47.70&	50.40&	49.93, 50.87&	0.0001\\\hline
O&	52.18&	50.72, 53.64&	49.77&	49.30, 50.25&	0.0022\\\hline
A&	46.57&	44.94, 48.21&	50.36&	49.89, 50.83&	0.0001\\\hline
C&	46.20&	44.75, 47.65&	50.40&	49.93, 50.87&	0.0001\\\hline

\multicolumn{6}{|c|}{\bf{Cannabis}}\\\hline
 \# &\multicolumn{2}{c}{648}&\multicolumn{2}{|c|}{1237}& \\\hline

N&	50.62&	49.82, 51.42&	49.68&	49.13, 50.22&	0.0555\\\hline
E&	50.17&	49.36, 50.97&	49.91&	49.37, 50.46&	0.6056\\\hline
O&	54.70&	54.05, 55.35&	47.54&	46.99, 48.09&	0.0001\\\hline
A&	48.78&	48.00, 49.56&	50.64&	50.09, 51.19&	0.0001\\\hline
C&	47.45&	46.67, 48.23&	51.34&	50.80, 51.88&	0.0001\\\hline

\multicolumn{6}{|c|}{\bf{Chocolate}}\\\hline
 \# &\multicolumn{2}{c}{1490}&\multicolumn{2}{|c|}{395}& \\\hline

N&	49.95&	49.44, 50.45&	50.20&	49.18, 51.23&	0.6601\\\hline
E&	50.22&	49.72, 50.73&	49.16&	48.14, 50.17&	0.0648\\\hline
O&	49.89&	49.37, 50.40&	50.42&	49.48, 51.36&	0.3270\\\hline
A&	50.20&	49.70, 50.70&	49.23&	48.19, 50.27&	0.0988\\\hline
C&	50.13&	49.62, 50.63&	49.51&	48.50, 50.53&	0.2864\\\hline

\multicolumn{6}{|c|}{\bf{Cocaine}}\\\hline
\# &\multicolumn{2}{c}{60}&\multicolumn{2}{|c|}{1825}& \\\hline

N&	53.24&	50.50, 55.99&	49.89&	49.44, 50.35&	0.0191\\\hline
E&	52.03&	49.22, 54.84&	49.93&	49.48, 50.39&	0.1464\\\hline
O&	51.40&	48.83, 53.97&	49.95&	49.49, 50.41&	0.2727\\\hline
A&	43.73&	40.79, 46.68&	50.21&	49.75, 50.66&	0.0001\\\hline
C&	46.72&	44.32, 49.11&	50.11&	49.65, 50.57&	0.0072\\\hline

\multicolumn{6}{|c|}{\bf{Caffeine}}\\\hline
 \# &\multicolumn{2}{c}{1658}&\multicolumn{2}{|c|}{227}& \\\hline

N&	50.07&	49.59, 50.55&	49.46&	48.10, 50.81&	0.3998\\\hline
E&	50.20&	49.72, 50.68&	48.54&	47.27, 49.81&	0.0166\\\hline
O&	50.04&	49.56, 50.52&	49.71&	48.40, 51.03&	0.6450\\\hline
A&	49.93&	49.45, 50.41&	50.48&	49.15, 51.81&	0.4462\\\hline
C&	49.95&	49.47, 50.43&	50.37&	49.01, 51.72&	0.5690\\\hline

\multicolumn{6}{|c|}{\bf{Crack}}\\\hline
 \# &\multicolumn{2}{c}{11}&\multicolumn{2}{|c|}{1874}& \\\hline
N&	55.26&	45.53, 64.99&	49.97&	49.52, 50.42&	0.2540\\\hline
E&	46.46&	38.43, 54.50&	50.02&	49.57, 50.47&	0.3480\\\hline
O&	48.01&	43.74, 52.28&	50.01&	49.56, 50.47&	0.3242\\\hline
A&	39.61&	30.12, 49.11&	50.06&	49.61, 50.51&	0.0343\\\hline
C&	44.28&	36.73, 51.83&	50.03&	49.58, 50.49&	0.1211\\\hline

\multicolumn{6}{|c|}{\bf{Ecstasy}}\\\hline
\# &\multicolumn{2}{c}{84}&\multicolumn{2}{|c|}{1801}& \\\hline

N&	50.28	&47.99, 52.58&	49.99	&49.53, 50.45&	0.8031\\\hline
E&	53.37&	50.71, 56.03&	49.84&	49.39, 50.30&	0.0110\\\hline
O&	56.15&	54.15, 58.15&	49.71&	49.25, 50.17&	0.0001\\\hline
A&	48.56&	46.38, 50.75&	50.07&	49.61, 50.53&	0.1837\\\hline
C&	46.98&	44.78, 49.18	&50.14&	49.68, 50.60&	0.0065\\\hline

\multicolumn{6}{|c|}{\bf{Heroin}}\\\hline
\# &\multicolumn{2}{c}{29}&\multicolumn{2}{|c|}{1856}& \\\hline

N&	58.66&	55.84, 61.47&	49.86&	49.41, 50.32&	0.0001\\\hline
E&	44.77&	39.80, 49.73&	50.08&	49.63, 50.53&	0.0376\\\hline
O&	52.41&	48.00, 56.81&	49.96&	49.51, 50.42&	0.2681\\\hline
A&	41.48&	38.25, 44.70&	50.13&	49.68, 50.59&	0.0001\\\hline
C&	43.04&	38.95, 47.12&	50.11&	49.66, 50.56&	0.0015\\\hline

\multicolumn{6}{|c|}{\bf{Ketamine}}\\\hline
\# &\multicolumn{2}{c}{37}&\multicolumn{2}{|c|}{1848}& \\\hline
N&	50.47&	46.78, 54.16&	49.99&	49.54, 50.45&	0.7952\\\hline
E&	47.23&	42.01, 52.46&	50.06&	49.61, 50.51&	0.2822\\\hline
O&	54.39&	50.87, 57.90&	49.91&	49.46, 50.37&	0.0148\\\hline
A&	44.50&	40.15, 48.85&	50.11&	49.66, 50.56&	0.0133\\\hline
C&	44.99&	41.13, 48.85&	50.10&	49.65, 50.55&	0.0113\\\hline

\multicolumn{6}{|c|}{\bf{Legal highs}}\\\hline
\# &\multicolumn{2}{c}{131}&\multicolumn{2}{|c|}{1754}& \\\hline	
N&	53.13&	51.32, 54.94&	49.77&	49.30, 50.23&	0.0005\\\hline
E&	47.12&	45.06, 49.18&	50.22&	49.76, 50.67&	0.0044\\\hline
O&	53.16&	51.53, 54.79&	49.76&	49.30, 50.23&	0.0001\\\hline
A&	46.25&	44.62, 47.87&	50.28&	49.81, 50.75&	0.0001\\\hline
C&	44.50&	42.88, 46.11&	50.41&	49.95, 50.88&	0.0001\\\hline

\multicolumn{6}{|c|}{\bf{LSD}}\\\hline
\# &\multicolumn{2}{c}{69}&\multicolumn{2}{|c|}{1816}& \\\hline

N&	50.28&	47.82, 52.73&	49.99&	49.53, 50.45&	0.8196\\\hline
E&	52.70&	49.90, 55.51&	49.90&	49.44, 50.35&	0.0531\\\hline
O&	57.57&	55.78, 59.35&	49.71&	49.25, 50.17&	0.0001\\\hline
A&	50.03&	47.88, 52.17&	50.00&	49.54, 50.46&	0.9793\\\hline
C&	46.98&	44.90, 49.06&	50.11&	49.65, 50.58&	0.0045\\\hline

\multicolumn{6}{|c|}{\bf{Methadone}}\\\hline
\# &\multicolumn{2}{c}{121}&\multicolumn{2}{|c|}{1764}& \\\hline

N&	54.99&	53.14, 56.84&	49.66&	49.20, 50.12&	0.0001\\\hline
E&	45.27&	43.10, 47.43&	50.32&	49.87, 50.78&	0.0001\\\hline
O&	51.87&	50.02, 53.72&	49.87&	49.41, 50.34&	0.0406\\\hline
A&	46.00&	43.84, 48.15&	50.27&	49.82, 50.73&	0.0002\\\hline
C&	45.74&	43.89, 47.59&	50.29&	49.83, 50.76&	0.0001\\\hline

 \multicolumn{6}{|c|}{\bf{Magic Mushrooms}}\\\hline
 \# &\multicolumn{2}{c}{44}&\multicolumn{2}{|c|}{1841}& \\\hline
N&	49.79&	46.82, 52.75&	50.01&	49.55, 50.46&	0.8844\\\hline
E&	53.71&	50.31, 57.12&	49.91&	49.46, 50.37&	0.0309\\\hline
O&	57.89&	55.73, 60.05&	49.81&	49.35, 50.27&	0.0001\\\hline
A&	50.14&	46.71, 53.57&	50.00&	49.54, 50.45&	0.9352\\\hline
C&	48.03&	45.33, 50.74&	50.05&	49.59, 50.51&	0.1459\\\hline
	
 \multicolumn{6}{|c|}{\bf{Nicotine}}\\\hline
 \# &\multicolumn{2}{c}{767}&\multicolumn{2}{|c|}{1118}& \\\hline

N&	51.32&	50.59, 52.04&	49.10&	48.52, 49.67&	0.0001\\\hline
E&	49.91&	49.17, 50.65&	50.06&	49.49, 50.63&	0.7485\\\hline
O&	51.57&	50.85, 52.28&	48.92&	48.35, 49.50&	0.0001\\\hline
A&	49.04&	48.31, 49.78&	50.66&	50.09, 51.23&	0.0007\\\hline
C&	47.69&	46.99, 48.38&	51.59&	51.01, 52.16&	0.0001\\\hline

 \multicolumn{6}{|c|}{\bf{VSA}}\\\hline
 \# &\multicolumn{2}{c}{21}&\multicolumn{2}{|c|}{1864}& \\\hline

N&	50.92&	46.02, 55.82&	49.99&	49.54, 50.44&	0.6974\\\hline
E&	52.60&	47.66, 57.53&	49.97&	49.52, 50.42&	0.2828\\\hline
O&	56.30&	53.63, 58.97&	49.93&	49.47, 50.38&	0.0001\\\hline
A&	46.36&	41.38, 51.34&	50.04&	49.59, 50.49&	0.1407\\\hline
C&	49.99&	45.49, 54.49&	50.00&	49.55, 50.45&	0.9955\\\hline
\end{longtable}
\end{adjustwidth}
\end{center}

\newpage
\paragraph*{S2 Appendix.}
\label{S2_Appendix}
{\bf PCCs between drug consumptions.} In this section we show PCCs between drug consumptions for  decade and year based user/non user separation.
\begin{landscape}
\begin{table}[!ht]
\begin{adjustwidth}{-.44in}{0in}
\footnotesize{
\caption {PCCs between drug consumptions with decade based user/non user separation}
\label{tab:14}
\begin{tabular}{|l@{\;}|@{\;}r@{\;}|r@{\;}|r@{\;}|r@{\;}|r@{\;}|@{\;}r@{\;}|r@{\;}|r@{\;}|r@{\;}|r@{\;}|@{\;}r@{\;}|r@{\;}|r@{\;}|r@{\;}|@{\;}r@{\;}|r@{\;}|r@{\;}|@{\;}r@{\;}|}
\hline
\multicolumn{1}{|c}{\rotatebox{90}{Drug }}&\multicolumn{1}{|c}{\rotatebox{90}{Alcohol}}&\multicolumn{1}{|c}{\rotatebox{90}{Amphetamines}}&\multicolumn{1}{|c}{\rotatebox{90}{Amyl nitrite}}&\multicolumn{1}{|c}{\rotatebox{90}{Benzodiazepines}}&\multicolumn{1}{|c}{\rotatebox{90}{Cannabis}}&\multicolumn{1}{|c}{\rotatebox{90}{	Chocolate}}&\multicolumn{1}{|c}{\rotatebox{90}{Cocaine}}&\multicolumn{1}{|c}{\rotatebox{90}{	Caffeine}}&\multicolumn{1}{|c}{\rotatebox{90}{	Crack}}&\multicolumn{1}{|c}{\rotatebox{90}{	Ecstasy}}&	\multicolumn{1}{|c}{\rotatebox{90}{Heroin}}&\multicolumn{1}{|c}{\rotatebox{90}{	Ketamine}}&\multicolumn{1}{|c}{\rotatebox{90}{	Legal highs}}&\multicolumn{1}{|c}{\rotatebox{90}{	LSD}}&\multicolumn{1}{|c}{\rotatebox{90}{	Methadone}}&\multicolumn{1}{|c}{\rotatebox{90}{MMushrooms}}&\multicolumn{1}{|c}{\rotatebox{90}{	Nicotine}}&\multicolumn{1}{|c|}{\rotatebox{90}{	VSA}} \\\hline
Alcohol &        	&$0.074^1$&	$0.074^1$&	$0.051^2$&	$0.119^1$&	$0.099^1$&	$0.111^1$&	$0.157^1$&	$0.027^4$&	$0.105^1$&	$0.033^4$&	$0.078^2$&	$0.061^2$&	$0.069^2$&	$-0.007^4$&	$0.071^1$	&$0.113^1$&	$0.046^3$ \\ \hline
Amphetamines  & $0.074^1$&	        &	$0.372^1$	&$0.463^1$	&$0.469^1$&	$0.013^4$&$0.580^1$&	$0.106^1$&	$0.323^1$& $0.597^1$&	$0.359^1$& 	$0.412^1$ &	 $0.481^1$ &	$0.490^1$ & $0.415^1$ &	$0.481^1$ &	$0.343^1$&	$0.304^1$\\ \hline
Amyl nitrite &$0.074^1$&	$0.372^1$&	        &	$0.226^1$&	$0.292^1$&	$0.028^4$&	$0.381^1$	&$0.060^2$&	$0.144^1$&	$0.392^1$&	$0.137^1$&	$0.345^1$&	$0.268^1$&	$0.213^1$&	$0.084^1$&	$0.271^1$&	$0.196^1$&	$0.130^1$\\ \hline
Benzodiazepines&$0.051^2$& 	$0.463^1$ 	&$0.226^1$&	        &	$0.354^1$&	$0.006^4$& $0.428^1$ & $0.055^2$	&$0.326^1$&	$0.383^1$&	$0.395^1$	&$0.303^1$&	$0.348^1$&	$0.352^1$& $0.468^1$ &	$0.366^1$&	$0.260^1$&	$0.294^1$ \\ \hline
Cannabis &$0.119^1$& $0.469^1$ &	$0.292^1$&	$0.354^1$&	        &	$0.046^3$&	$0.453^1$&	$0.113^1$&	$0.216^1$& $0.521^1$&	$0.217^1$& 	$0.302^1$ & $0.526^1$& $0.421^1$ &	$0.299^1$& 	$0.497^1$ & 	$0.533^1$&$0.237^1$\\ \hline
Chocolate	&$0.099^1$&	$0.013^4$&	$0.028^4$&	$0.006^4$&	$0.046^3$&	        &	$0.006^4$&	$0.122^1$&	$0.032^4$&	$0.040^4$&	$-0.026^4$&	$0.035^4$&	$0.017^4$&	$0.029^4$&	$0.007^4$&	$0.024^4$&	$0.037^4$&	$-0.021^4$\\ \hline
Cocaine	&$0.111^1$& $0.580^1$&	$0.381^1$& $0.428^1$ & $0.453^1$ &	$0.006^4$&	        &	$0.099^1$&	$0.396^1$& $0.633^1$&	 $0.414^1$ & $0.454^1$ & 	$0.445^1$ &	 $0.442^1$ &	$0.354^1$&	$0.480^1$ &	$0.362^1$&	$0.277^1$\\ \hline
Caffeine	&$0.157^1$&	$0.106^1$&	$0.060^2$&	$0.055^3$&	$0.113^1$&	$0.122^1$&	$0.099^1$&	        &	$0.035^4$&	$0.107^1$&	$0.026^4$&	$0.058^3$&	$0.085^1$&	$0.075^1$&	$0.039^4$&	$0.100^1$&	$0.145^3$&	$0.053^3$\\ \hline
Crack&$0.027^4$&	$0.323^1$&	$0.144^1$&	$0.326^1$&	$0.216^1$&	$0.032^4$&	$0.396^1$&	$0.035^4$&         &	$0.280^1$& $0.509^1$&	$0.255^1$&	$0.203^1$&	$0.268^1$&	$0.367^1$&	$0.276^1$&	$0.191^1$&	$0.278^1$\\ \hline
Ecstasy	&$0.105^1$& $0.597^1$&	$0.392^1$&	$0.383^1$& 	$0.521^1$&	$0.040^4$& $0.633^1$&	$0.107^1$&	$0.280^1$&	        &	$0.301^1$& $0.511^1$& $0.586^1$& $0.599^1$ &	$0.315^1$& 	$0.599^1$ &	$0.370^1$&	$0.289^1$\\ \hline
Heroin&$0.033^4$&	$0.359^1$&	$0.137^1$&	$0.395^1$&	$0.217^1$&	$-0.026^4$& $0.414^1$ &	$0.026^4$& 	$0.509^1$ &	$0.301^1$&	        &	$0.274^1$&	$0.237^1$&	$0.347^1$&	$0.494^1$ &	$0.306^1$&	$0.185^1$&	$0.293^1$\\ \hline
Ketamine&$0.078^1$& $0.412^1$ &	$0.345^1$&	$0.303^1$&	$0.302^1$&	$0.035^4$& $0.454^1$ &	$0.058^3$&	$0.255^1$& 	$0.511^1$ &	$0.274^1$&	&	$0.393^1$& 	$0.462^1$ &	$0.246^1$& $0.436^1$ &	$0.243^1$&	$0.192^1$\\ \hline
Legal highs	&$0.061^2$& $0.481^1$ &	$0.268^1$&	$0.348^1$& 	$0.526^1$ &	$0.017^4$& $0.445^1$ &	$0.085^1$&	$0.203^1$& 	$0.586^1$ &	$0.237^1$&	$0.393^1$&	        & 	$0.519^1$ &	$0.334^1$& 	$0.575^1$ &	$0.364^1$&	$0.314^1$\\\hline
LSD	&$0.069^2$&	 $0.490^1$ &	$0.213^1$&	$0.352^1$& 	$0.421^1$ &	$0.029^4$& $0.442^1$ &	$0.075^1$&	$0.268^1$& 	$0.599^1$ &	$0.347^1$&	 $0.462^1$ & 	$0.519^1$ &	        &	$0.343^1$& 	$0.680^1$  &	$0.289^1$&	$0.299^1$\\ \hline
Methadone	&$-0.007^4$& $0.415^1$ &	$0.084^1$& $0.468^1$ &	$0.299^1$&	$0.007^4$&	$0.354^1$&	$0.039^4$&	$0.367^1$&	$0.315^1$& 	$0.494^1$&	$0.246^1$&	$0.334^1$&	$0.343^1$&	        &	$0.343^1$&	$0.234^1$&	$0.277^1$\\ \hline
MMushrooms	&$0.071^1$& $0.481^1$&	$0.271^1$&	$0.366^1$& $0.497^1$&	$0.024^4$& $0.480^1$&	$0.100^1$&	$0.276^1$& 	$0.599^1$ &	$0.306^1$& 	$0.436^1$& 	$0.575^1$ & 	$0.680^1$ &	$0.343^1$&	        &	$0.324^1$&	$0.253^1$\\ \hline
Nicotine	&$0.113^1$&	$0.343^1$&	$0.196^1$&	$0.260^1$& 	$0.533^1$ &	$0.037^4$&	$0.362^1$&	$0.145^1$&	$0.191^1$&	$0.370^1$&	$0.185^1$&	$0.243^1$&	$0.364^1$&	$0.289^1$&	$0.234^1$&	$0.324^1$&	        &	$0.221^1$\\ \hline
VSA &$0.046^3$&	$0.304^1$&	$0.130^1$&	$0.294^1$&	$0.237^1$&	$-0.021^4$&	$0.277^1$&	$0.053^3$&	$0.278^1$&	$0.289^1$&	$0.293^1$&	$0.192^1$&	$0.314^1$& $0.299^1$&	$0.277^1$&	$0.253^1$&	$0.221^1$&	        \\\hline
\end{tabular}
\begin{flushleft}   Note: $^1p$-value$<0.001$, $^2p$-value$<0.01$, $^3p$-value$<0.05$, $^4p$-value$>0.05$. \emph{p}-value is the probability to observe by chance the same or greater correlation for uncorrelated variables.
\end{flushleft}}
\end{adjustwidth}
\end{table}
\end{landscape}

\begin{landscape}
\begin{table}[!ht]
\begin{adjustwidth}{-.44in}{0in}
\footnotesize{
\caption {PCCs between drug consumptions with year based user definition}
\label{tab:14a}
\begin{tabular}{|l@{\;}|@{\;}r@{\;}|r@{\;}|r@{\;}|r@{\;}|r@{\;}|@{\;}r@{\;}|r@{\;}|r@{\;}|r@{\;}|r@{\;}|@{\;}r@{\;}|r@{\;}|r@{\;}|r@{\;}|@{\;}r@{\;}|r@{\;}|r@{\;}|@{\;}r@{\;}|}
\hline
\multicolumn{1}{|c}{\rotatebox{90}{Drug }}&\multicolumn{1}{|c}{\rotatebox{90}{Alcohol}}&\multicolumn{1}{|c}{\rotatebox{90}{Amphetamines}}&\multicolumn{1}{|c}{\rotatebox{90}{Amyl nitrite}}&\multicolumn{1}{|c}{\rotatebox{90}{Benzodiazepines}}&\multicolumn{1}{|c}{\rotatebox{90}{Cannabis}}&\multicolumn{1}{|c}{\rotatebox{90}{	Chocolate}}&\multicolumn{1}{|c}{\rotatebox{90}{Cocaine}}&\multicolumn{1}{|c}{\rotatebox{90}{	Caffeine}}&\multicolumn{1}{|c}{\rotatebox{90}{	Crack}}&\multicolumn{1}{|c}{\rotatebox{90}{	Ecstasy}}&	\multicolumn{1}{|c}{\rotatebox{90}{Heroin}}&\multicolumn{1}{|c}{\rotatebox{90}{	Ketamine}}&\multicolumn{1}{|c}{\rotatebox{90}{	Legal highs}}&\multicolumn{1}{|c}{\rotatebox{90}{	LSD}}&\multicolumn{1}{|c}{\rotatebox{90}{	Methadone}}&\multicolumn{1}{|c}{\rotatebox{90}{MMushrooms}}&\multicolumn{1}{|c}{\rotatebox{90}{	Nicotine}}&\multicolumn{1}{|c|}{\rotatebox{90}{	VSA}} \\\hline

Alcohol	&        	&$0.046^4$&	$0.061^3$&	$0.048^2$&	$ 0.078^3$&	$0.077^1$&	$0.124^1$&	$0.111^1$	&$0.058^1$&	$0.107^1$&	$0.030^1$&	$0.072^4$&	$0.093^1$&	$0.084^1$&	$-0.005^1$&	$0.075^4$&	$0.081^1$&$	0.055^3$\\\hline
Amphetamines&	$0.046^3$&	        &	$0.222^1$&	$0.436^1$&	$0.421^1$&	$0.003^4$	&$0.453	^1$&$0.072^2$&	$0.193^1$&	$0.461^1$&	$0.305^1$&	$0.325^1$&	$0.471^1$&	$0.392^1$&	$0.382^1$&	$0.375^1$&	$0.311^1$&	$0.173^1$\\\hline
Amyl nitrite&$	0.061^2$&	$0.222^1$&	        &	$0.199^1$&	$0.185^1$&	$0.016^4$&	$0.262^1$&	$0.050^3$&	$0.077^1$&	$0.276^1$&	$0.100^1$&	$0.280^1$&	$0.277^1$&	$0.120^1$&	$0.091^1$&	$0.159^1$&	$0.139^1$&	$0.107^1$\\\hline
Benzodiazepines&	$0.048^3$&	$0.436^1$&	$0.199^1$&	        &	$0.334^1$&	$-0.009^4$&	$0.365^1$&	$0.062^2$&	$0.232^1$&	$0.304^1$&	$0.318^1$&	$0.263^1$&	$0.318^1$&	$0.212^1$&	$0.464^1$&	$0.271^1$&	$0.261^1$&	$0.183^1$\\\hline
Cannabis&	$0.078^1$&	$0.421^1$&	$0.185^1$&	$0.334^1$&	        &	$0.020^4$&	$0.392^1$&	$0.074^2$&	$0.165^1$&	$0.484^1$&	$0.199^1$&	$0.277^1$&	$0.516^1$&	$0.433^1$&	$0.301^1$&	$0.470^1$&	$0.517^1$&	$0.164^1$\\\hline
Chocolate&	$0.077^1$&	$0.003^4$&	$0.016^4$&	$-0.009^4$&	$0.020^4$&	        &	$0.008^4$&	$0.089^1$&	$-0.037^4$&	$0.057^3$&	$-0.003^4$&	$0.000^4$&	$0.026^4$&	$0.044^4$&	$0.006^4$&	$0.019^4$&	$0.009^4$&	$-0.012^4$\\\hline
Cocaine&	$0.124^1$&	$0.453^1$&	$0.262^1$&	$0.365^1$&	$0.392^1$&	$0.008^4$&	        &	$0.069^2$&	$0.322^1$&	$0.535^1$&	$0.358^1$&	$0.379^1$&	$0.394^1$&	$0.302^1$&	$0.314^1$&	$0.346^1$&	$0.329^1$&	$0.187^1$\\\hline
Caffeine&	$0.111^1$&	$0.072^2$&	$0.050^3$&	$0.062^2$&	$0.074^2$&	$0.089^1$&	$0.069^2$&	        &	$0.023^4$&	$0.059^3$&	$0.023^4$&	$0.026^4$&	$0.067^3$&	$0.032^4$&	$0.027^4$&	$0.050^3$&	$0.105^1$	& $0.042^4$\\\hline
Crack&	$0.058^3$&	$0.193^1$&	$0.077^1$&	$0.232^1$&	$0.165^1$&	$-0.037^4$&	$0.322^1$&	$0.023^4$&		        &$0.156^1$&	$0.350^1$&	$0.180^1$&	$0.147^1$&	$0.139^1$&	$0.265^1$&	$0.181^1$&	$0.126^1$&	$0.145^1$\\\hline
Ecstasy&	$0.107^1$&	$0.461^1$&	$0.276^1$&	$0.304^1$&	$0.484^1$&	$0.057^3$&	$0.535^1$&	$0.059^1$&	$0.156^1$&	        &	$0.190^1$&	$0.455^1$&	$0.502^1$&	$0.509^1$&	$0.245^1$&	$0.480^1$&	$0.343^1$&	$0.174^1$\\\hline
Heroin&	$0.030^4$&	$0.305^1$&	$0.100^1$&	$0.318^1$&	$0.199^1$&	$-0.003^4$&	$0.358^1$&	$0.023^4$& $0.350^1$&	$0.190^1$&	        &$	0.217^1$&	$0.185^1$&$	0.170^1$&	$0.385^1$&$	0.171^1$&$	0.149^1$&	$0.121^1$\\\hline
Ketamine&	$0.072^2$&	$0.325^1$&	$0.280^1$&	$0.263^1$&	$0.277^1$&	$0.000^4$&	$0.379^1$&	$0.026^4$&	$0.180^1$&	$0.455^1$&	$0.217^1$&		&$0.373^1$&	$0.351^1$&	$0.202^1$&	$0.362^1$&	$0.222^1$	& $0.151^1$\\\hline
Legal highs&	$0.093^1$&	$0.471^1$&	$0.277^1$&	$0.318^1$&	$0.516^1$&	$0.026^4$&	$0.394^1$&	$0.067^2$&	$0.147^1$&	$0.502^1$&	$0.185^1$&	$0.373^1$&	        &	$0.434^1$&	$0.309^1$&	$0.485^1$&	$0.348^1$&	$0.220^1$\\\hline
LSD	&$0.084^1$&	$0.392^1$&$	0.120^1$&	$0.212^1$&$	0.433^1$&	$0.044^4$&	$0.302^1$&$	0.032^4$&	$0.139^1$&	$0.509^1$&	$0.170^1$&	$0.351^1$&	$0.434^1$	&	        &$0.234^1$&$0.627^1$&	$0.267^1$&	$0.174^1$\\\hline
Methadone&	$-0.005^4$&	$0.382^1$&$	0.091^1$&	$0.464^1$&	$0.301^1$&	$0.006^4$&$	0.314^1$&$	0.027^4$&$	0.265^1$&$	0.245^1$&$	0.385^1$&$	0.202^1$&$	0.309^1$&$	0.234^1$&        &$0.253^1$&$	0.211^1$&$	0.167^1$\\\hline
MMushrooms&$	0.075^2$&$	0.375^1$&$	0.159^1$&$	0.271^1$&$	0.470^1$&$	0.019^4$&$	0.346^1$&$	0.050^3$&$	0.181^1$&$	0.480^1$&$	0.171^1$&$	0.362^1$&$	0.485^1$&$	0.627^1$&$	0.253^1$&	        &$	0.282^1$	&$ 0.174^1$\\\hline
Nicotine&$	0.081^1$&$	0.311^1$&$	0.139^1$&$	0.261^1$&$	0.517^1$&$	0.009^4$&$	0.329^1$&$	0.105^1$&$	0.126^1$&$	0.343^1$&$	0.149^1$&$	0.222^1$&$	0.348^1$&$	0.267^1$&$	0.211^1$	&$0.282	^1$&        &$	0.145^1$\\\hline
VSA&$0.055^3$&	$0.173^1$&$	0.107^1$&$	0.183^1$&$	0.164^1$&	$-0.012^4$&$	0.187^1$&	$0.042^4$&$	0.145^1$&$	0.174^1$&	$0.121^1$&$	0.151^1$&$	0.220^1$&$	0.174^1$&$	0.167^1$&$	0.174^1$&$	0.145 ^1$&        \\\hline
\end{tabular}
\begin{flushleft}   Note: $^1p$-value$<0.001$, $^2p$-value$<0.01$, $^3p$-value$<0.05$, $^4p$-value$>0.05$. \emph{p}-value is the probability to observe by chance the same or greater correlation for uncorrelated variables.
\end{flushleft}}
\end{adjustwidth}
\end{table}
\end{landscape}


\end{document}